\documentclass[12pt,letterpaper]{article}
\pdfoutput=1
\usepackage{amsmath,amssymb,calc, amsthm,bbm, epsfig,psfrag, mathtools,comment}
\usepackage{mathrsfs}
\usepackage{graphicx, enumerate}
\usepackage{float}
 \usepackage[all,cmtip]{xy}
\usepackage[numbers,sort&compress]{natbib}
\usepackage[dvipsnames]{xcolor}
\usepackage{tikz}
\usepackage{tikz-cd}
\usepackage{physics}
\usepackage{tensor}
\usepackage[compat=1.1.0]{tikz-feynman}
\usepackage[utf8]{inputenc} 
\definecolor{azure}{rgb}{0.0, 0.5, 1.0}
\definecolor{darkblue}{rgb}{0.15,0.35,0.7}
\definecolor{reddish}{rgb}{0.65, 0.2, 0.2}
\definecolor{brandeisblue}{rgb}{0.0, 0.44, 1.0}
\definecolor{ceruleanblue}{rgb}{0.16, 0.32, 0.75}
\definecolor{indigo(dye)}{rgb}{0.0, 0.25, 0.42}
\usepackage[linktocpage=true]{hyperref}
\hypersetup{
colorlinks=true,
citecolor=ceruleanblue,
linkcolor=ceruleanblue,
urlcolor=ceruleanblue,
pdfauthor={},
pdftitle={},
pdfsubject={}
}

\definecolor{indigo(dye)}{rgb}{0.0, 0.25, 0.42}

\newcommand{\overbar}[1]{\mkern 1.5mu\overline{\mkern-1.5mu#1\mkern-1.5mu}\mkern 1.5mu}

\def\TT{{T\overbar{T}}}

\newcommand{\ppp}{{+++}} 

\newcommand{\mmp}{{--+}}

\newcommand{\mmmm}{{----}}
\newcommand{\ppmm}{{++--}}

\newcommand{\pp}{{++}}
\newcommand{\mm}{{--}}
\newcommand{\Th}{T^{(\text{Hilb})}} 

\newcommand\thetab{\overbar{\theta}}
\newcommand\psib{\overbar{\psi}}
\newcommand\Psib{\overbar{\Psi}}
\newcommand\Dbar{\overbar{D}}
\newcommand\etat{\widetilde{\eta}}

\newcommand\calQbar{\overbar{\mathcal{Q}}}
\newcommand\calTt{\widetilde{\mathcal{T}}}

\usepackage{amsthm}

\usepackage{cleveref}
\usepackage{bm}
\crefname{lem}{lemma}{lemmas}
\crefname{thm}{theorem}{theorems}
\crefname{cor}{corollary}{corollaries}
\crefname{rem}{remark}{remarks}
\crefname{prop}{proposition}{propositions}

\makeatletter
\renewcommand\section{\@startsection {section}{1}{\z@}%
                               {-3.5ex \@plus -1ex \@minus -.2ex}%nn
                               {2.3ex \@plus.2ex}%
                               {\normalfont\large\bfseries}}
\renewcommand\subsection{\@startsection{subsection}{2}{\z@}%
                                 {-3.25ex\@plus -1ex \@minus -.2ex}%
                                 {1.5ex \@plus .2ex}%
                                 {\normalfont\bfseries}}
\makeatother

\let\non\nonumber

\def\bea#1\eea{\begin{align}#1\end{align}}

\def\bes #1\ees{\begin{split}#1\end{split}}

\newcommand{\be}{\begin{equation}}
\newcommand{\ee}{\end{equation}}

\newfont{\goth}{ygoth.tfm scaled 1200}                   % gothic font (usual)
\renewcommand{\thesection}{\arabic{section}}

\numberwithin{equation}{section}

\newcommand{\ul}{\underline}

\begin{document}
\begin{titlepage}

\begin{center}

%{\today}
%\today
\hfill         \phantom{xxx}  %EFI--20-5

\vskip 2 cm {\Large \bf $\TT$ Deformations of Supersymmetric \\ Quantum Mechanics} 
%\vskip 2 cm {\Large \bf A Superspace $\TT$ Deformation for $\mathcal{N} = 2$ Quantum Mechanics} 

\vskip 1.25 cm {\bf Stephen Ebert$^1$, Christian Ferko$^{2}$, Hao-Yu Sun$^3$ and Zhengdi Sun$^4$}\non\\

\vskip 0.2 cm
{\it $^1$ Mani L. Bhaumik Institute for Theoretical Physics,	
\\ University of California, Los Angeles, CA, 90095, USA}

\vskip 0.2 cm
 {\it $^2$ Center for Quantum Mathematics and Physics (QMAP), \\ Department of Physics \& Astronomy,  University of California, Davis, CA 95616, USA}
 
\vskip 0.2 cm
 {\it $^3$ Theory Group, Department of Physics,
University of Texas, Austin, TX 78712, USA}

\vskip 0.2 cm
 {\it $^4$ Department of Physics, University of California, San Diego, CA 92093, USA}
\end{center}
\vskip 1.5 cm

\begin{abstract}
\noindent We define a manifestly supersymmetric version of the $\TT$ deformation appropriate for a class of $(0+1)$-dimensional theories with $\mathcal{N} = 1$ or $\mathcal{N} = 2$ supersymmetry, including one presentation of the super-Schwarzian theory which is dual to JT supergravity. These deformations are written in terms of Noether currents associated with translations in superspace, so we refer to them collectively as $f(\mathcal{Q})$ deformations. We provide evidence that the $f(\mathcal{Q})$ deformations of $\mathcal{N} = 1$ and $\mathcal{N} = 2$ theories are on-shell equivalent to the dimensionally reduced supercurrent-squared deformations of $2d$ theories with $\mathcal{N} = (0,1)$ and $\mathcal{N} = (1,1)$ supersymmetry, respectively.
In the $\mathcal{N} = 1$ case, we present two forms of the $f(\mathcal{Q})$ deformation which drive the same flow, and clarify their equivalence by studying the analogous equivalent deformations in the non-supersymmetric setting.

\baselineskip=18pt

\end{abstract}

\end{titlepage}

\tableofcontents
%\newpage 

\section{Introduction} \label{intro}

Supersymmetric quantum mechanics is a fruitful playground for exploring the consequences of supersymmetry in a setting which is simpler than quantum field theory. In particular, since quantum mechanics is a theory with one time dimension and no space dimensions, almost all complications involving Lorentz structure disappear.\footnote{In what follows, we will use the phrases ``supersymmetric quantum mechanics,'' ``SUSY-QM,'' and ``supersymmetric $(0+1)$-dimensional theory'' interchangeably.}

Despite this apparent simplicity, SUSY-QM exhibits great mathematical depth including rich connections to geometry and topology. Perhaps the most famous example is the relationship between the index of a SUSY-QM theory, which encodes information about the spectrum of bosonic and fermionic ground states, and the Euler characteristic of the target space on which the quantum-mechanical particle moves \cite{WITTEN1982253}. A related well-known example is the connection between supersymmetric quantum mechanics and Morse theory \cite{Witten:1982im}. For surveys of supersymmetric quantum mechanics, see \cite{Cooper:1994eh,Frohlich:1995mr,Bellucci:2006ts,Gaiotto:2015aoa,Castellani:2017ycm,clay_notes}.

In addition to its surprisingly deep mathematical structure, there are at least two senses in which supersymmetric quantum mechanics is somehow ``generic'' or ``universal'':

\begin{enumerate}

    \item  Such theories encode the worldline dynamics of a supersymmetric particle, like the Brink-Schwarz superparticle and related models which are pointlike analogues of the superstring \cite{Brink:1976sz,BRINK1981310}. But in fact, the worldline theory of \emph{any} spinning particle is (locally) supersymmetric, even if the target spacetime does not possess any supersymmetries  \cite{Brink:1976uf,CASALBUONI197649,Barducci:1976qu,BEREZIN1977336,vanHolten:1995qt,Gagne:1996gq}. In some sense, SUSY-QM is relevant for any pointlike particle with spin.
    
    \item Supersymmetric quantum mechanics generically arises as the zero-energy sector of supersymmetric QFTs. Thus, although SUSY-QM is a simple $(0+1)$-dimensional theory, it carries information about the vacuum structure of more complicated $(d+1)$-dimensional theories for $d > 0$.
\end{enumerate}

Another reason to be interested in SUSY-QM theories is that one such theory describes the low-energy limit of a collection of D$0$-branes in type IIA string theory \cite{Polchinski:1995mt,Witten:1995im}. This matrix model possesses maximal supersymmetry and descends, via dimensional reduction, from super Yang-Mills in ten spacetime dimensions \cite{CLAUDSON1985689,FLUME1985189,doi:10.1063/1.526539}. An analysis of constraints arising from maximal supersymmetry in the corresponding SUSY-QM theory has provided data about terms arising in the effective D$0$-brane Lagrangian \cite{Paban:1998ea,Paban:1998qy,Sethi:2004np}. Bound states of D$0$-branes (viewed as instantons in D$4$-branes) required by the IIA/M-theory duality, and the closely-related bound states of instantons in the maximally supersymmetric $5d$ gauge theory, can be analyzed using SUSY-QM on the instanton moduli space as well \cite{Lee:1999xb,Dorey:2001ym}.

These examples illustrate that one can often gain additional insights into phenomena in (SUSY) quantum field theory by finding their analogues in (SUSY) quantum mechanics. One particularly interesting topic about which one might hope to learn something using this strategy is the $\TT$ deformation of two-dimensional quantum field theories \cite{Zamolodchikov:2004ce, Smirnov:2016lqw,Cavaglia:2016oda}. The $\TT$ operator is constructed from a certain combination of bilinears of the stress tensor $T_{\mu \nu}$ which can be written as
\begin{align}\label{stress-tensor-bilinear}
    \det ( T_{\mu \nu} ) = \frac{1}{2} \left( \left( \tensor{T}{^\mu_\mu} \right)^2 - T^{\mu \nu} T_{\mu \nu} \right) \, .
\end{align}
Despite the fact that (\ref{stress-tensor-bilinear}) involves products of local operators -- which are generically divergent in the coincident-point limit -- it was proved in \cite{Zamolodchikov:2004ce} that this combination unambiguously defines a local operator as the insertion points are brought together (up to total derivative terms, which can be neglected), and that the resulting operator exhibits remarkable properties such as factorization. The proofs of these facts only require translation invariance but not conformal invariance.\footnote{Here we only discuss the $\TT$ deformation for theories on flat spacetimes. A general metric will break translation invariance and make this analysis more difficult, although one can make some progress in highly symmetrical cases like $\mathrm{AdS}_2$ \cite{Jiang:2019tcq,Brennan:2020dkw}. For another approach to defining $\TT$ on curved $2d$ spaces using an auxiliary $3d$ bulk, see \cite{Mazenc:2019cfg}.} Therefore, this so-called $\TT$ operator furnishes a universal deformation of any translation-invariant $2d$ quantum field theory.\footnote{See \cite{Taylor:2018xcy,Hartman:2018tkw,Bonelli:2018kik} for proposed generalizations of the $\TT$ operator in higher dimensions. 
However, above $d=2$ there is no known procedure to unambiguously define a local $\TT$-like operator by point-splitting.}

At the classical level, this flow is described by the differential equation
\begin{align}\label{TT_flow_det}
    \frac{\partial \mathcal{L}^{(\lambda)}}{\partial \lambda} = - 2 \det \left( T_{\mu \nu}^{(\lambda)} \right) \, ,
\end{align}
where the overall factor of $- 2$ is a choice of conventions. Here the superscript in $T_{\mu \nu}^{(\lambda)}$ is meant to emphasize that, at each step along the flow, the stress tensor must be recomputed from the Lagrangian $\mathcal{L}^{(\lambda)}$ rather than using the stress tensor of the undeformed theory.

The mass dimension of the stress tensor $T_{\mu \nu}$ is equal to $d$ for a theory defined in $d$ spacetime dimensions. Therefore any product of stress tensors such as those appearing in (\ref{stress-tensor-bilinear}) -- and hence the local $\TT$ operator defined by their coincident point limit -- has dimension $2d$, and is irrelevant. Usually we expect that the addition of an irrelevant operator to the Lagrangian will turn on infinitely many other operators and lead to a loss of predictive power. However, contrary to this expectation, the $\TT$ deformation is ``solvable'' and the deformed theory remains under some analytic control. More precisely, by ``solvable'' we mean that certain quantities in a $\TT$-deformed theory can be expressed in terms of the corresponding quantities in the undeformed (``seed'') theory.

One example of such a controlled quantity is the spectrum of energy levels $E_n ( R, \lambda )$ for a $\TT$-deformed theory defined on a cylinder of radius $R$ and which has been deformed by a total flow parameter $\lambda$. This finite-volume spectrum obeys
\begin{align}\label{burgers}
    \frac{\partial E_n}{\partial \lambda} = E_n \frac{\partial E_n}{\partial R} + \frac{1}{R} P_n^2 \, , 
\end{align}
which is the inviscid Burgers' equation. Here $P_n = P_n ( R )$ is the spatial momentum along the circular direction, which is quantized and does not flow with $\lambda$. If the seed theory enjoys conformal symmetry, then the flow equation (\ref{burgers}) can be solved exactly to yield
\begin{align}\label{square_root_energies}
    E_n ( R , \lambda ) = \frac{R}{2 \lambda} \left( \sqrt{ 1 + \frac{4 \lambda E_n^{(0)}}{R} + \frac{4 \lambda^2 P_n^2}{R^2} } - 1 \right) \, .
\end{align}
Other quantities in the deformed theory -- such as the torus partition function \cite{Cardy:2018sdv,Datta:2018thy,Aharony:2018bad}, the flat space $S$-matrix \cite{Dubovsky:2013ira,Dubovsky:2017cnj}, torus one-point functions \cite{Cardy:2022mhn}, and (to some degree) correlation functions \cite{Kraus:2018xrn,Cardy:2019qao,Aharony:2018vux,He:2019vzf,He:2019ahx,He:2020udl,Hirano:2020nwq,Ebert:2020tuy,Ebert:2022gyn,1836175,Ebert:2022cle} -- can also be related to quantities in the undeformed theory. One interesting consequence of these results is that the high-energy density of states in a $\TT$-deformed theory follows Hagedorn rather than Cardy growth, which means that it cannot be a local quantum field theory and instead shares some properties with string theories. Connections between $\TT$ and little string theory have been discussed in \cite{Giveon:2017myj,Asrat:2017tzd}.

We will mention one other link between $\TT$ and string theory since similar equations in the supersymmetric setting will appear later in this work. Suppose we deform the seed theory of a free massless boson $\phi$, beginning from the undeformed Lagrangian
\begin{align}
    \mathcal{L}_0 = \partial^\mu \phi \partial_\mu \phi \, .
    \label{eq:masslessscalar}
\end{align}
It was shown in \cite{Cavaglia:2016oda} that the solution to the $\TT$ flow equation (\ref{TT_flow_det}) with this initial condition (\ref{eq:masslessscalar}) is given by
\begin{align}
    \mathcal{L}_\lambda = \frac{1}{2 \lambda} \left( \sqrt{1 + 4 \lambda \partial_\mu \phi \partial^\mu \phi } - 1 \right) \, ,
\end{align}
which is the Lagrangian for the Nambu-Goto string in static gauge with a three-dimensional target space. This provides a separate hint that the $\TT$ deformation is related to string theory, besides the Hagedorn density of states which we mentioned above. We will see similar square-root-type expressions appearing in the solutions to superspace flow equations in subsequent sections, for instance in equations (\ref{dimensionally_reduced_answer}) and (\ref{susy_qm_1_scalar_soln}).

The list of references related to $\TT$ that we have presented here is necessarily incomplete, since the collection of work on $\TT$ and related deformations now includes hundreds of papers. We refer the reader to \cite{Jiang:2019epa} and references therein for a more comprehensive overview of the subject.

We have argued earlier that it is desirable to learn more about phenomena in field theory, such as $\TT$, by studying their analogues in (SUSY) quantum mechanics. In particular, we are interested in a SUSY-QM presentation of the $\TT$ operator. Such an endeavor requires first understanding how the usual $2d$ $\TT$ interacts with supersymmetry, and second understanding how to dimensionally reduce from $(1+1)$-dimensions to $(0+1)$-dimensions.\footnote{Another natural $(0+1)$-dimensional system is a spin chain. For discussions of $\TT$-like deformations in spin chain models, see \cite{Marchetto:2019yyt,Pozsgay:2019ekd}.}

The first of these problems has been studied in some detail. Manifestly supersymmetric versions of the $\TT$ operator has been presented for $2d$ field theories with $(1,1)$, $(0,1)$, $(2,2)$, or $(0,2)$ supersymmetry \cite{Chang:2018dge,Baggio:2018rpv,
Coleman:2019dvf,Chang:2019kiu,Jiang:2019hux}. This formalism was also extended to supersymmetric versions of the Lorentz-breaking $J \overbar{T}$ deformation in \cite{Jiang:2019trm}, and used to compute correlation functions for SUSY theories in \cite{He:2019ahx,Ebert:2020tuy}. A four-dimensional version of the supercurrent-squared operator was studied in \cite{Ferko:2019oyv} and showed to be related to non-linearly realized supersymmetry and the Born-Infeld action (for a related study of $\TT$ in the $\mathcal{N} = 2$ SUSY-BI theory, see \cite{Babaei-Aghbolagh:2020kjg}). This result was later extended to the ModMax theory \cite{Bandos:2020jsw,Bandos:2020hgy,Bandos:2021rqy} and its ModMax-BI extension in \cite{Babaei-Aghbolagh:2022uij,Ferko:2022iru}. A two-dimensional analogue of the non-abelian DBI action, which is compatible with maximal supersymmetry and defined using $\TT$, was obtained in \cite{Brennan:2019azg}. It was also shown that the $2d$ $\mathcal{N}=(2,2)$ Volkov-Akulov model, related to spontaneous supersymmetry breaking, is a supercurrent-squared flow in \cite{Cribiori:2019xzp}. A similar result for the $4d$ Volkov-Akulov model appeared in \cite{Ferko:2019oyv}.

The second question, about dimensional reduction, has been addressed in the non-manifestly supersymmetric context. In \cite{Gross:2019ach,Gross:2019uxi}, it was shown that one can solve for the spatial component $T_{xx}$ of the two-dimensional stress tensor using the $\TT$ trace flow equation, which holds for deformations of conformally invariant seed theories. Doing this allows one to dimensionally reduce along the spatial direction and obtain a flow equation for the Euclidean action $S_E$ of the reduced theory, which takes the form

\begin{align}\label{gross_flow_eqn}
    \frac{\partial S_E}{\partial \lambda} = \int \, d t \, \frac{T^2}{\frac{1}{2} - 2 \lambda T} \, .
\end{align}
Here $T$ is the single diagonal component of the ``stress tensor'' (or ``stress scalar'') in the $(0+1)$-dimensional theory. The solution for deformed worldline actions of this form with canonical kinetic terms, including an arbitrary number of fermionic fields $\psi^i$, was also presented in \cite{Gross:2019ach}. This result can be used to understand the deformed versions of a class of supersymmetric quantum mechanical theories, at least in component form. 

However, the additional control provided by supersymmetry is most powerful when the symmetry is made manifest, for example by a superspace construction that geometrizes the supersymmetry transformations. Thus it is desirable to have a superfield analogue of this deformation. The goal of the present paper is to find such an analogue: that is, we wish to combine the two ingredients described above in order to find a manifestly supersymmetric version of the dimensionally reduced $\TT$ operator for SUSY-QM theories.

In particular, we will obtain versions of the flow equation (\ref{gross_flow_eqn}) which are presented directly in superspace. These deformations will be written in terms of superspace Noether currents, which contain the Hamiltonian and we will typically represent with variables like $\mathcal{Q}$. For this reason, we will refer to this class of operators as $f(\mathcal{Q})$ deformations.

For $\mathcal{N} = 2$ theories, the corresponding Noether currents $\mathcal{Q}, \calQbar$ are complex. Thus we will also refer to the $\mathcal{N} = 2$ version of the $f(\mathcal{Q})$ operator as the $f(\mathcal{Q}, \calQbar)$ deformation. We will eventually see that it takes the form
\begin{align}\label{susy_qq_deformation_definition}
    \frac{\partial S_E}{\partial \lambda} = \int \, d t \, d^2 \theta \, \frac{\mathcal{Q} \calQbar}{\frac{1}{2} - 2 \lambda \Dbar \mathcal{Q}} \, ,
\end{align}
where the precise definition of the superfield $\mathcal{Q}$ will be given later.
%
% We will refer to the integrand appearing in (\ref{susy_qq_deformation_definition}) as $f(\mathcal{Q}, \calQbar)$:
% %
% \begin{align}
%     f(\mathcal{Q}, \calQbar) \equiv \frac{\mathcal{Q} \calQbar}{\frac{1}{2} - 2 \lambda \Dbar \mathcal{Q}} \, .
% \end{align}
% %
For $\mathcal{N} = 1$ theories, we will present two equivalent forms of the appropriate $f(\mathcal{Q})$ flow,
\begin{align}\label{n_equals_one_intro}
    \frac{\partial S_E}{\partial \lambda}  = \int \, dt \, d \theta \, \frac{\widetilde{\mathcal{Q}_\theta} \mathcal{Q}_t}{1 + 2 \lambda \mathcal{Q}_t} \quad \text{ and } \quad \frac{\partial S_E}{\partial \lambda} = \frac{1}{2} \int \, dt \, d \theta \, \mathcal{Q}_\theta \mathcal{Q}_t \, ,
\end{align}
which will likewise be defined later. Due to the second expression in (\ref{n_equals_one_intro}), the $\mathcal{N} = 1$ version of the $f(\mathcal{Q})$ operator will also be referred to as the $\mathcal{Q}_\theta \mathcal{Q}_t$ deformation. Although the two forms of the deformation in (\ref{n_equals_one_intro}) look very different, we will see that they lead to the same superspace flow equation for a free scalar. This surprising equivalence holds by virtue of a rewriting of the non-supersymmetric deformation (\ref{gross_flow_eqn}). In particular, for a certain class of quantum mechanical theories, it turns out that
\begin{align}\label{HT_equivalence_intro}
    \frac{H^2}{\frac{1}{2} - 2 \lambda H} = - \frac{1}{2} H \Th \, ,
\end{align}
where $\Th$ is the (Euclidean) Hilbert stress tensor computed from the Euclidean Lagrangian $H$. Therefore it is equivalent to deform by either the rational function of $H$ appearing on the left side of (\ref{HT_equivalence_intro}) (whose $\mathcal{N} = 1$ superspace version is $\frac{\widetilde{\mathcal{Q}_\theta} \mathcal{Q}_t}{1 + 2 \lambda \mathcal{Q}_t}$) or to the simple product on the right side of (\ref{HT_equivalence_intro}) (whose $\mathcal{N} = 1$ superspace version is $\mathcal{Q}_\theta \mathcal{Q}_t$), as we explain later.

Besides making the supersymmetry of the deformed theory manifest, this procedure has the additional advantage that the supercharges of the deformed theory continue to act in the canonical way on superfields, whereas in a component presentation of the deformed quantum mechanics, the supercharges $Q_i$ must be corrected order-by-order in $\lambda$. 

A final piece of motivation for performing this analysis is the relationship between certain $(0+1)$-dimensional theories and higher-dimensional gauge and gravity theories. For instance, the two-dimensional JT gravity theory (which descends via dimensional reduction from $3d$ gravity on $\mathrm{AdS}_3$ \cite{Maxfield:2020ale,Gross:2019ach,Achucarro:1993fd}) is related to the Schwarzian theory as suggested in \cite{Jensen:2016pah,Maldacena:2016upp,Engelsoy:2016xyb}; the Schwarzian itself can be written as the theory of a particle moving on an $SL(2, \mathbb{R})$ group manifold \cite{Mertens:2017mtv,Mertens:2018fds}. JT gravity can also be written in BF variables as a two-dimensional gauge theory \cite{PhysRevLett.63.834,Chamseddine:1989yz}, and the interpretation of the $\TT$ deformation in this setting is explored in a companion paper \cite{us:gravity}. One would like to understand the action of $\TT$ deformations in all of these related theories, both with and without supersymmetry. The present work represents one step towards such an understanding, where we study the manifestly supersymmetric version of the deformation in the simplest member of this family of related theories, i.e., $(0+1)$-dimensional quantum mechanics.

The layout of this paper is as follows. In Section \ref{sec:rev_conv}, we outline our conventions, review some salient aspects of previous works in order to make the present manuscript self-contained, and describe the class of theories we will consider in later sections. Section \ref{method1} pursues one method of obtaining deformed SUSY-QM theories, namely first solving the superspace flow equation for a simple class of models in $2d$ and then dimensionally reducing the result to quantum mechanics. In Section \ref{method2}, we instead dimensionally reduce the supercurrent-squared operator itself to produce a candidate superspace deformation for theories in $(0+1)$-dimensions. The main result of our paper is Section \ref{method3}, where we use a Noether procedure to construct a superspace deformation directly in the superspace of an $\mathcal{N}=2$ quantum mechanics theory, and check that this deformation is consistent with the dimensional reductions of the preceding sections. In Section \ref{sec:n_equals_one}, we present an abridged version of this analysis for theories with $\mathcal{N} = 1$ supersymmetry, including defining two equivalent forms of the appropriate deformation and comparing the deformed theory of a single scalar to the dimensional reduction of the corresponding deformed $2d$ $\mathcal{N} = (0, 1)$ theory. Section \ref{sec: Discussion} concludes with a summary of our results and identifies some directions for future research. We have relegated certain details to Appendices, including a change of conventions from real to complex supercurrents in Appendix \ref{app:change_to_complex} and an example of a non-supersymmetric dimensional reduction of a theory with a potential in Appendix \ref{app:no_trace_flow}.

\section{Preliminaries}\label{sec:rev_conv}

In this initial section, we will lay the groundwork for the analyses in the rest of the paper. 

% Sections \ref{subsec:gross_review} and \ref{subsec:bf_review} review some known results about $\TT$ in $(0+1)$-dimensions -- and the relationship of such theories to BF theory -- in the non-supersymmetric setting, which will be relevant for the later parts of this paper. We then establish our conventions for the various superspaces that will be used in subsequent calculations in Section \ref{sec:conventions}. Finally, in Section \ref{subsec:description} we motivate the class of SUSY-QM theories which we will consider in this work and outline the three ways in which one might attempt to $\TT$-deform such theories in a manifestly supersymmetric way. These three methods will be investigated in more detail in Sections \ref{method1}, \ref{method2}, \ref{method3}.

\subsection{Conventions}\label{sec:conventions}

In this subsection, we outline our notation for the superspaces used in the remainder of this work. Although the main focus of our analysis is on a supersymmetric quantum mechanics theory in $(0+1)$-dimensions, we will obtain some expressions for $\TT$-type deformations by dimensionally reducing previous results for $(1+1)$-dimensional supersymmetric theories. For this reason, we begin with an overview of the conventions for $\mathcal{N} = (1,1)$ supersymmetry in two spacetime dimensions following \cite{Chang:2018dge}.

We begin by discussing two-dimensional Lorentzian field theories. We will assume that such theories have coordinates $(t, x)$, and when we perform dimensional reduction we will assume that the spatial coordinate $x$ parameterizes a circle with some radius $R$ so that $x \sim x + R$.

It will often be convenient to change coordinates from $(t, x)$ to light-cone coordinates:
\begin{align}\label{bispinor_light_cone}
    x^{\pm \pm} = \frac{1}{\sqrt{2}} \left( t \pm x \right) \, .
\end{align}
Here we have adopted the bi-spinor convention where a vector index is written as a pair of spinor indices. The derivatives with respect to the coordinates (\ref{bispinor_light_cone}) are
\begin{align}
    \partial_{\pm \pm} = \frac{1}{\sqrt{2}} \left( \partial_t \pm \partial_x \right) \, , 
\end{align}
which satisfy
\begin{align}
    \partial_{\pm \pm} x^{\pm \pm} = 1 \, , \qquad \partial_{\pm \pm} x^{\mp \mp} = 0 \, .
\end{align}
Spinor indices, which are written with early Greek letters, are raised or lowered with the epsilon tensor as
\begin{align}
    \psi_\beta = \epsilon_{\beta \alpha} \psi^{\alpha} \, ,
\end{align}
where we take $\epsilon_{+-} = 1$ so $\epsilon_{-+} = -1$, $\epsilon^{+-} = -1$, $\epsilon^{-+} = 1$. For instance, this implies that
\begin{align}\label{spinor_raise_lower}
    \psi^- = \psi_+ \, , \qquad \psi^+ = - \psi_- \, . 
\end{align}
For two-dimensional theories with $\mathcal{N} = (1,1)$ supersymmetry, we write the Grassmann coordinates as $\theta^{\pm}$. The supercovariant derivatives with respect to these anticommuting coordinates are
\begin{align}
    D_{\pm} = \frac{\partial}{\partial \theta^{\pm}} + \theta^{\pm} \partial_{\pm \pm} \, .
\end{align}
These satisfy
\begin{align}
    D_\pm D_\pm = \partial_{\pm \pm} \, , \qquad \{ D_+ , D_- \} = 0 \, .
\end{align}
We will also be interested in discussing theories of supersymmetric quantum mechanics in $(0+1)$-dimensions, so next we describe how to perform this reduction and match conventions between the two theories.

When we reduce from $(1+1)$-dimensional field theory to $(0+1)$-dimensional quantum mechanics we shall assume that all quantities are independent of the spatial direction $x$. Operationally, one can achieve this by setting $\partial_x \equiv 0$ everywhere, which amounts to making the replacement $\partial_{\pm \pm} = \frac{1}{\sqrt{2}} \partial_t$. We will re-scale our time coordinate $t$ to eliminate the factor of $\frac{1}{\sqrt{2}}$ and instead write $\partial_{\pm \pm} = \partial_t$

We note that making this replacement leads to expressions which have unbalanced numbers of $+$ and $-$ indices, like $D_+ = \frac{\partial}{\partial \theta^+} + \theta^+ \partial_t$. Although such an expression would not exhibit the correct properties under Lorentz transformation in a $(1+1)$-dimensional theory, in our reduced $(0+1)$-dimensional theory, there is no notion of spin nor of Lorentz symmetry. Performing the dimensional reduction in this way therefore yields a consistent set of conventions.

It will be convenient to write the superspace of the $\mathcal{N} = 2$ supersymmetric quantum mechanics theory in complex coordinates, which more closely matches the conventions in the literature. We first Wick-rotate our time coordinate,\footnote{We will be somewhat cavalier about real versus imaginary time. All formulas in $2d$ field theory will be Lorentzian and involve real times $t$, but upon dimensional reduction to quantum mechanics, we eventually rotate $t \to it$ in order to match more common conventions. However we will continue to use the symbol $t$ rather than $\tau$ in this context for simplicity.} sending $t \to i t$, so that the supercovariant derivatives are
\begin{align}
    D_{\pm} = \frac{\partial}{\partial \theta^{\pm}} - i \theta^{\pm} \partial_t \, .
\end{align}
Next we perform the change of variables
\begin{align}\label{theta_complex_change_of_variables}
    \theta = \frac{1}{\sqrt{2}} \left( \theta^+ - i \theta^- \right) \, , \qquad \thetab = \frac{1}{\sqrt{2}} \left( \theta^+ + i \theta^- \right) \, ,
\end{align}
so that
\begin{align}
    D = \frac{1}{\sqrt{2}} \left( D_+ + i D_- \right) = \frac{\partial}{\partial \theta} - i \thetab \partial_t \, , \quad \Dbar = \frac{1}{\sqrt{2}} \left( D_+ - i D_- \right) = \frac{\partial}{\partial \thetab} - i \theta \partial_t \, .
\end{align}
The new supercovariant derivatives satisfy the canonical algebra
\begin{align}\label{ddbar_algebra}
    \{ D , \Dbar \} = - 2 i \partial_t \, , 
\end{align}
with $D^2 = \Dbar^2 = 0$.

The rotation from real to complex Grassmann coordinates will introduce a factor of $i$ in the measure, since
\begin{align}
    d \theta \, d \thetab = i \, d \theta^+ \, d \theta^- \, ,
\end{align}
but this is compensated by the factor of $i$ arising from the Wick rotation $d t \to i \, d t$. 

Finally, in Section \ref{sec:n_equals_one} we will briefly also discuss the $\mathcal{N} = 1$ version of our deformation. In $\mathcal{N} = 1$ superspace we have a single anticommuting coordinate $\theta$, along with a corresponding supercovariant derivative
\begin{align}
    D = \frac{\partial}{\partial \theta} - i \theta \frac{\partial}{\partial t} \, ,
\end{align}
which satisfies the algebra
\begin{align}
    \{ D, D \} = - 2 i \partial_t \, .
\end{align}

\subsection{Review of $\TT$ in Quantum Mechanics}\label{subsec:gross_review}

We now recall certain facts about the dimensional reduction of the $\TT$ operator from $(1+1)$ dimensions to $(0+1)$ dimensions, but without manifest supersymmetry. We follow the discussion in \cite{Gross:2019ach}, where these results first appeared.

Although the $\TT$ deformation can be defined for any translationally-invariant QFT, here we restrict to the case of a conformally invariant seed theory with Lagrangian $\mathcal{L}_0$. The flow equation (\ref{TT_flow_det}), which can be written as
\begin{align}\label{tt_flow_comparison}
    \frac{\partial \mathcal{L}^{(\lambda)}}{\partial \lambda} = T^{\mu \nu} T_{\mu \nu} - \left( \tensor{T}{^\mu_\mu} \right)^2  \, ,
\end{align}
therefore determines a one-parameter family of Lagrangians $\mathcal{L}^{(\lambda)}$ with the initial condition that $\mathcal{L}^{(0)}$ matches the CFT Lagrangian $\mathcal{L}_0$. We note that $\lambda$ has length dimension $2$, which means that there is an effective energy scale $\Lambda$ set by
\begin{align}
    \Lambda = \frac{1}{\sqrt{\lambda}} \, .
\end{align}
Because the seed theory was conformal and hence had no dimensionful parameters, in the deformed theories the quantity $\Lambda$ is the only scale in the problem. This means that an infinitesimal change in $\Lambda$ is equivalent to an infinitesimal scale transformation of the theory, and on general grounds we know that the response of the action $S^{(\lambda)} = \int d^2 x \, \mathcal{L}^{(\lambda)}$ to such a scale transformation is controlled by the trace of the stress tensor. We therefore have
\begin{align}
    \Lambda \frac{d}{d \Lambda} S^{(\lambda)} = \int d^2 x \, \tensor{T}{^\mu_\mu} \, .
\end{align}
On the other hand, by comparing to equation (\ref{tt_flow_comparison}), we see that by definition the response of the action $S^{(\lambda)}$ to a change in $\lambda$ is given by the integral of the $\TT$ operator. That is,
\begin{align}
    \Lambda \frac{d}{d \Lambda} S^{(\lambda)} &= \frac{1}{\sqrt{\lambda}} \frac{d}{d \left( \frac{1}{\sqrt{\lambda}} \right)} S^{(\lambda)} \nonumber \\
    &= - 2 \lambda \int d^2 x \, \left( T^{\mu \nu} T_{\mu \nu} - \left( \tensor{T}{^\mu_\mu} \right)^2 \right) \, .
\end{align}
We therefore conclude that
\begin{align}
\label{trace_flow}
    \tensor{T}{^\mu_\mu} = 2 \lambda \, \left( \left( \tensor{T}{^\mu_\mu} \right)^2 - T^{\mu \nu} T_{\mu \nu} \right) \, ,
\end{align}
according to our normalization of $\TT$. We emphasize that this relationship, namely the trace flow equation, between the components of the stress tensor holds at any point along the trajectory of $\TT$-flow, without imposing any equations of motion or conservation equations. We may therefore solve (\ref{trace_flow}) for a component of the stress tensor and use the resulting equation to eliminate this component in the definition of the deformation.

Suppose that our $2d$ theory has coordinates $x$ and $t$, where we take $x$ to parameterize a circular direction.\footnote{Note that \cite{Gross:2019ach} uses the symbol $\tau$ rather than $t$ to emphasize the Euclidean signature. As discussed in Section \ref{sec:conventions}, we will be cavalier about signature and use the symbol $t$ regardless.} We wish to dimensionally reduce along this circle in order to get an effective deformation in the resulting $(0+1)$-dimensional theory. To do this, it is natural to solve for the spatial component $T_{xx}$. Writing out both sides of (\ref{trace_flow}) in components yields
\begin{align}
    \tensor{T}{^x_x} + \tensor{T}{^t_t} = 2 \lambda \left( \left( \tensor{T}{^x_x} + \tensor{T}{^t_t} \right)^2 - \left( T^{xx} T_{xx} + 2 T^{xt} T_{xt} + T_{tt} T_{tt} \right) \right) \, ,
\end{align}
which can be solved to find
\begin{align}\label{Txx_soln}
    \tensor{T}{^x_x} = \frac{\tensor{T}{^t_t} + 4 \lambda T^{xt} T_{xt}}{-1 + 4 \lambda \tensor{T}{^t_t}} \, .
\end{align}
In order to dimensionally reduce, we will assume that the mixed component $T_{xt}$ of the stress tensor vanishes. We would then like to replace the spatial component $\tensor{T}{^x_x}$ and write a flow equation which depends only on the ``stress scalar'' $T \equiv \tensor{T}{^t_t}$ for the $(0+1)$-dimensional theory. After replacing $\tensor{T}{^x_x}$ and setting $T_{xt} = 0$ in this way, the flow equation for the Lagrangian becomes
\begin{align}\label{gross_reduction_intermediate}
    \frac{\partial \mathcal{L}^{(\lambda)}}{\partial \lambda} &= \left(  \left( \frac{\tensor{T}{^t_t}}{-1 + 4 \lambda \tensor{T}{^t_t}} \right)^2 + \left( \tensor{T}{^t_t} \right)^2 - \left( \tensor{T}{^t_t} + \frac{\tensor{T}{^t_t}}{-1 + 4 \lambda \tensor{T}{^t_t}} \right)^2 \right) \nonumber \\
    &= \frac{\left( \tensor{T}{^t_t} \right)^2}{\frac{1}{2} - 2 \lambda \tensor{T}{^t_t}} \, .
\end{align}
We can now assume that $T \equiv \tensor{T}{^t_t}$ is independent of the spatial coordinate $x$ and perform the integral over the $x$ circle. This will introduce an irrelevant length factor, which can be scaled away. The result is a deformation for the action of the $(0+1)$-dimensional quantum mechanics theory:
\begin{align}\label{reduced_qm_flow_no_susy}
    \frac{\partial S_E}{\partial \lambda} = \int \, dt \, \frac{T^2}{\frac{1}{2} - 2 \lambda T} \, .
\end{align}
Interpreting the Euclidean Lagrangian as the Hamiltonian, we can evaluate (\ref{reduced_qm_flow_no_susy}) in an energy eigenstate $| n \rangle$ to find
\begin{align}
    \frac{\partial \langle n \, \vert \, H \, \vert \, n \rangle}{\partial \lambda} = \frac{ \left( \langle n \, \vert \, H \, \vert \, n \rangle \right)^2 }{\frac{1}{2} - 2 \lambda \langle n \, \vert \, H \, \vert \, n \rangle} \, ,
\end{align}
or more simply writing $E_n$ for $\langle n \, \vert \, H \, \vert \, n \rangle$,
\begin{align}
    \frac{\partial E_n}{\partial \lambda} = \frac{E_n^2}{\frac{1}{2} - 2 \lambda E_n} \, .
\end{align}
This differential equation has a solution
\begin{align}
    E_n ( \lambda ) = \frac{1 - \sqrt{1 - 8 \lambda E_n^{(0)}}}{4 \lambda} \, ,
\end{align}
which is reminiscent of (\ref{square_root_energies}). Another perspective on deriving the deformed energy spectrum is provided by the Wheeler-DeWitt equation (see \cite{McGough:2016lol} for 3d and \cite{Hartman:2018tkw} for general dimensions). Following \cite{Hartman:2018tkw}, the Wheeler-DeWitt equation was specialized to JT gravity and dilaton gravity by \cite{Gross:2019ach} at a finite radial cutoff and correctly derived the deformed energy spectrum.\footnote{For further studies of $\TT$-like flows in dilaton gravity, see \cite{Grumiller:2020fbb}. A different perspective on $\TT$ and cutoff geometries using path integral optimization is discussed in \cite{Jafari:2019qns}.} 

This analysis of \cite{Gross:2019ach} demonstrates that such $f(T)$ or $f(H)$ deformations in quantum mechanics are in some sense simple, as energy eigenstates of the undeformed theory remain energy eigenstates of the deformed theory albeit with shifted eigenvalues. However, we point out the important caveat that this derivation relied upon the trace flow equation \eqref{trace_flow}, which only holds if the parent $2d$ theory is conformally invariant. For a quantum mechanical theory which descends from a non-conformal $2d$ QFT, such as one with a potential, the flow equation (\ref{reduced_qm_flow_no_susy}) is \emph{not} equivalent to $\TT$, and it is not obvious that the energy eigenstates are unchanged; indeed, we will comment in Section \ref{sec: Discussion} that a na\"ive classical analysis suggests that the eigenstates must be modified due to the presence of poles in the potential.

Since we cannot use the trace flow equation \eqref{trace_flow} in analyzing such non-conformal theories, our only option is to solve the $\TT$ flow equation in two dimensions and then dimensionally reduce the solution at the end (rather than reducing the $\TT$ operator itself). A simple example of this procedure for a single boson subject to an arbitrary potential is presented in Appendix \ref{app:no_trace_flow}.

Our eventual goal will be to find a supersymmetric analogue of this dimensional reduction procedure for $\TT$, which proceeds by eliminating $T_{xx}$, directly in superspace. This will be the subject of Section \ref{method2}. First we will turn to a different piece of motivation for studying $(0+1)$-dimensional theories of the form considered here.

\subsection{Review of Relationship to BF Theory}\label{subsec:bf_review}

In the previous subsection, we have seen that quantum mechanical models which descend from conformally invariant $2d$ QFTs can be directly deformed in the $(0+1)$-dimensional theory using a convenient dimensional reduction of the $\TT$ operator that relies upon the trace flow equation \eqref{trace_flow}. We have emphasized that this class of models is special and does not, for example, include theories of a particle subject to a potential. However, one class of models for which this deformation \emph{is} applicable is that of a purely kinetic Lagrangian,
\begin{align}\label{bosonic_kinetic_lagrangian}
    L = \frac{1}{2} g_{ij} ( x ) \, \dot{x}^i \, \dot{x}^j \, ,
\end{align}
for a collection of bosonic degrees of freedom $x^i$. This theory is of course interpreted as a particle moving on a target space geometry with metric $g_{ij}$ and coordinates $x^i$. 

Although theories of the form (\ref{bosonic_kinetic_lagrangian}) are somewhat simplistic, they are especially of interest in the case where the target space parameterized by the $x^i$ is a Lie group $G$. In this case, the theory is dual to BF gauge theory with gauge group $G$ \cite{Mertens:2018fds}. By way of motivation, we will now review this correspondence, although in the remainder of this work we will not explore the connections with BF theory further; see \cite{us:gravity} for an analysis of $\TT$ deformations from this perspective.

The fields of two-dimensional BF gauge theory are a scalar field $\phi$ and a spacetime gauge field $A_\mu$. We fix a gauge group $G$ and assume that both $\phi$ and $A_\mu$ take values in the adjoint representation of $G$. Writing $\Tr$ for the trace in the adjoint, the action (including boundary term) on a two-dimensional surface $M$ is
\begin{align}\label{bf_action_non_susy}
    S_{BF} = \int_M d^2 x \, \Tr \left( \phi F \right) + \frac{1}{2} \oint_{\partial M} \, dt \, \Tr \left( \phi A_t \right) \, .
\end{align}
The bulk equations of motion are determined from the first term of (\ref{bf_action_non_susy}); the field $\phi$ acts as a Lagrange multiplier which imposes the flatness condition $F = 0$. In particular, this means that the connection $A_\mu$ is pure gauge and can be written as
\begin{align}\label{Amu_is_pure_gauge}
    A_\mu = g^{-1} \partial_\mu g
\end{align}
for some group element $g\in G$.

In order to obtain a good variational principle, we must also cousider the boundary variation of (\ref{bf_action_non_susy}). Varying with respect to the fields and discarding the bulk equation of motion term yields
\begin{align}\label{BF_boundary_term_variational}
    \delta S_{BF} \Big\vert_{\text{on-shell}} = \frac{1}{2} \int_{\partial M} \, dt \, \Tr \left( A_t \, \delta \phi - \phi \delta A_t \right) \, .
\end{align}
To force the boundary term appearing in (\ref{BF_boundary_term_variational}) to vanish, we choose the Dirichlet boundary condition
\begin{align}\label{BF_boundary_constraint}
    \phi \big\vert_{\partial M} = A_t \big\vert_{\partial M} \, .
\end{align}
Since $A_\mu$ transforms as a boundary one-form on $\partial M$, the constraint (\ref{BF_boundary_constraint}) should be read as the statement that the one-form $\phi \, dt$ is set equal to the one-form $A_\mu$ on $\partial M$. 

We now consider the on-shell boundary action arising from imposing this constraint. Since $\phi$ is related to $A_t$ by (\ref{BF_boundary_constraint}) and $A_\mu$ itself is pure gauge according to (\ref{Amu_is_pure_gauge}), the boundary term in (\ref{bf_action_non_susy}) can be written as
\begin{align}\label{BF_boundary_term_to_group_intermediate}
    S_{BF} \Big\vert_{\text{bdry}} = \frac{1}{2} \oint_{\partial M} \, dt \, \Tr \left( ( g^{-1} \partial_t g) ( g^{-1} \partial_t g) \right) \, .
\end{align}
Next we recall that, for a general mapping from a spacetime $M$ to a Lie group $G$, the expression $g^{-1} \partial_\mu g$ is simply the pullback of the Maurer-Cartan form on $G$ to $M$.\footnote{The Maurer-Cartan form itself is a push-forward from $T_gG$ into $\mathfrak{g}$.} By writing the group element $g$ as the exponential of some linear combination of basis elements $T_i$ for the Lie algebra $\mathfrak{g}$ of $G$,
\begin{align}
    g = \exp \left( x^i T_i \right) \, ,
\end{align}
one finds that the expression $g^{-1} \partial_\mu g$ can be related to the metric $g_{ij}$ on the Lie group $G$ as
\begin{align}
    g_{ij} \, dx^i \, dx^j = \mathrm{Tr} \left( ( g^{-1} \partial_\mu g ) ( g^{-1} \partial^\mu g ) \right) \, .
\end{align}
Therefore the boundary term (\ref{BF_boundary_term_to_group_intermediate}) can be expressed as
\begin{align}
    S_{BF} \Big\vert_{\text{bdry}} = \frac{1}{2} \oint_{\partial M} \, dt \, g_{ij} ( x ) \, \dot{x}^i \dot{x}^j \, ,
\end{align}
where $x^i$ are coordinates on $G$.

The upshot of this discussion is that the boundary term for BF gauge theory with gauge group $G$ is equivalent to the theory of a particle moving on a Lie group $G$ with the appropriate metric $g_{ij}$ (the Cartan metric tensor induced by the Killing form) and a purely kinetic Lagrangian.\footnote{For this particle-on-a-group theory, the Hamiltonian can be interpreted as a certain quadratic Casimir $J_a J_b \mathrm{tr} ( T^a T^b )$ of the target-space group, where $J = g^{-1} \partial_t g$. The $\TT$ flow can then be viewed as deforming this Casimir. There is a conceptually similar interpretation of the deformation of $2d$ Yang-Mills, which is a quasi-topological theory, as investigated in \cite{Conti:2018jho,Ireland:2019vvj,Griguolo:2022xcj}.}

Thus, although $\TT$ deformations of purely-kinetic $(0+1)$-dimensional theories such as (\ref{bosonic_kinetic_lagrangian}) may seem trivial, one might be motivated to study them due to this connection to two-dimensional BF theory (in addition to the connections to other theories like the Schwarzian theory, JT gravity, and SYK, which we have already alluded to). In particular, one might be especially interested in the manifestly supersymmetric versions of these theories where one expects to have some additional control. We will take this as part of our motivation for studying the class of theories which we consider in later sections.

\subsection{Description of Models and Deformation Methods}\label{subsec:description}

In the previous subsections, we have motivated the study of kinetic Lagrangians of the form
\begin{align}\label{later_purely_kinetic}
    S = \frac{1}{2} \int \, d t \, g_{ij} ( x ) \dot{x}^i \dot{x}^j \, , 
\end{align}
and noted that they can be deformed by the dimensionally-reduced $\TT$ operator via
\begin{align}
    \frac{\partial S_E}{\partial \lambda} = \int \, d t \, \frac{T^2}{\frac{1}{2} - 2 \lambda T} \, .
\end{align}
Next we will recall how to embed theories whose bosonic parts take the form (\ref{later_purely_kinetic}) into superfields. For concreteness, we will focus on $\mathcal{N} = 2$ supersymmetric quantum mechanics (i.e., $2$ real supercharges or $1$ complex supercharge). Consider a collection of $\mathcal{N} = 2$ superfields with expansions
\begin{align}
    X^i ( t, \theta, \thetab ) = x^i ( t ) + \theta \, \psi^i ( t ) - \thetab \psib^i ( t ) + \theta \thetab F^i ( t ) \, .
\end{align}
The superspace action whose bosonic part reduces to (\ref{later_purely_kinetic}) is simply
\begin{align}\label{susy_particle_on_group}
    S = \frac{1}{2} \int \, dt \, d \thetab \, d \theta \, g_{ij} ( X ) \, \left( D X^i ( t, \theta , \thetab ) \right)^{\ast} \, D X^j ( t , \theta , \thetab ) \, .
\end{align}
By performing the integration over the anticommuting coordinates $\theta, \thetab$, one can show that the superspace action (\ref{susy_particle_on_group}) reduces to the component form
\begin{align}\label{susy_kinetic_components}
    S = \frac{1}{2} \int \, dt \, &\Big( g_{ij} \dot{x}^i \dot{x}^j + g_{ij} F^i F^j + 2 i g_{ij} \psib^i \dot{\psi}^j + \psib^i \psi^j \dot{x}^k \left( \partial_k g_{ij} + \partial_j g_{ik} - \partial_i g_{jk} \right) \nonumber \\
    &\quad + \psib^i \psi^j F^k \left( \partial_k g_{ij} - \partial_j g_{ik} - \partial_i g_{jk} \right) - \psib^i \psi^j \psib^k \psi^l \partial_k \partial_l g_{ij} \Big) \, .
\end{align}
From (\ref{susy_kinetic_components}) we see that the equations of motion for the auxiliary fields $F^i$ are purely algebraic. On-shell, they can be eliminated in terms of fermions via the equation of motion
\begin{align}
    F^i = \tensor{\Gamma}{^i_j_k} \psib^j \psi^k \, ,
\end{align}
where $\tensor{\Gamma}{^i_j_k}$ are the Christoffel symbols associated with the metric $g_{ij}$. Similarly, it is convenient to define a covariant derivative $\nabla_t$ with the property that $\psi^i$ transform as vectors:
\begin{align}
    \nabla_t \psi^j = \dot{\psi}^j + \tensor{\Gamma}{^j_l_m} \psi^l \dot{x}^m \, .
\end{align}
The terms involving four fermions can be written in terms of the Riemann curvature tensor $R_{ijkl}$ which is computed from the metric $g_{ij}$ in the usual way. In terms of these new quantities, the action can be written more compactly as
\begin{align}\label{susy_kinetic_components_final}
    S = \frac{1}{2} \int \, dt \, \left( g_{ij} \dot{x}^i \dot{x}^j + 2 i g_{ij} \psib^i \nabla_t \psi^j + \frac{1}{2} R_{ikjl} \psib^i \psi^j \psib^k \psi^l \right) \, .
\end{align}
This theory therefore reduces to the theory of a collection of bosonic degrees of freedom $x^i$ and their fermionic superpartners $\psi^i$. The $x^i$ are subject to the purely kinetic Lagrangian (\ref{later_purely_kinetic}) as desired whereas the fermions have both kinetic terms and four-fermion couplings determined by the Riemann curvature of the target space.

In the remainder of this work, we will restrict our attention to supersymmetric $\TT$-type deformations of seed theories which take the form (\ref{susy_particle_on_group}). There are three, na\"ively different, ways in which one could study supersymmetric current-squared deformations of this $(0+1)$-dimensional theory:
\begin{enumerate}
    \item Write a flow equation for a $(1+1)$-dimensional field theory which reduces to (\ref{susy_particle_on_group}) using the supercurrent-squared operator. Solve this flow equation in the parent $(1+1)$-dimensional theory, and only after finding the full deformed solution, dimensionally reduce the result to quantum mechanics. This will be explored in Section \ref{method1}.

    \item Begin with the definition of the supercurrent-squared operator in a $(1+1)$-dimensional theory. Apply dimensional reduction to this operator itself, thus defining a deformation of the $(0+1)$-dimensional theory. We perform this procedure in Section \ref{method2}.

    \item Work directly in the superspace of the quantum mechanics theory. Construct a conserved superfield which contains the Hamiltonian and then define an appropriate superspace deformation using bilinears in this superfield with the property that this flow equation reduces to (\ref{gross_flow_eqn}) after integrating out the anticommuting directions (and imposing on-shell conditions). This is done in Section \ref{method3}.
\end{enumerate}
A priori, it is not clear that these three procedures are equivalent, since one might imagine that the process of performing dimensional reduction does not commute with the process of deforming by a supercurrent-squared operator. However in following sections we will provide evidence that the three approaches yield the same deformation on-shell.

\section{Dimensional Reduction of Solution to $2d$ Flow}\label{method1}

In this section, we will directly solve the supercurrent-squared flow equation in the $2d , \, \mathcal{N} = (1, 1)$ field theory and then dimensionally reduce the result. This is a slight generalization of the analysis for a single $\mathcal{N} = (1,1)$ superfield whose flow equation was studied in \cite{Baggio:2018rpv,Chang:2018dge,Ferko:2021loo}. Although much of this analysis has appeared before, we review it here to make the present work self-contained and to provide a check on our results in Section \ref{method3}.

\subsection{Definition of Supercurrents}

Consider a general superspace Lagrangian $\mathcal{A}$ which depends on a collection of superfields $\Phi^i$ and their derivatives as
\begin{align}
    \mathcal{A} = \mathcal{A}\left( \Phi^i , D_+ \Phi^i, D_- \Phi^i, \partial_{++} \Phi^i, \partial_{--} \Phi^i, D_+ D_- \Phi^i \right) \, .
\end{align}
A general variation $\delta \mathcal{A}$ of this superspace Lagrangian can be written as
\begin{equation}\label{general_superspace_variation} 
\begin{split}
%\begin{align}
	\delta \mathcal{A} &= D_+ \left( \delta \Phi^i \frac{\delta \mathcal{A}}{\delta (D_+ \Phi^i)} \right) + D_- \left( \delta \Phi^i \frac{\delta \mathcal{A}}{\delta (D_- \Phi^i)} \right) + \partial_{++} \left( \delta \Phi^i \frac{\delta \mathcal{A}}{\delta (\partial_{++} \Phi^i)} \right) \cr
    &+ \partial_{--} \left( \delta \Phi^i \frac{\delta \mathcal{A}}{\delta (\partial_{--} \Phi^i)} \right) + \frac{1}{2} \left( D_+ \left( \frac{\delta \mathcal{A}}{\delta (D_+ D_- \Phi^i)} D_- \delta \Phi^i \right) + D_- \left( \delta \Phi^i D_+ \frac{\delta \mathcal{A}}{\delta (D_+ D_- \Phi^i)} \right) \right) \cr
    &- \frac{1}{2} \left( D_- \left( \frac{\delta \mathcal{A}}{\delta (D_+ D_- \Phi^i)} D_+ \delta \Phi^i \right) + D_+ \left( \delta \Phi^i D_- \frac{\delta \mathcal{A}}{\delta (D_+ D_- \Phi^i)} \right)  \right) \cr
    &- \delta \Phi^i \left( - \frac{\delta \mathcal{A}}{\delta \Phi^i} + D_+ \frac{\delta \mathcal{A}}{\delta (D_+ \Phi^i)} + D_- \frac{\delta \mathcal{A}}{\delta (D_- \Phi^i)} + \partial_{++} \frac{\delta \mathcal{A}}{\delta (\partial_{++} \Phi^i)} + \partial_{--} \frac{\delta \mathcal{A}}{\delta (\partial_{--} \Phi^i)}  \right. \cr & \left. - D_+ D_- \frac{\delta \mathcal{A}}{\delta (D_+ D_- \Phi^i)} \right) . 
%\end{align}
\end{split} 
\end{equation}
First, this general variation (\ref{general_superspace_variation}) can be used to derive the superspace equations of motion for each of the $\Phi^i$. After performing some superspace integrations by parts and collecting the terms proportional to each $\delta \Phi^i$, we find that the overall variation $\delta \mathcal{A}$ will vanish for a general variation of the superfield $\Phi^i$ if
\begin{align}\label{susy_eom}
	\frac{\delta \mathcal{A}}{\delta \Phi^i} & =  D_+ \left( \frac{\delta \mathcal{A}}{\delta (D_+ \Phi^i)} \right) + D_- \left( \frac{\delta \mathcal{A}}{\delta (D_- \Phi^i)} \right) + \partial_{++} \left( \frac{\delta \mathcal{A}}{\delta (\partial_{++} \Phi^i)} \right) \cr & + \partial_{--} \left( \frac{\delta \mathcal{A}}{\delta (\partial_{--} \Phi^i)} \right) - D_+ D_- \left( \frac{\delta \mathcal{A}}{\delta (D_+ D_- \Phi^i)} \right) ,
\end{align}
which is exactly the $\Phi^i$ equation of motion. A related calculation can be used to find the superspace Noether current for spatial translations. Consider a spacetime translation $\delta x^{\pm \pm} = a^{\pm \pm}$ where the parameters $a^{\pm \pm}$ are constants. For such a translation, the variations appearing in (\ref{general_superspace_variation}) can be replaced as $\delta \mathcal{A} = a^{++} \partial_{++} \mathcal{A} + a^{--} \partial_{--} \mathcal{A}$ and likewise for $\delta \Phi^i$, $D_{\pm} \delta \Phi^i$, and so on. Restricting to the case of on-shell variations, so that we can discard the term proportional to the superspace equations of motion, one finds that the resulting equation can be written as
\begin{align}
    0 = a^{++} \left( D_+ \mathcal{T}_{++-} + D_- \mathcal{T}_{+++} \right) + a^{--} \left( D_+ \mathcal{T}_{---} + D_- \mathcal{T}_{--+} \right) \, , 
\end{align}
where the components of $\mathcal{T}$ are given by
\begin{align}
	\mathcal{T}_{++-} &= \partial_{++} \Phi^i \frac{\delta \mathcal{A}}{\delta (D_+ \Phi^i)} + D_+ \left(  \partial_{++} \Phi^i \frac{\delta \mathcal{A}}{\delta (\partial_{++} \Phi^i)} \right) + \frac{1}{2} \frac{\delta \mathcal{A}}{\delta (D_+ D_- \Phi^i)} D_- \left( \partial_{++} \Phi^i  \right)  \nonumber \\
	&\quad - \frac{1}{2} \partial_{++} \Phi^i D_- \left( \frac{\delta \mathcal{A}}{\delta (D_+ D_- \Phi^i)} \right) - D_+ \mathcal{A}  , \nonumber \\
    \mathcal{T}_{+++} &= \partial_{++} \Phi^i \frac{\delta \mathcal{A}}{\delta (D_- \Phi^i)} + D_- \left(  \partial_{++} \Phi^i  \frac{\delta \mathcal{A}}{\delta (\partial_{--} \Phi^i)} \right)  - \frac{1}{2} \frac{\delta \mathcal{A}}{\delta (D_+ D_- \Phi^i)} D_+ \left( \partial_{++} \Phi^i \right) \nonumber \\
    &\quad + \frac{1}{2} \partial_{++} \Phi^i D_+ \left( \frac{\delta \mathcal{A}}{\delta (D_+ D_- \Phi^i)} \right) , \nonumber \\
    \mathcal{T}_{---} &= \partial_{--} \Phi^i \frac{\delta \mathcal{A}}{\delta (D_+ \Phi^i)} + D_+ \left( \partial_{--} \Phi^i \frac{\delta \mathcal{A}}{\delta (\partial_{++} \Phi^i)} \right) + \frac{1}{2} \frac{\delta \mathcal{A}}{\delta (D_+ D_- \Phi^i)} D_- \left( \partial_{--} \Phi^i \right) \label{final_ttbar_general} \\
    &\quad - \frac{1}{2} \partial_{--} \Phi^i D_- \left( \frac{\delta \mathcal{A}}{\delta (D_+ D_- \Phi^i)} \right) , \nonumber \\
    \mathcal{T}_{--+} &=  \partial_{--} \Phi^i \frac{\delta \mathcal{A}}{\delta (D_- \Phi^i)} + D_- \left( \partial_{--} \Phi^i \frac{\delta \mathcal{A}}{\delta (\partial_{--} \Phi^i)} \right) - \frac{1}{2} \frac{\delta \mathcal{A}}{\delta (D_+ D_- \Phi^i)} D_+ \left( \partial_{--} \Phi^i \right) \nonumber \\
    &\quad + \frac{1}{2} \partial_{--} \Phi^i D_+ \left( \frac{\delta \mathcal{A}}{\delta (D_+ D_- \Phi^i)} \right) - D_- \mathcal{A} \nonumber . 
\end{align}
We interpret the superfield $\mathcal{T}$ as a conserved superspace supercurrent, since it satisfies the conservation equations
\begin{align}\label{susy_conservation}
    D_+ \mathcal{T}_{++-} + D_- \mathcal{T}_{+++} = 0 \, , \qquad D_+ \mathcal{T}_{---} + D_- \mathcal{T}_{--+} = 0 \, .
\end{align}
Writing the superfield equation (\ref{susy_conservation}) in components reduces to the usual conservation equation for the stress tensor, $\partial^\mu T_{\mu \nu} = 0$, along with other equations related to this one by supersymmetry.

\subsection{Supercurrent-Squared flow for $n$ Scalars}

Next we define a superspace deformation which is built from bilinears in $\mathcal{T}$. If we write the superspace Lagrangian as $\mathcal{A}$, so that
\begin{align}
    S = \int d^2 x \, d^2 \theta \, \mathcal{A} \, , 
\end{align}
then the flow equation generated by the supercurrent-squared operator is
\begin{align}\label{supercurrent_squared_our_notation}
    \frac{\partial \mathcal{A} ( \lambda )}{\partial \lambda} = \mathcal{T}_{+++}^{(\lambda)} \mathcal{T}_{---}^{(\lambda)} - \mathcal{T}_{--+}^{(\lambda)} \mathcal{T}_{++-}^{(\lambda)} \, , 
\end{align}
where the superscript $(\lambda)$ is meant to emphasize that the supercurrent $\mathcal{T}^{(\lambda)}$ must be re-computed from $\mathcal{A} ( \lambda )$ at each point along the flow, rather than using the supercurrent $\mathcal{T}^{(0)}$ of the undeformed theory, as with the ordinary $\TT$ flow.

To get intuition for the structures in the superspace Lagrangian which will be generated by this deformation, it is helpful to write out the deforming operator to leading order in $\lambda$ in a particular example. We will focus on the $2d$ field theory whose dimensional reduction produces an undeformed superspace action of the form (\ref{susy_particle_on_group}), which is the theory of a collection of superfields $\Phi^i$ with the superspace Lagrangian
\begin{align}\label{undeformed_2d_susy_theory}
    S = \int d^2 x \, d^2 \theta \, g_{ij} ( \Phi ) \, D_+ \Phi^i D_- \Phi^j \, .
\end{align}
Computing the supercurrent components for this theory, one finds
\begin{align}
	\mathcal{T}_{++-}^{(0)} &=  \left( \partial_{++} \Phi^i \right) g_{ij} D_- \Phi^j - D_+ \left( g_{ij} D_+ \Phi^i D_- \Phi^j \right) , \nonumber \\
    \mathcal{T}_{+++}^{(0)} &= - \left( \partial_{++} \Phi^i \right) g_{ij} D_+ \Phi^j , \nonumber \\
    \mathcal{T}_{---}^{(0)} &= \left( \partial_{--} \Phi^i \right) g_{ij} D_- \Phi^j, \nonumber \\
    \mathcal{T}_{--+}^{(0)} &=  - \left( \partial_{--} \Phi^i \right) g_{ij} D_+ \Phi^j - D_- \left( g_{ij} D_+ \Phi^i D_- \Phi^j \right) \, .
\end{align}
Therefore, we see that the leading correction to $\mathcal{A}$ from the supercurrent-squared flow equation is $\mathcal{A}^{(0)} \longrightarrow \mathcal{A}^{(0)} + \lambda \mathcal{A}^{(1)}$ where
\begin{align}\label{leading_sc2_2d}
    \mathcal{A}^{(1)} &= \mathcal{T}_{+++}^{(0)} \mathcal{T}_{---}^{(0)} - \mathcal{T}_{--+}^{(0)} \mathcal{T}_{++-}^{(0)} \nonumber \\
    &= - g_{ij} g_{kl} \left( \partial_{++} \Phi^i \right) \left( \partial_{--} \Phi^k \right) D_+ \Phi^j D_- \Phi^l + g_{ij} g_{kl} ( \partial_{--} \Phi^i ) (\partial_{++} \Phi^k ) D_+ \Phi^j D_- \Phi^l \nonumber \\
    &\quad - ( \partial_{--} \Phi^i ) g_{ij} D_+ \Phi^j D_+ \left( g_{kl} D_+ \Phi^k D_- \Phi^l \right) - ( \partial_{++} \Phi^i ) g_{ij} D_- \Phi^j D_- \left( g_{kl} D_+ \Phi^k D_- \Phi^l \right) \nonumber \\
    &\quad + D_+ \left(  g_{ij} D_+ \Phi^i D_- \Phi^j \right) D_- \left( g_{kl} D_+ \Phi^k D_- \Phi^l \right)  \, .
\end{align}
The leading deformation (\ref{leading_sc2_2d}) contains terms proportional to the undeformed Lagrangian (\ref{undeformed_2d_susy_theory}) in addition to new terms which have more than two fermions. For instance, terms involving $D_+ \Phi^i D_- \Phi^j D_+ \Phi^k D_- \Phi^l$ will be generated. The full solution for the finite-$\lambda$ deformed superspace Lagrangian will therefore take the schematic form
\begin{align}\label{finite_lambda_2d_schematic}
    \mathcal{A}^{(\lambda)} = F_1 ( D \Phi )^2 + F_2 ( D \Phi )^4 + \cdots + F_n ( D \Phi )^{2n} \, ,
\end{align}
where each of the functions $F_i$ depends on a collection of Lorentz scalars built from the $\Phi^j$ and their derivatives, and the expression $(D \Phi)^{2k}$ is shorthand for a product of the form $D_+ \Phi^{i_1} \cdots D_- \Phi^{i_{2k}}$. This expansion is only schematic; for instance, there can be multiple inequivalent ways of constructing a term $( D \Phi )^{2k}$ by changing which fields in the product are acted on by $D_+$ and which are acted on by $D_-$, and all such inequivalent combinations can appear in principle. Three examples of Lorentz scalars on which the functions $F_i$ can depend are
\begin{align}\label{xyz_def}
    x &= \lambda g_{ij} ( \Phi ) \partial_{++} \Phi^i \partial_{--} \Phi^j \, , \nonumber \\
    y &= \lambda g_{ij} ( \Phi ) \left( D_+ D_- \Phi^i \right) \left( D_+ D_- \Phi^j \right) \, , \nonumber \\
    z &= \lambda^2 \left( g_{ij} \partial_{++} \Phi^i \partial_{++} \Phi^j \right) \left( g_{mn} \partial_{--} \Phi^m \partial_{--} \Phi^n \right) \, .
\end{align}
The number $n$ appearing in the highest term of (\ref{finite_lambda_2d_schematic}) is the same as the number of scalars $\Phi^i$, since the $2n$ possible derivatives of the form $D_{\pm} \Phi^i$ are all fermionic quantities and thus any term with a product of more than $2n$ such factors must vanish by nilpotency. 

The general flow equation for $\mathcal{A}^{(\lambda)}$ induced by supercurrent-squared will therefore yield a complicated set of partial differential equations relating the various $F_i$ and their derivatives with respect to the several independent scalars. We will not undertake an analysis of this general case here. However, we can make some comments about the most fermionic term in the action. First note that there is only one independent term that one can write down involving $2n$ copies of $\Phi$, which is simply
\begin{align}
    D_+ \Phi^1 D_- \Phi^1 \, \cdots \, D_+ \Phi^n D_- \Phi^n \, .
\end{align}
This is in contrast to other terms like $( D \Phi )^4$ for which \emph{a priori} it appears that multiple inequivalent expressions can be written down, like
\begin{align}
    D_+ \Phi^i D_- \Phi^j D_+ \Phi^k D_- \Phi^l \text{ and } D_+ \Phi^i D_+ \Phi^j D_- \Phi^k D_- \Phi^l \, , 
\end{align}
which need not yield the same contribution when contracted against a general $f_{ijkl}$ without any special symmetry properties. 

Next, we claim that -- if we are willing to go partially on-shell by imposing one implication of the equations of motion in the Lagrangian -- the function $F_n$ multiplying the unique term $(D \Phi)^{2n}$ can be taken to be independent of the scalar $y$ in (\ref{xyz_def}). To see this, we will begin with the superspace equation of motion (\ref{susy_eom}) and then multiply both sides by the most fermionic term $(D \Phi)^{2n}$. The left side of the equation of motion is $\frac{\delta \mathcal{A}}{\delta \Phi^i}$, which is a sum of terms of the form
\begin{align}\label{dA_dPhi_eqn}
    \frac{\delta \mathcal{A}}{\delta \Phi^i} = \sum_k \left[ \sum_j  \frac{\partial F_k}{\partial x_j} \frac{\partial x_j}{\partial \Phi^i} ( D \Phi )^{2k} + F_k \cdot \frac{\partial ( D \Phi )^{2k}}{\partial g_{nm}} \frac{\partial g_{nm}}{\partial \Phi^i} \right] \, .
\end{align}
Here $x_j$ are the collection of Lorentz scalars that the coefficient functions $F_k$ can depend on. This equation is again only schematic and the details of these scalars $x_j$ are not important. The only important point is that every term in (\ref{dA_dPhi_eqn}) contains at least two fermions, since taking the derivative of any term in $\mathcal{A}$ with respect to some $\Phi^i$ will not change the number of fermions in that term. Therefore, when we multiply by the maximally fermionic term $(D \Phi)^{2n}$, all terms in (\ref{dA_dPhi_eqn}) vanish by nilpotency. Similarly, the two terms
\begin{align}\label{spacetime_deriv_terms}
    \partial_{++} \left( \frac{\delta \mathcal{A}^{(\lambda)}}{\delta (\partial_{++} \Phi^i)} \right) \, , \qquad  \partial_{--} \left( \frac{\delta \mathcal{A}^{(\lambda)}}{\delta (\partial_{--} \Phi^i)} \right)
\end{align}
appearing on the right side of the equation of motion will also vanish when multiplied by $(D \Phi)^{2n}$. This is because every term in either of (\ref{spacetime_deriv_terms}) will be proportional either to some product $D_+ \Phi^i D_- \Phi^j$, or to a factor of the form $ \left( \partial_{++} D_+ \Phi^i \right) D_- \Phi^j$, and in either case such a term is proportional to at least one of the $2n$ fermions $D_{\pm} \Phi^i$.

Dropping these terms that do not contribute, we can write
\begin{align}\label{susy_eom_multiplied_intermediate}
	0 & =  ( D \Phi )^{2n} \, \left[ D_+ \left( \frac{\delta \mathcal{A}^{(\lambda)}}{\delta (D_+ \Phi^i)} \right) + D_- \left( \frac{\delta \mathcal{A}^{(\lambda)}}{\delta (D_- \Phi^i)} \right) - D_+ D_- \left( \frac{\delta \mathcal{A}^{(\lambda)}}{\delta ( D_+ D_- \Phi^i)} \right) \right] \, .
\end{align}
Furthermore, we claim that the only term in the superspace Lagrangian which affects the right side of (\ref{susy_eom_multiplied_intermediate}) is the lowest term involving $F_1$. For any term involving four or more fermions, the three combinations inside the brackets of (\ref{susy_eom_multiplied_intermediate}) will all contain at least two fermions and therefore vanish when multiplying $(D \Phi)^{2n}$. The only term which survives is the one arising from $F_1$, which gives
\begin{align}\label{susy_eom_multiplied_intermediate_two}
	0 & = 2 ( D \Phi )^{2n} \, ( g_{il} ( D_+ D_- 
\Phi^l ) ) \,  \left[ F_1  + \lambda \frac{\partial F_1}{\partial y}  g_{km} D_+ D_- \Phi^k D_+ D_- \Phi^m \right] \, .
\end{align}
Thus, when multiplying $(D \Phi)^{2n}$ and on-shell, either the combination $F_1  + y \frac{\partial F_1}{\partial y}$ vanishes or the object $g_{il} ( D_+ D_- 
\Phi^l )$ vanishes. The former cannot hold identically since it fails near the free theory where $F_1 = 1$. Therefore we conclude that the combination $g_{il} ( D_+ D_- 
\Phi^l )$ can be set to zero when multiplying $(D \Phi)^{2n}$ as a consequence of the equations of motion, and as a result the scalar $y$ (which is proportional to this combination) can be set to zero in this context as well. In particular, since we may view the most fermionic term in the Lagrangian as a Taylor series in $y$ via
\begin{align}\label{most_fermionic_taylor}
    F_n ( D \Phi )^{2n} = \left( F_n  \Big\vert_{y=0}  + y \cdot \frac{\partial F_n}{\partial y} \Big\vert_{y=0} + \cdots \right) ( D \Phi )^{2n} \, ,
\end{align}
we see that all terms but the first can be set to zero on-shell. Thus we are free to impose that the function $F_n$ be independent of $y$ when the equations of motion are satisfied. We note that this trick of simplifying $\TT$-like flows by going partially on-shell using the superspace equations of motion was first used in the series of works \cite{Baggio:2018rpv,Chang:2019kiu,Ferko:2019oyv}. In terms of component fields, imposing this implication of the superspace equations of motion is equivalent to integrating out the auxiliary fields using their (algebraic) equations of motion. 

\subsection{Solution for One Scalar}

Finally we will specialize to a case where we can explicitly solve the flow equation and dimensionally reduce the result, which will provide a check for the $(0+1)$-dimensional deformation that we will introduce in Section \ref{method3}. This is the case of a single scalar field $\Phi$. The undeformed Lagrangian is simply
\begin{align}
    \mathcal{A}^{(0)} = g ( \Phi ) \, D_+ \Phi \, D_- \Phi \, .
\end{align}
Following the definitions (\ref{xyz_def}) in the general case, we define
\begin{align}\label{xyz_def_later}
    x &= \lambda g ( \Phi ) \partial_{++} \Phi \partial_{--} \Phi \, , \nonumber \\
    y &= \lambda g ( \Phi ) \left( D_+ D_- \Phi \right) \left( D_+ D_- \Phi \right) \, , 
\end{align}
and make an ansatz for the finite-$\lambda$ solution of the form
\begin{align}
    \mathcal{A}^{(\lambda)} = F ( x, y ) \, g ( \Phi ) \, D_+ \Phi \, D_- \Phi \, .
\end{align}
In the case of a single scalar, the two-fermion term $D_+ \Phi D_- \Phi$ is also the most fermionic term that one can construct. Therefore, in view of the general result discussed around equation (\ref{most_fermionic_taylor}), we can assume that the function $F(x, y)$ is independent of $y$ up to terms which vanish on-shell.

Next we compute the components of the supercurrents $\mathcal{T}_{\pm \pm \pm}$, $\mathcal{T}_{\pm \pm \mp}$. Since we will ultimately drop dependence on the variable $y$, we will omit any terms proportional to $y$ in the supercurrents.\footnote{The full flow equation, including dependence on $y$, can be found in \cite{Chang:2018dge}.} We will also drop terms which that are proportional to $D_+ \Phi D_- \Phi$, since every term in the supercurrents is at least proportional to either $D_+ \Phi$ or $D_- \Phi$, and therefore terms which contain both fermionic quantities will not contribute to bilinears because they vanish by nilpotency when multiplied against another component of $\mathcal{T}$. For instance, $\mathcal{T}_{++-}$ contains a term
\begin{align}
    D_+ \left(  \partial_{++} \Phi \frac{\delta \mathcal{A}}{\delta \partial_{++} \Phi} \right) = D_+ \left( \left( \partial_{++} \Phi \right) \lambda \frac{\partial F}{\partial x} \partial_{--} \Phi g(\Phi)^2 D_+ \Phi D_- \Phi \right) \, .
\end{align}
Although in principle this generates several terms when the $D_+$ acts on each factor, the only relevant one for bilinears is the term where it acts on $D_+ \Phi$ to give $D_+^2 \Phi = \partial_{++} \Phi$. Every other term will either be proportional to $D_+ \Phi D_- \Phi$ or to $y$ and therefore can be ignored.

Using the symbol $\sim$ to mean ``equal up to terms which are either proportional to $y$ or do not contribute to bilinears,'' we find
\begin{align}\label{supercurrents_for_bilinears}
    \mathcal{T}_{++-} &\sim ( \partial_{++} \Phi ) F g(\Phi) D_- \Phi + x \frac{\partial F}{\partial x} ( \partial_{++} \Phi ) g ( \Phi ) D_- \Phi - F g ( \Phi ) \partial_{++} \Phi D_- \Phi \, , \nonumber \\
    \mathcal{T}_{+++} &\sim - ( \partial_{++} \Phi ) F g ( \Phi )  D_+ \Phi - x \frac{\partial F}{\partial x} ( \partial_{++} \Phi ) g ( \Phi ) D_+ \Phi \, , \nonumber \\
    \mathcal{T}_{---} &\sim ( \partial_{--} \Phi ) F g ( \Phi ) D_- \Phi + x \frac{\partial F}{\partial x} ( \partial_{--} \Phi ) g ( \Phi ) D_- \Phi \, , \nonumber \\
    \mathcal{T}_{--+} &\sim - ( \partial_{--} \Phi ) F g ( \Phi ) D_+ \Phi - x \frac{\partial F}{\partial x} ( \partial_{--} \Phi ) g ( \Phi ) D_+ \Phi + F g ( \Phi ) ( \partial_{--} \Phi ) D_+ \Phi \, .
\end{align}
On the other hand, when we ignore dependence on $y$ the $\lambda$-derivative of $\mathcal{A}$ is simply
\begin{align}
    \frac{\partial \mathcal{A}^{(\lambda)}}{\partial \lambda} = g ( \Phi )^2  ( \partial_{++} \Phi ) ( \partial_{--} \Phi ) \frac{\partial F}{\partial x} \, D_+ \Phi \, D_- \Phi \, .
\end{align}
Equating this with the combination of supercurrents appearing on the right side of (\ref{supercurrent_squared_our_notation}), using their expressions (\ref{supercurrents_for_bilinears}), we arrive at the simple differential equation
\begin{align}
    0 = \frac{\partial F}{\partial x} + F^2 + 2 x F \frac{\partial F}{\partial x} \, .
\end{align}
The solution is
\begin{align}
    F(x) = \frac{1}{2x} \left( -1 + \sqrt{ 1 + 4 x } \right) \, .
\end{align}
Therefore the finite-$\lambda$ solution to the flow equation is on-shell equivalent to
\begin{align}
    \mathcal{A}^{(\lambda)} = \frac{1}{2 x} \left( -1 + \sqrt{ 1 + 4 x } \right) g ( \Phi ) D_+ \Phi D_- \Phi \, .
\end{align}
For $g(\Phi) = 1$, this solution was first obtained in \cite{Baggio:2018rpv}. It is easy to dimensionally reduce this result to quantum mechanics. We assume that the superfield $\Phi ( x, t )$ is independent of the spatial coordinate $x$. It is convenient to express the spacetime derivatives $\partial_{\pm \pm}$ acting on $\Phi$ in terms of supercovariant derivatives using the algebra $D_{\pm} D_{\pm} = \partial_{\pm \pm}$, so that
\begin{align}
    x = \lambda g ( \Phi ) ( D_+ D_+ \Phi ) ( D_- D_- \Phi ) \, .
\end{align}
Following our conventions for dimensional reduction in Section \ref{sec:conventions}, we will rotate to complex supercovariant derivatives defined by
\begin{align}
    D = \frac{1}{\sqrt{2}} \left( D_+ + i D_- \right) \, , \quad \Dbar = \frac{1}{\sqrt{2}} \left( D_+ - i D_- \right) \, , 
\end{align}
Then one finds that
\begin{align}
    D_+ \Phi D_- \Phi &= i \, D \Phi \, \Dbar \Phi \, , \nonumber \\
    D_+ D_+ \Phi = D_-D_- \Phi &= \frac{1}{2} ( D \Dbar + \Dbar D ) \Phi \, , \nonumber \\
    D_+ D_- \Phi &= \frac{i}{2} ( \Dbar D - D \Dbar ) \Phi \, .
\end{align}
In this notation, the on-shell condition which allows us to set $D_+ D_- \Phi = 0$ in terms which multiply $D_+ \Phi D_- \Phi$ means that we can replace $D \Dbar \Phi$ with $\Dbar D \Phi$ (and vice-versa) in terms which multiply $D \Phi \Dbar \Phi$. Therefore we can write $x$ in several on-shell equivalent ways as
\begin{align}
    x = \lambda g ( \Phi ) \left( D \Dbar \Phi \right)^2 = \lambda g ( \Phi ) \left( \Dbar D \Phi \right)^2 = \lambda g ( \Phi ) \left( D \Dbar \Phi \right) \left( \Dbar D \Phi \right) \, .
\end{align}
We will choose the last of these rewritings because it is more symmetrical. After dimensionally reducing and absorbing some irrelevant constant factors, we arrive at a deformed $(0+1)$-dimensional theory with the action
\begin{align}\label{dimensionally_reduced_answer}
    S = \int \, dt \, d \theta \, d \thetab \, \frac{1}{2 \lambda g ( \Phi ) \left( D \Dbar \Phi \right) \left( \Dbar D \Phi \right)} \left( -1 + \sqrt{ 1 + 4 \lambda g ( \Phi ) \left( D \Dbar \Phi \right) \left( \Dbar D \Phi \right) } \right) g ( \Phi ) \, D \Phi \, \Dbar \Phi \, .
\end{align}

\section{Reduction of $2d$ Supercurrent-Squared Operator}\label{method2}

In this subsection, we will follow a slightly different strategy. Rather than solving the flow driven by supercurrent-squared in two dimensions, and then dimensionally reducing the solution, we will aim to dimensionally reduce the supercurrent-squared operator itself. This will suggest a supersymmetric version of the $f(H)$ operator which can be applied directly in the superspace of a $(0+1)$-dimensional theory. Later in Section \ref{method3} we will see how to identify this dimensionally reduced operator as a function of certain conserved superfields that can be obtained from a Noether procedure.

\subsection{Trace Flow Equation}

In order to dimensionally reduce supercurrent-squared, we would like to eliminate some of the components of the superfield $\mathcal{T}$. This process is analogous to that reviewed in Section \ref{subsec:gross_review} for the dimensional reduction of the non-supersymmetric $\TT$. In that context, it was convenient to use the trace flow equation
\begin{align}\label{trace_flow_later}
    \tensor{T}{^\mu_\mu} = - 2 \lambda \left( T^{\mu \nu} T_{\mu \nu} - \left( \tensor{T}{^\mu_\mu} \right)^2 \right) \, , 
\end{align}
in order to solve for the spatial component $T_{xx}$ of the stress tensor as
\begin{align}\label{solve_trace}
    \tensor{T}{^x_x} = \frac{\tensor{T}{^t_t} + 4 \lambda T^{tx} T_{tx} }{4 \lambda \tensor{T}{^t_t} - 1 } \, , 
\end{align}
where the coordinates of the $2d$ spacetime $(t, x)$ are related to the light-cone coordinates by $x^{\pm \pm} = \frac{1}{\sqrt{2}} \left( t \pm x \right)$. We note that the trace flow equation only holds if the seed theory is conformal.

Next we will motivate a superspace analogue of this trace flow relation. First we recall the interpretation of the components in the expansion of the supercurrents $\mathcal{T}_{\pm \pm \pm}$, $\mathcal{T}_{\mp \mp \pm}$. It was argued in \cite{Chang:2018dge} that, on-shell, these superfields can be written as
\begin{align}
\begin{split}
    \mathcal{T}_{+++} &= - S_{+++} - \theta^+ T_{++++} - \theta^- Z_{++} + \theta^+ \theta^- \partial_{++} S_{-++} , \label{s-multiplet} \\
    \mathcal{T}_{---} &= S_{---} + \theta^+ Z_{--} + \theta^- T_{----} + \theta^+ \theta^- \partial_{--} S_{+--} , \\
    \mathcal{T}_{++-} &= S_{-++} + \theta^+ Z_{++} + \theta^- T_{++--} - \theta^+ \theta^- \partial_{++} S_{+--} , \\
    \mathcal{T}_{--+} &= - S_{+--} - \theta^+ T_{++--} - \theta^- Z_{--} - \theta^+ \theta^- \partial_{--} S_{-++} . 
\end{split}
\end{align}
Here $T_{\mu \nu}$ is the stress tensor, $S_{\mu \alpha}$ is the conserved current associated with supersymmetry transformations,\footnote{$S_{\mu \alpha}$ is often called the supercurrent, although we reserve that term for superfields like $\mathcal{T}_{\pm \pm \pm}$.} and $Z_\mu$ is a vector  which is associated with a scalar central charge.

Because we will reduce along the spatial coordinate $x$, it will be convenient to change from $x^{\pm \pm}$ to $x, t$ coordinates. First we want to act with various $D$ operators in order to construct superfields whose lowest components are stress tensors. If we define
\begin{align}
    \widetilde{\mathcal{T}}_{++++} = - D_+ \mathcal{T}_{+++} \, , \qquad \widetilde{\mathcal{T}}_{----} = D_- \mathcal{T}_{---} \, , \nonumber \\
    \widetilde{\mathcal{T}}_{++--} = D_- \mathcal{T}_{++-} \, , \qquad \widetilde{\mathcal{T}}_{--++} = - D_+ \mathcal{T}_{--+} \, ,
\end{align}
then the lowest components of these superfields are simply
\begin{align}\label{calt_4_index_defn}
    \widetilde{\mathcal{T}}_{++++} \Big\vert_{\theta =0 } = T_{++++} \, , \qquad \widetilde{\mathcal{T}}_{----} \Big\vert_{\theta =0 } = T_{----} \, , \nonumber \\
    \widetilde{\mathcal{T}}_{++--} \Big\vert_{\theta = 0 } = T_{++--} \, , \qquad \widetilde{\mathcal{T}}_{--++} \Big\vert_{\theta =0 } = T_{++--} \, .
\end{align}
Note that symmetry of the stress tensor implies $\calTt_{--++} = \calTt_{++--}$. Another way to see this is to note that the supercurrents are related to fields of the $S$ multiplet by $\mathcal{T}_{++-} = \chi_+$, $\mathcal{T}_{--+} = - \chi_-$, and the $S$ multiplet fields satisfy the constraint $D_- \chi_+ = D_+ \chi_-$. One can then change coordinates to $(t, x)$ to find
\begin{gather}
    \calTt_{tt} = \frac{1}{2} \left( \calTt_{++++} + 2 \calTt_{++--}  + \calTt_{----} \right) \, , \qquad \calTt_{tx} = \frac{1}{2} \left( \calTt_{++++} - \calTt_{----} \right) \, , \nonumber \\
    \calTt_{xx} = \frac{1}{2} \left( \calTt_{++++} - 2 \calTt_{++--}  + \calTt_{----} \right) \, .
\end{gather}
The superfield equation whose lowest component is (\ref{solve_trace}) is
\begin{align}\label{elim_cal_Txx}
    \calTt_{xx} = \frac{\calTt_{tt} + 4 \lambda \calTt_{tx}^2}{4 \lambda \calTt_{tt} - 1 } \, .
\end{align}
This is the desired superspace analogue of the trace flow equation. We will assume that it holds as an exact superfield expression, at least on-shell. Furthermore, as in the dimensional reduction of the non-supersymmetric $\TT$ operator, we will assume that $T_{tx} = 0$ and therefore $\calTt_{tx} = 0$, which implies that
\begin{align}\label{no_Ttx_condition}
    \calTt_{++++} = \calTt_{----} \, .
\end{align}
We note that from this point on, the number of $+$ and $-$ indices in our equations need no longer match because we are explicitly breaking Lorentz invariance by forcing $\calTt_{tx} = 0$. However, this is unproblematic since our goal is to single out the $x$ direction for dimensional reduction, which is also not a Lorentz-invariant procedure.

As a consequence of the condition (\ref{no_Ttx_condition}), we may write the other components of $\calTt_{\mu \nu}$ as
%s
\begin{align}\label{calTt_simplified_consequence}
    \calTt_{tt} &= \calTt_{++++} + \calTt_{++--} = - D_+ \left( \mathcal{T}_{+++} + \mathcal{T}_{--+} \right) = D_- \left( \mathcal{T}_{---} + \mathcal{T}_{++-} \right)  \, , \nonumber \\
    \calTt_{xx} &= \calTt_{++++} - \calTt_{++--} = D_+ \left( - \mathcal{T}_{+++} + \mathcal{T}_{--+} \right) = D_- \left( \mathcal{T}_{---} - \mathcal{T}_{++-} \right) \, .
\end{align}

\subsection{Rewriting of Supercurrent-Squared}

Next we would like to explain a relationship between the product $\calTt_{tt} \calTt_{xx}$ and the supercurrent-squared operator, which holds under our assumptions thus far. To organize this calculation, it is helpful to list the on-shell constraints relating the various objects created from one supercovariant derivative acting on one component of $\mathcal{T}$. There are na\"ively $8$ such objects, but there are four constraints.
\begin{align}
    \label{relation1} &\text{By conservation, } D_- \mathcal{T}_{+++} = - D_+ \mathcal{T}_{++-} \text{ and } D_+ \mathcal{T}_{---} = - D_- \mathcal{T}_{--+} \, . \\
    \label{relation2} &\text{By the assumption that } \calTt_{tx} = 0 \text{, we have } D_+ \mathcal{T}_{+++} = - D_- \mathcal{T}_{---} \, , \\
    \label{relation3} &\text{By symmetry of the stress tensor, } D_- \mathcal{T}_{++-} = - D_+ \mathcal{T}_{--+}
\end{align}
Therefore there are, in fact, only four independent objects of the form $D \mathcal{T}$ after imposing these conditions. Further, by acting with a second supercovariant derivative and using the algebra $D_\pm D_{\pm} = \partial_{\pm \pm}$, $\{ D_+, D_- \} = 0$,  we obtain the added constraints that
\begin{align}\begin{split}\label{spacetime_deriv_constraints}
    \partial_{++} \mathcal{T}_{+++} + \partial_{--} \mathcal{T}_{--+} = 0 \, , &\qquad \partial_{++} \mathcal{T}_{++-} + \partial_{--} \mathcal{T}_{---} = 0 \, ,  \\
    \partial_{--} \mathcal{T}_{+++} + \partial_{++} \mathcal{T}_{--+} = 0 \, , &\qquad \partial_{++} \mathcal{T}_{---} + \partial_{--} \mathcal{T}_{++-} = 0 \, .
\end{split}\end{align}
Next recall that, since $D_{\pm} = \frac{\partial}{\partial \theta^{\pm}} + \theta^{\pm} \partial_{\pm \pm}$, we can exchange a superspace integral for a supercovariant derivative, up to a total spacetime derivative. This allows us to write
\begin{align}\label{scsquare_reduction_intermediate}
    &\int \, d^2 x \, d^2 \theta \, \left( \mathcal{T}_{+++} \mathcal{T}_{---} - \mathcal{T}_{--+} \mathcal{T}_{++-} \right) \nonumber \\
    &\quad \sim \int \, d^2 x \, D_+ D_- \, \left( \mathcal{T}_{+++} \mathcal{T}_{---} - \mathcal{T}_{--+} \mathcal{T}_{++-} \right)  \, .
    % &\quad \sim \int d^2 x \, \left( D_- \mathcal{T}_{+++} D_+ \mathcal{T}_{---} - D_+ \mathcal{T}_{+++} D_- \mathcal{T}_{---} - D_- \mathcal{T}_{--+} D_+ \mathcal{T}_{++-} + D_+ \mathcal{T}_{--+} D_- \mathcal{T}_{++-} \right) ,
\end{align}
where $\sim$ means equivalence assuming we can ignore boundary terms. When we write expressions like those in the last line of (\ref{scsquare_reduction_intermediate}), which involve a superfield expression that is integrated over the spacetime coordinates $d^2 x$ but not over the superspace coordinates $d^2 \theta$, it is always implied that we mean to take the lowest component of the superfield expression.

When the combination $D_+ D_-$ acts on the combination in parentheses, we generate two types of terms: (I) terms with both supercovariant derivatives acting on a single superfield $\mathcal{T}$, and (II) terms involving one supercovariant derivative acting on each factor $\mathcal{T}$. We claim that all terms of type (I) can be ignored using our constraint equations (\ref{relation1}) - (\ref{relation2}) and (\ref{spacetime_deriv_constraints}) above, up to integration by parts. We will show this by considering each term explicitly. The type I terms are
\begin{align}\label{type_I_terms}
    \hspace{-5pt}\int d^2 x \, \Big[ ( D_+ D_- \mathcal{T}_{+++} ) \mathcal{T}_{---} + \mathcal{T}_{+++} ( D_+ D_- \mathcal{T}_{---} ) - ( D_+ D_- \mathcal{T}_{--+} ) \mathcal{T}_{++-} - \mathcal{T}_{--+} ( D_+ D_- \mathcal{T}_{++-} ) \Big] \, .
\end{align}
The first term in (\ref{type_I_terms}) can be rewritten as
\begin{align}
    \int d^2 x \, ( D_+ D_- \mathcal{T}_{+++} ) \mathcal{T}_{---} &= - \int d^2 x \, ( D_+ D_+ \mathcal{T}_{++-} ) \mathcal{T}_{---} \nonumber \\
    &= - \int d^2 x \, ( \partial_{++} \mathcal{T}_{++-} ) ( \mathcal{T}_{---} ) \, .
\end{align}
In the first step, we have used the conservation equation $D_- \mathcal{T}_{+++} = - D_+ \mathcal{T}_{++-}$; in the second step we use the algebra $D_+ D_+ = \partial_{++}$.

On the other hand, by a similar manipulation the third term can be written as
\begin{align}
    - \int d^2 x \left( D_+ D_- \mathcal{T}_{--+} \right) \mathcal{T}_{++-} &= \int d^2 x \, \left( D_+ D_+ \mathcal{T}_{---} \right) \mathcal{T}_{++-} \nonumber \\
    &= \int d^2 x \, \left( \partial_{++} \mathcal{T}_{---} \right) \mathcal{T}_{++-} \, .
\end{align}
In the first step we have used $D_- \mathcal{T}_{--+} = - D_+ \mathcal{T}_{---}$ and in the second step we have again used the algebra $D_+ D_+ = \partial_{++}$. Therefore, the sum of the first and third terms is
\begin{align}
    &\int d^2 x \Big[ ( \partial_{++} \mathcal{T}_{---} ) \mathcal{T}_{++-} + \mathcal{T}_{---} ( \partial_{++} \mathcal{T}_{++-} )  \Big] \nonumber \\
    &\qquad= \int d^2 x \, \partial_{++} \left( \mathcal{T}_{---} \mathcal{T}_{++-}  \right) \, , 
\end{align}
which is a total spacetime derivative and can be ignored.

We repeat this procedure for the remaining terms. Using similar arguments the second term can be written as
\begin{align}
    \int d^2 x \, \mathcal{T}_{+++} ( D_+ D_- \mathcal{T}_{---} ) &= - \int d^2 x \mathcal{T}_{+++} ( D_+ D_+ \mathcal{T}_{+++} ) \nonumber \\
    &= - \int d^2 x \mathcal{T}_{+++} ( \partial_{++} \mathcal{T}_{+++} ) \nonumber \\
    &= \int d^2 x \mathcal{T}_{+++} ( \partial_{--} \mathcal{T}_{--+} )
\end{align}
Likewise the fourth term is on-shell equivalent to
\begin{align}
    - \int d^2 x \, \mathcal{T}_{--+} ( D_+ D_- \mathcal{T}_{++-} ) &= \int d^2 x \, \mathcal{T}_{--+} ( D_- D_+ \mathcal{T}_{++-} ) \nonumber \\
    &= - \int d^2 x \, \mathcal{T}_{--+} ( D_- D_- \mathcal{T}_{+++} ) \nonumber \\
    &= - \int d^2 x \, \mathcal{T}_{--+} \partial_{--} \mathcal{T}_{+++} \, .
\end{align}
Adding the second and fourth terms gives
\begin{align}
    &\int d^2 x \Big[ \mathcal{T}_{+++} ( \partial_{--} \mathcal{T}_{--+} ) + (\partial_{--} \mathcal{T}_{+++} ) \mathcal{T}_{--+} \Big] \nonumber \\
    \qquad &= \int d^2 x \, \partial_{--} \left(  \mathcal{T}_{+++} \mathcal{T}_{--+} \right) \, ,
\end{align}
which is again a total spacetime derivative that we will drop.

The upshot is that all terms which involve two supercovariant derivatives acting on a single $\mathcal{T}$ will drop out, up to equations of motion and total derivatives. Therefore we are left with
\begin{align}\label{scsquare_reduction_intermediate_two}
    &\int \, d^2 x \, d^2 \theta \, \left( \mathcal{T}_{+++} \mathcal{T}_{---} - \mathcal{T}_{--+} \mathcal{T}_{++-} \right) \nonumber \\
    &\quad \sim \int d^2 x \, \left( D_- \mathcal{T}_{+++} D_+ \mathcal{T}_{---} - D_+ \mathcal{T}_{+++} D_- \mathcal{T}_{---} - D_- \mathcal{T}_{--+} D_+ \mathcal{T}_{++-} + D_+ \mathcal{T}_{--+} D_- \mathcal{T}_{++-} \right) .
\end{align}
The first and third terms of (\ref{scsquare_reduction_intermediate_two}) cancel after after using the conservation equation (\ref{relation1}):
\begin{align}
    &\int d^2 x \, \Big( D_- \mathcal{T}_{+++} D_+ \mathcal{T}_{---} - D_- \mathcal{T}_{--+} D_+ \mathcal{T}_{++-} \Big) \nonumber \\
    &\quad \sim \int d^2 x \, \Big( D_+ \mathcal{T}_{++-} D_- \mathcal{T}_{--+} - D_- \mathcal{T}_{--+} D_+ \mathcal{T}_{++-} \Big) = 0 \, .
\end{align}
We then arrive at the conclusion
\begin{align}\label{scsquare_reduction_intermediate_three}
    \int \, d^2 x \, d^2 \theta \, \left( \mathcal{T}_{+++} \mathcal{T}_{---} - \mathcal{T}_{--+} \mathcal{T}_{++-} \right) \sim \int d^2 x \, \left( D_+ \mathcal{T}_{--+} D_- \mathcal{T}_{++-}  - D_+ \mathcal{T}_{+++} D_- \mathcal{T}_{---} \right) .
\end{align}
We now compare this to the combination
\begin{align}\label{caltsquared_on_shell}
    \calTt_{tt} \calTt_{xx} &= \left( - D_+ \mathcal{T}_{+++} - D_+ \mathcal{T}_{--+} \right) \left( D_- \mathcal{T}_{---} - D_- \mathcal{T}_{++-} \right) \, \nonumber \\
    &= - D_+ \mathcal{T}_{+++} D_- \mathcal{T}_{---} + D_+ \mathcal{T}_{+++} D_- \mathcal{T}_{++-} - D_+ \mathcal{T}_{--+} D_- \mathcal{T}_{---} + D_+ \mathcal{T}_{--+} D_- \mathcal{T}_{++-} \, \nonumber \\
    &= D_+ \mathcal{T}_{--+} D_- \mathcal{T}_{++-}  - D_+ \mathcal{T}_{+++} D_- \mathcal{T}_{---} \, , 
\end{align}
where in the last step we have used that the second and third terms in the second line again cancel after using (\ref{relation1}) - (\ref{relation2}):
\begin{align}
    D_+ \mathcal{T}_{+++} D_- \mathcal{T}_{++-} - D_+ \mathcal{T}_{--+} D_- \mathcal{T}_{---} &= - D_- \mathcal{T}_{---} D_- \mathcal{T}_{++-} - D_+ \mathcal{T}_{--+} D_- \mathcal{T}_{---} \nonumber \\
    &= - D_- \mathcal{T}_{---} \left( D_- \mathcal{T}_{++-} + D_+ \mathcal{T}_{--+} \right) \nonumber \\
    &= 0 \, .
\end{align}
In the last step we have used symmetry of the stress tensor (\ref{relation3}). Therefore, comparing (\ref{scsquare_reduction_intermediate_three}) to (\ref{caltsquared_on_shell}), we see that on-shell one has
\begin{align}
    \int d^2 x \, d^2 \theta \, \left( \mathcal{T}_{+++} \mathcal{T}_{---} - \mathcal{T}_{--+} \mathcal{T}_{++-} \right) \sim \int d^2 x \, \calTt_{tt} \calTt_{xx} \, .
\end{align}
If we now impose the superfield analogue of the trace flow relation (\ref{elim_cal_Txx}), setting $\calTt_{tx} = 0$ as we have already assumed, then on-shell we can write supercurrent-squared as
\begin{align}\label{scsquare_reduction_intermediate_four}
    &\int d^2 x \, d^2 \theta \, \left( \mathcal{T}_{+++} \mathcal{T}_{---} - \mathcal{T}_{--+} \mathcal{T}_{++-} \right) \nonumber\\
    &\quad = \int d^2 x \, \frac{\calTt_{tt}^2}{ 4 \lambda \calTt_{tt} - 1 } \, \nonumber \\
    &\quad = \int d^2 x \, \frac{D_+ ( - \mathcal{T}_{+++} - \mathcal{T_{--+}} ) D_- ( \mathcal{T}_{---} + \mathcal{T}_{++-} ) }{ 4 \lambda D_+ ( - \mathcal{T}_{+++} - \mathcal{T_{--+}} ) - 1 } \, ,\nonumber \\
    &\quad = \int d^2 x \, D_+ D_- \Bigg[ \frac{(\mathcal{T}_{+++} + \mathcal{T_{--+}} ) ( \mathcal{T}_{---} + \mathcal{T}_{++-} ) }{ 4 \lambda D_+ ( \mathcal{T}_{+++} + \mathcal{T_{--+}} ) + 1 } \Bigg] \, ,\nonumber \\
    &\quad = \int d^2 x \, d^2 \theta \, \frac{ ( \mathcal{T}_{+++} + \mathcal{T_{--+}} ) ( \mathcal{T}_{---} + \mathcal{T}_{++-} ) }{ 4 \lambda D_+ (  \mathcal{T}_{+++} + \mathcal{T_{--+}} ) + 1 } \, .
\end{align}
In the second step we have used the expressions (\ref{calTt_simplified_consequence}) for $\calTt_{tt}$. In the middle step, we have pulled out an overall pair of supercovariant derivatives; this manipulation relies on the fact that additional type I terms which may have been generated when two supercovariant derivatives hit a single factor of $\mathcal{T}$ in the numerator all drop out by a similar calculation as the one presented above, where we saw that such type I terms in (\ref{type_I_terms}) did not contribute. There are also no additional terms generated when the supercovariant derivatives act on the denominator. We will see a simple way to understand why the denominator does not generate additional terms when we present the interpretation of this combination in terms of (complex) conserved charges in the dimensionally reduced theory.

\subsection{Dimensional Reduction and Interpretation}

We have now written the supercurrent-squared operator in a form which is suitable for dimensional reduction, since the combination appearing in (\ref{scsquare_reduction_intermediate_three}) is a function only of $\calTt_{tt}$ and not of any $x$-components of the supercurrents. We may therefore assume that all superfields are independent of $x$ and perform the $dx$ integral, which yields a constant. We then arrive at an expression for a supercurrent-squared operator in the $(0+1)$-dimensional quantum mechanics theory:
\begin{align}\label{scsquare_reduction_intermediate_five}
    \int d t \, d^2 \theta \, \frac{ ( \mathcal{T}_{+++} + \mathcal{T_{--+}} ) ( \mathcal{T}_{---} + \mathcal{T}_{++-} ) }{ 4 \lambda D_+ ( \mathcal{T}_{+++} + \mathcal{T_{--+}} ) + 1 } \, .
\end{align}
To aid interpretation, we will define the auxiliary quantities
\begin{align}
    \mathcal{Q}_+ = \mathcal{T}_{+++} + \mathcal{T}_{--+} \, , \qquad \mathcal{Q}_- = \mathcal{T}_{---} + \mathcal{T}_{++-} \, .
\end{align}
These satisfy the conservation equation
\begin{align}
    D_- \mathcal{Q}_+ + D_+ \mathcal{Q}_- = D_- \mathcal{T}_{+++} + D_- \mathcal{T}_{--+} + D_+ \mathcal{T}_{---} + D_+ \mathcal{T}_{++-} = 0 \, ,
\end{align}
as a consequence of the conservation equations for $\mathcal{T}$. They also obey
\begin{align}
    D_+ \mathcal{Q}_+ = D_+ \mathcal{T}_{+++} + D_+ \mathcal{T}_{--+} = - D_- \mathcal{T}_{---} - D_- \mathcal{T}_{++-} = - D_- \mathcal{Q}_- \, ,
\end{align}
due to the conditions (\ref{relation2}) and (\ref{relation3}). In terms of the $\mathcal{Q}_{\pm}$, the deformation (\ref{scsquare_reduction_intermediate_five}) is
\begin{align}\label{scsquare_reduction_final}
    \int d t \, d^2 \theta \, \frac{ \mathcal{Q}_+ \mathcal{Q}_- }{ 4 \lambda D_+ \mathcal{Q}_+ + 1 } \, .
\end{align}
We define the combination under the integral in (\ref{scsquare_reduction_final}) as $f ( \mathcal{Q}_+ , \mathcal{Q}_- ) = \frac{ \mathcal{Q}_+ \mathcal{Q}_- }{ 4 \lambda D_+ \mathcal{Q}_+ + 1 }$. This is a manifestly supersymmetric deformation of the $(0+1)$-dimensional supersymmetric quantum mechanics theory constructed from objects $\mathcal{Q}_{\pm}$ which satisfy certain conservation equations and therefore resemble conserved currents.

In order to compare with the common conventions for supersymmetric quantum mechanics, which use a complex Grassmann coordinate $\theta, \thetab$ rather than $\theta^{\pm}$, we will now translate (\ref{scsquare_reduction_final}) into this new notation. The details of this change of variables are described in Appendix \ref{app:change_to_complex}; here we will simply summarize the results. In complex coordinates, the operator (\ref{scsquare_reduction_intermediate_five}) is on-shell proportional to
\begin{align}\label{scsquare_reduction_final_complex}
    \int \, dt \, d^2 \theta \, \frac{\mathcal{Q} \calQbar}{\frac{1}{2} - 2 \lambda \Dbar \mathcal{Q}} \, .
\end{align}
Similarly, we call this combination $f(\mathcal{Q}, \calQbar) = \frac{\mathcal{Q} \calQbar}{\frac{1}{2} - 2 \lambda \Dbar \mathcal{Q}}$. The new supercurrents $\mathcal{Q}$ and $\calQbar$ satisfy the conservation equation $\Dbar \mathcal{Q} + D \calQbar = 0$ as shown in Appendix \ref{app:change_to_complex}. Since the complex supercovariant derivatives obey the algebra $D^2 = \Dbar^2 = 0$, one also has $D \Dbar \mathcal{Q} = 0$ and $\Dbar D \calQbar = 0$. This presentation makes it more transparent that no additional terms are generated when the overall $D_+$ and $D_-$ derivatives act on the denominator of (\ref{scsquare_reduction_intermediate_four}). In complex notation, this is simply the statement that
\begin{align}
    D \Dbar \left( \frac{1}{4 \lambda \Dbar \mathcal{Q} - 1} \right) = \Dbar  D \left( \frac{1}{4 \lambda \Dbar \mathcal{Q} - 1} \right) = 0 \, ,
\end{align}
which is clear since $\Dbar^2 \mathcal{Q} = D \Dbar \mathcal{Q} = 0$ as we have pointed out.

Because $\mathcal{Q}, \calQbar$ contain the single component $T$ of the stress tensor in their component expansion, the combination (\ref{scsquare_reduction_final}) can be viewed as a manifestly supersymmetric extension of the deforming operator (\ref{gross_flow_eqn}), which is now written directly in superspace. This gives a prescription for deforming any theory of supersymmetric quantum mechanics which descends via dimensional reduction from a $2d$ superconformal field theory (we have assumed that the seed theory in $2d$ is conformal in order to use the trace flow equation).

One shortcoming of this presentation is that we have not provided any operational method for computing the objects $\mathcal{Q}, \calQbar$ within a given theory of supersymmetric quantum mechanics. In order to construct these objects using the procedure described in this section, one would need to lift such a $(0+1)$-dimensional theory to theory in $(1+1)$-dimensions, construct the supercurrents of this parent theory, and then assemble the appropriate combination of the supercurrents which appear upon reducing back down to quantum mechanics. It is of course desirable to have a complementary view of $\mathcal{Q}, \calQbar$ which facilitates direct computation of these conserved superfields in a $(0+1)$-dimensional theory, ideally via some Noether procedure which provides an interpretation of these objects as conserved charges associated with time translations. We turn to this issue next.

\section{Direct Definition of Deformation in $1d$}\label{method3}

In this section, we will define a manifestly supersymmetric deformation directly in the superspace of a $(0+1)$-dimensional quantum mechanics theory. We first develop some formalism for defining a conserved Noether ``current,'' denoted $\mathcal{Q}$, associated with time translation invariance. Because the theory has no spatial dimensions, this conserved quantity is really a charge rather than a current; however we will still use the term ``supercurrent'' rather than ``supercharge'' for this object in order to avoid confusion with the supercharges $Q_i$ which represent the action of the SUSY algebra on superfields.

\subsection{Noether Currents in $\mathcal{N} = 2$ Theories}

We will follow the strategy of using a superspace Noether procedure which closely parallels the discussion of Section \ref{method1}; we note that a similar analysis for supersymmetric quantum mechanics appeared in \cite{Clark:2001zv} for a different class of superspace Lagrangians.

Begin by considering a theory for a collection of real scalars $X^i$ described by the action
\begin{align}\label{general_lag_susy_qm}
    S = \int \, dt \, d \theta \, d \thetab \, \mathcal{A} \left( X^i, D X^i , \Dbar X^i , D \Dbar X^i , \Dbar D X^i \right) \, ,
\end{align}
Although we have not allowed the superspace Lagrangian $\mathcal{A}$ to explicitly depend on the time derivatives $\dot{X}^i = \partial_t X^i$, such dependence is implicitly allowed since
\begin{align}
    D \Dbar X^i + \Dbar D X^i = - 2 i \dot{X}^i
\end{align}
according to our conventions for the supersymmetry algebra which are described in Section \ref{sec:conventions}. Therefore, since $\mathcal{A}$ depends on both $D \Dbar X^i$ and $\Dbar D X^i$, arbitrary dependence on $\dot{X}^i$ can also be accommodated.

For the moment, we will make no additional assumptions about the superfields $X^i$ besides the reality condition $\left( X^i \right)^{\ast} = X^i$. We first consider an arbitrary variation of the superspace Lagrangian under a field fluctuation $\delta X^i$:
\begin{align}\label{general_susy_qm_variation}
    \hspace{-15pt} \delta \mathcal{A} &= \delta X^i \, \frac{\delta \mathcal{A}}{\delta X^i}  + \delta ( D X^i ) \, \frac{\delta \mathcal{A}}{\delta ( D X^i ) }  + \delta ( \Dbar X^i ) \, \frac{\delta \mathcal{A}}{\delta ( \Dbar X^i ) } + \delta ( D \Dbar X^i ) \, \frac{\delta \mathcal{A}}{\delta ( D \Dbar X^i ) } + \delta ( \Dbar D  X^i ) \, \frac{\delta \mathcal{A}}{\delta ( \Dbar D  X^i ) }  \, .
\end{align}
It will be convenient to re-express (\ref{general_susy_qm_variation}) by writing each term as the derivative of a product, minus an appropriate correction. For instance,
\begin{align}\label{derivative_rewriting}
    \delta ( D X^i ) \, \frac{\delta \mathcal{A}}{\delta ( D X^i ) } &= D \left( \delta X^i  \, \frac{\delta \mathcal{A}}{\delta ( D X^i ) } \right) - \left( \delta X^i \right) D \left( \frac{\delta \mathcal{A}}{\delta ( D X^i ) } \right) \, ,  \nonumber \\
	\delta \left( D \Dbar X^i \right) \frac{\delta \mathcal{A}}{\delta ( D \Dbar X^i ) } &= D \left( \frac{\delta \mathcal{A}}{\delta (D \Dbar X^i)} \Dbar \delta X^i \right) + \Dbar \left( \delta X^i D \frac{\delta \mathcal{A}}{\delta (D \Dbar X^i)} \right) - ( \delta X^i ) \Dbar D  \left( \frac{\delta \mathcal{A}}{\delta (D \Dbar X^i)} \right) . 
\end{align}
This gives
\begin{align}\label{susy_variation_ibp}
    \hspace{-10pt} &\delta \mathcal{A} = D \left( \delta X^i  \, \frac{\delta \mathcal{A}}{\delta ( D X^i ) } \right) + \Dbar \left( \delta X^i  \, \frac{\delta \mathcal{A}}{\delta ( \Dbar X^i ) } \right)  \nonumber \\
    &+ D \left( \frac{\delta \mathcal{A}}{\delta (D \Dbar X^i)} \Dbar \delta X^i \right) + \Dbar \left( \delta X^i D \frac{\delta \mathcal{A}}{\delta (D \Dbar X^i)} \right) + \Dbar \left( \frac{\delta \mathcal{A}}{\delta  (\Dbar D  X^i)} D \delta X^i \right) + D \left( \delta X^i \Dbar \frac{\delta \mathcal{A}}{\delta (\Dbar D X^i)} \right) \nonumber \\
    &- \delta X^i \left( - \frac{\delta \mathcal{A}}{\delta X^i} + D \left( \frac{\delta \mathcal{A}}{\delta ( D X^i ) } \right) + \Dbar \left( \frac{\delta \mathcal{A}}{\delta ( \Dbar X^i ) } \right) + D \Dbar \left( \frac{\delta \mathcal{A}}{\delta ( D \Dbar X^i ) } \right) + \Dbar D  \left( \frac{\delta \mathcal{A}}{\delta (  \Dbar D  X^i ) } \right) \right) \, .
\end{align}
One advantage of the form (\ref{susy_variation_ibp}) is that we can immediately read off the superspace equations of motion. Suppose we consider a linearized fluctuation $\delta X^i$ around a configuration $X^i$ which satisfies the equations of motion, and demand that $\delta S = \int \, dt \, d \theta \, d \thetab \, \delta \mathcal{A} = 0$. Since the terms in the first two lines are total superspace derivatives, and the final line must vanish for any $\delta X^i$, we obtain the on-shell condition
\begin{align}\label{susy_qm_eom}
    \frac{\delta \mathcal{A}}{\delta X^i} = D \left( \frac{\delta \mathcal{A}}{\delta ( D X^i ) } \right) + \Dbar \left( \frac{\delta \mathcal{A}}{\delta ( \Dbar X^i ) } \right) + D \Dbar \left( \frac{\delta \mathcal{A}}{\delta ( D \Dbar X^i ) } \right) + \Dbar D  \left( \frac{\delta \mathcal{A}}{\delta (  \Dbar D  X^i ) } \right) \, .
\end{align}
Next we would like to study the conserved charge associated with time translations $t \to t' = t + \delta t$. Under such a transformation, the superspace Lagrangian varies as
\begin{align}
    \delta \mathcal{A} = \left( \delta t \right) \partial_t \mathcal{A} = \frac{i}{2} \left( \delta t \right) \left( D \Dbar \mathcal{A} + \Dbar D \mathcal{A} \right) \, ,
\end{align}
where we have again used the algebra $\{ D , \Dbar \} = - 2 i \partial_t$. Meanwhile, each superfield $X^i$ also transforms as
\begin{align}
    \delta X^i = ( \delta t ) \dot{X}^i = \frac{i}{2} \left( \delta t \right) \left( D \Dbar X^i + \Dbar D X^i \right) \, .
\end{align}
We use these expressions in equation (\ref{susy_variation_ibp}) and also restrict to the case of on-shell variations, which means that the equations of motion are satisfied and we can discard the term proportional to $\delta X^i$ in the final line. This gives
\begin{align}
    0 &= - \frac{i}{2} \left( \delta t \right) \left( D \Dbar \mathcal{A} + \Dbar D \mathcal{A} \right) + \left( \delta t \right) D \left( \frac{i}{2}  \left( D \Dbar X^i + \Dbar D X^i \right)  \, \frac{\delta \mathcal{A}}{\delta ( D X^i ) } \right) \nonumber \\
    &\qquad + \frac{i}{2} \left( \delta t \right) \Dbar \left(  \left( D \Dbar X^i + \Dbar D X^i \right)  \, \frac{\delta \mathcal{A}}{\delta ( \Dbar X^i ) } \right)  + \frac{i}{2} \left( \delta t \right) D \left( \frac{\delta \mathcal{A}}{\delta (D \Dbar X^i )} \Dbar \left( D \Dbar X^i  \right) \right) \nonumber \\
    &\qquad + \frac{i}{2} ( \delta t ) \Dbar \left( \frac{\delta \mathcal{A}}{\delta (\Dbar D  X^i)} D ( \Dbar D X^i ) \right) + \frac{i}{2} ( \delta t ) D \left( \left( D \Dbar X^i + \Dbar D X^i \right) \Dbar \frac{\delta \mathcal{A}}{\delta (\Dbar D X^i)} \right)  \nonumber \\
    &\qquad + \frac{i}{2} ( \delta t ) \Dbar \left( \left( D \Dbar X^i + \Dbar D X^i \right) D \frac{\delta \mathcal{A}}{\delta (D \Dbar X^i)} \right) \, .
\end{align}
To ease notation, we define $\eta^i = D \Dbar X^i$, $\etat^i = \Dbar D X^i$. After simplifying and collecting terms, the resulting equation can be written as
\begin{align}\label{direct_1d_Q_cons_result}
    0 &= \frac{i}{2} ( \delta t ) \, D \left[ ( \eta^i + \etat^i ) \left( \frac{\delta \mathcal{A}}{\delta ( D X^i ) } + \Dbar \left( \frac{\delta \mathcal{A}}{\delta \etat^i} \right) \right) + \frac{\delta \mathcal{A}}{\delta \eta^i} \Dbar \eta^i  - \Dbar \mathcal{A} \right] \nonumber \\
    &\qquad + \frac{i}{2} ( \delta t ) \, \Dbar \left[ ( \eta^i + \etat^i ) \left( \frac{\delta \mathcal{A}}{\delta ( \Dbar X^i ) } + D \left( \frac{\delta \mathcal{A}}{\delta \eta^i} \right) \right) + \frac{\delta \mathcal{A}}{\delta \etat^i} D \etat^i - D \mathcal{A} \right] \, .
\end{align}
This can be interpreted as a superspace conservation equation of the form
\begin{align}\label{susy_conservation_Q}
    D \calQbar + \Dbar \mathcal{Q} = 0 \, ,
\end{align}
where
\begin{align}\label{susy_Q_def}
    \mathcal{Q} &= ( \eta^i + \etat^i ) \left( \frac{\delta \mathcal{A}}{\delta ( \Dbar X^i ) } + D \left( \frac{\delta \mathcal{A}}{\delta \eta^i} \right) \right) + \frac{\delta \mathcal{A}}{\delta \etat^i} D \etat^i - D \mathcal{A} \,  , \nonumber \\
    \calQbar &=  ( \eta^i + \etat^i ) \left( \frac{\delta \mathcal{A}}{\delta ( D X^i ) } + \Dbar \left( \frac{\delta \mathcal{A}}{\delta \etat^i} \right) \right) + \frac{\delta \mathcal{A}}{\delta \eta^i} \Dbar \eta^i  - \Dbar \mathcal{A} \, .
\end{align}
We note that there is an overall factor of $i$ multiplying each term in equation (\ref{direct_1d_Q_cons_result}), but we have chosen to strip off this factor in defining the charges (\ref{susy_Q_def}). From the perspective of conservation properties, this is of course irrelevant because any scalar multiple of the combination $D \calQbar + \Dbar \mathcal{Q}$ still vanishes. Thus we are free to rescale $\mathcal{Q}$ and $\calQbar$ by any constant. However, this means that there will be a relative factor of $i$ when comparing $\mathcal{Q}$, $\calQbar$ to $\mathcal{Q}_+$, $\mathcal{Q}_-$, in which case there was no factor of $i$ naturally appearing in the conservation equation. We will account for this re-scaling when converting between conventions in Appendix \ref{app:change_to_complex}.

In summary, the Noether procedure leading to (\ref{susy_Q_def}) provides a direct definition of the objects $\mathcal{Q}, \calQbar$ obtained in (\ref{scsquare_reduction_final}) without relying upon dimensional reduction. Since $D^2 = \Dbar^2 = 0$, the superspace conservation equation (\ref{susy_conservation_Q}) also implies that $D \Dbar \mathcal{Q} = \Dbar D \calQbar = 0$. As mentioned above, we note that the supercurrents $\mathcal{Q}, \calQbar$ are not to be confused with the supercharges $Q, \overbar{Q}$ defined by
\begin{align}
    Q = \frac{\partial}{\partial \theta} + i \thetab \frac{\partial}{\partial t} \, , \qquad \overbar{Q} = \frac{\partial}{\partial \thetab} + i \theta \frac{\partial}{\partial t} \, , 
\end{align}
which represent the action of the supersymmetry algebra on superfields.

\subsection{Definition of $f(\mathcal{Q}, \calQbar)$ Deformation}

To acquire some intuition for the objects $\mathcal{Q}, \calQbar$, it is useful to consider a simple example. The theory of a single real scalar is described by
\begin{align}
    L = \frac{m}{2} \int \, d \theta \, d \thetab \, D X \, \Dbar X \, .
\end{align}
Setting $m=2$ for simplicity, the supercurrent $\mathcal{Q}$ and its conjugate are given by
\begin{align}\label{first_order_free_Qs}
    \mathcal{Q} &= - \left( \eta + \etat \right) D X - D ( D X \, \Dbar X )  \nonumber \\
    &= - \etat D X \, , \nonumber \\
    \calQbar &= \left( \eta + \etat \right) \Dbar X - \Dbar ( D X \, \Dbar X ) \nonumber \\
    &= \eta \Dbar X \, ,
\end{align}
where we used $\Dbar D X = - 2 i \dot{X} - D \Dbar X$. The component expressions for these charges are
\begin{align}
    \mathcal{Q} &= \psi ( i \dot{x} - F ) - 2 i \theta \psi \dot{\psi} + \thetab \left( \dot{x}^2 + 2 i F \dot{x} - F^2 \right) + \theta \thetab \left( i \psi \dot{F} + \psi \ddot{x} - 3 \dot{x} \dot{\psi} - 3 i F \dot{\psi} \right) \, , \nonumber \\
    \calQbar &= \psib ( F + i \dot{x} ) + 2 i \thetab \psib \dot{\psib} + \theta \left( F^2 + 2 i F \dot{x} - \dot{x}^2 \right)  + \theta \thetab \left( i \psib \dot{F} - \psib \ddot{x} + 3 \dot{x} \dot{\psib} - 3 i F \dot{\psib}  \right) \, .
\end{align}
We can also compute the highest component of the product $\mathcal{Q} \calQbar$. For simplicity, we will set the auxiliary field to zero using its equation of motion. Then
\begin{align}
    \mathcal{Q} \calQbar \Big\vert_{\theta^2, F=0} = - 4 \psi \psib \dot{\psi} \dot{\psib} + 3 i \left( \psi \dot{\psib} + \psib \dot{\psi} \right) \dot{x}^2 + \dot{x}^4 \, .
\end{align}
We compare this with the Lagrangian of the theory written in components,
\begin{align}
    L = \dot{x}^2 + i \left( \psib \dot{\psi} - \dot{\psib} \psi \right) + F^2 \, .
\end{align}
To further develop our intuition we focus on the bosonic sector. Settting $F = 0$ and $\dot{\psi} = \dot{\psib} = 0$, we have the relation
\begin{align}
    L^2 = \mathcal{Q} \calQbar \, ,
\end{align}
Interpreting the Euclidean Lagrangian as the Hamiltonian, we see that deforming the bosonic sector by the product $\mathcal{Q} \calQbar$ is equivalent to a deformation by $H^2$. Next, the lowest component of $\Dbar \mathcal{Q}$ is given by
\begin{align}
    \Dbar \mathcal{Q} \Big\vert_{\theta = \thetab = 0} = - F^2 + 2 i F \dot{x} + \dot{x}^2 \, .
\end{align}
When the auxiliary field equation of motion is satisfied, the lowest component of $\Dbar \mathcal{Q}$ is therefore $\dot{x}^2$, which is the Hamiltonian for the bosonic degree of freedom.

Using the intuition that $\mathcal{Q} \calQbar$ has $H^2$ as its top component and $\Dbar \mathcal{Q}$ has $H$ as its bottom component, a natural guess for a combination of superfields which has the deforming operator (\ref{gross_flow_eqn}) as its top component is $\frac{\mathcal{Q} \calQbar}{\frac{1}{2} - 2 \lambda \Dbar \mathcal{Q}}$, which suggests the flow equation
\begin{align}\label{susy_qm_Qsquare_flow}
    \frac{\partial \mathcal{A}}{\partial \lambda} = \frac{\mathcal{Q} \calQbar}{\frac{1}{2} - 2 \lambda \Dbar \mathcal{Q}} \, .
\end{align}
This is exactly the form of the deformation (\ref{scsquare_reduction_final}) which we obtained by dimensionally reducing the supercurrent-squared operator in $(1+1)$-dimensions.

\subsection{Solution for One Scalar}

We now provide evidence that this is on-shell equivalent to the deformation (\ref{gross_flow_eqn}). In particular, we will check that the flow (\ref{susy_qm_Qsquare_flow}) generates the expected superspace Lagrangian on-shell for the case of a single real scalar field $X$. From the expressions (\ref{first_order_free_Qs}) for the conserved charges in the free theory, we see that the leading deformation is $- \lambda \eta \etat D X \Dbar X$. Motivated by this, we will make an ansatz for the finite-$\lambda$ solution of the form
\begin{align}\label{susy_qm_ansatz}
    \mathcal{A} = f ( \lambda \eta \etat ) \, DX \, \Dbar X \, ,
\end{align}
with $f(y) \to 1 - y + \mathcal{O} ( y^2 )$ as $y \to 0$. Using the definition (\ref{susy_Q_def}) we can compute the conserved superspace charges associated with a Lagrangian of this form, which gives
\begin{align}
    \mathcal{Q} &= - ( \eta + \etat ) \left( f D X + \lambda \eta \etat f' DX + \lambda^2 \etat f'' D ( \eta \etat ) DX \Dbar X \right) + \lambda \eta f' DX \, \Dbar X \, D \etat \nonumber \\
    &\qquad + f \eta D X - \lambda f' D ( \eta \etat ) DX \, \Dbar X \, , \nonumber \\
    \calQbar &= ( \eta + \etat ) \left( f \Dbar X + \lambda \eta \etat f' \, \Dbar X + \lambda^2 \eta f'' D ( \eta \etat ) DX \Dbar X \right) + \lambda \etat f' D X \, \Dbar X \, \Dbar \eta \nonumber \\
    &\qquad - f \etat \Dbar X - \lambda f' \Dbar ( \eta \etat ) DX \, \Dbar X \, .
\end{align}
The product of these is therefore
\begin{align}\label{qqbar_for_ansatz}
    \mathcal{Q} \calQbar = - \left( \eta \etat \left( f + f' \eta \lambda ( \eta + \etat ) \right) \left( f + f' \lambda \etat ( \eta + \etat \right) \right) DX \, \Dbar X \, .
\end{align}
We now pause to investigate an implication of the superspace equations of motion which will allow us to simplify this expression for $\mathcal{Q} \calQbar$ and therefore the flow equation. This will be the analogue of equation (\ref{susy_eom_multiplied_intermediate}), which we used to make a similar simplification in the field theory setting. The equation of motion (\ref{susy_qm_eom}) can be written as
\begin{align}
    0 = D \left( \frac{\delta \mathcal{A}}{\delta ( D X ) } \right) + \Dbar \left( \frac{\delta \mathcal{A}}{\delta ( \Dbar X ) } \right) + D \Dbar \left( \frac{\delta \mathcal{A}}{\delta \eta } \right) + \Dbar D  \left( \frac{\delta \mathcal{A}}{\delta \etat } \right) \, ,
\end{align}
which for our ansatz (\ref{susy_qm_ansatz}) is
\begin{align}\label{on_shell_intermediate_step}
    0 = D \left( f \Dbar X \right) - \Dbar \left( f D X \right) + D \Dbar \left( \etat \lambda f' D X \Dbar X \right) + \Dbar D \left( \eta \lambda f' DX \, \Dbar X \right) \, .
\end{align}
Suppose we multiply both sides of equation (\ref{on_shell_intermediate_step}) by $DX \Dbar X$. Since $(DX)^2 = ( \Dbar X )^2 = 0$, several terms vanish by nilpotency, and the surviving contributions are
\begin{align}
    f \eta D X \Dbar X - f \etat D X \Dbar X + \lambda \etat^2 \eta f' D X \Dbar X - \lambda \eta^2 \etat f' DX \Dbar X = 0 \, .
\end{align}
It follows that
\begin{align}
    \left( f - \lambda f' \eta \etat \right) \left( \eta - \etat \right) DX \Dbar X = 0 \, .
\end{align}
Therefore, either $\eta - \etat$ or $f - \lambda f' \eta \etat$ vanishes when multiplying $DX \Dbar X$ . The latter cannot hold identically unless $f(y) = \frac{c}{y}$ which is not consistent with the boundary condition $f(0) = 1$. We conclude that 
\begin{align}
    \left( \eta - \etat \right) D X \, \Dbar X = 0 \, .
\end{align}
In particular this means that, on-shell, we can replace $\eta$ with $\etat$ or vice-versa when either is multiplying $DX \, \Dbar X$. Making this replacement in the expression (\ref{qqbar_for_ansatz}) for the bilinear $\mathcal{Q} \calQbar$ gives
\begin{align}\label{QQbar_eom_replacement}
    \mathcal{Q} \calQbar = - \eta \etat \left( f + 2 f' \lambda \eta \etat \right)^2 DX \, \Dbar X \, .
\end{align}
Next, to construct our deforming operator (\ref{susy_qm_Qsquare_flow}), we consider the combinations $D \calQbar$ and $\Dbar \mathcal{Q}$ (which are of course related by the conservation equation). Any term appearing in these combinations which is proportional to $DX$ or $\Dbar X$ will not contribute to the deformation, since the function of $\Dbar \mathcal{Q}$ appearing in (\ref{susy_qm_Qsquare_flow}) comes multiplying $\mathcal{Q} \calQbar$, which is already proportional to $DX \, \Dbar X$. The only terms which we need to retain are therefore
\begin{align}
    \Dbar \mathcal{Q} &\sim - \eta \etat f - 2 \lambda \eta^2 \etat^2 f' \, , \nonumber \\
    D \calQbar &\sim \eta \etat f + 2 \lambda \eta^2 \etat^2 f' \, , 
\end{align}
where by ``$\sim$'' we mean equivalence up to terms proportional to $DX$ or $\Dbar X$, which vanish when multiplying $DX \, \Dbar X$ by nilpotency. We have thus found that, on-shell, the combination of superfields which drives our flow equation can be written as
\begin{align}
    \frac{\mathcal{Q} \calQbar}{\frac{1}{2} - 2 \lambda \Dbar Q} = - \frac{ \eta \etat ( f + 2 \lambda \eta \etat f' )^2 }{\frac{1}{2} + 2 \lambda \eta \etat ( f + 2 \lambda \eta \eta f' )} \, DX \, \Dbar X \, ,
\end{align}
In terms of the dimensionless variable $y = \lambda \eta \etat$, the flow equation then reduces to an ordinary differential equation for $f(y)$,
\begin{align}
    f'(y) = \frac{2 \left( f(y) + 2 y f'(y) \right)^2}{1 + 4 y \left( f(y) + 2 y f'(y) \right)} \, ,
\end{align}
whose solution is
\begin{align}
    f ( y ) = \frac{1}{4 y} \left( \sqrt{ 1 + 8 y } - 1 \right) \, .
\end{align}
We therefore conclude that the all-orders solution to the flow equation (\ref{susy_qm_Qsquare_flow}) is on-shell equivalent to the expression
\begin{align}
    \mathcal{A} ( \lambda ) = \frac{1}{4 \lambda \eta \etat} \left( \sqrt{ 1 + 8 \lambda \eta \etat } - 1 \right) \, DX \, \Dbar X \, .
\end{align}
In order to facilitate comparison with our earlier analysis, we re-scale $\lambda \to \frac{\lambda}{2}$ and replace $\eta, \etat$ with their explicit expressions. The resulting deformed quantum mechanics theory is
\begin{align}\label{susy_qm_1_scalar_soln}
    S = \int \, dt \, d \theta \, d \thetab \frac{1}{2 \lambda (D \Dbar X) ( \Dbar D X )} \left(  -1 + \sqrt{ 1 + 4 \lambda ( D \Dbar X ) ( \Dbar D X ) } \right) \, D X \, \Dbar X \, .
\end{align}
We see that this matches (\ref{dimensionally_reduced_answer}) on the nose after identifying $X$ with $\Phi$ and setting the metric to $g(\Phi) = 1$. The case with a non-trivial metric for the $(0+1)$-dimensional theory can be treated similarly.

One could also consider flows driven by other operators constructed from $\mathcal{Q}$ and $\calQbar$. These are supersymmetric versions of the $f(H)$ deformations considered in \cite{Gross:2019ach,Gross:2019uxi}. From the perspective of the quantum mechanics theory, there is no distinguished choice of $f(H)$ since any such function drives a qualitatively similar flow where all energy eigenstates remain eigenstates and their energy eigenvalues change in a prescribed way. The only reason for treating the particular $f(H)$ corresponding to $\TT$ as special is because of its connections to interesting deformations of higher dimensional theories.

As an example of a different supersymmetric $f(H)$ deformation, one could instead study the flow
\begin{align}
    \frac{\partial \mathcal{A}}{\partial \lambda} = \mathcal{Q} \calQbar \, ,
\end{align}
which is analogous to the deformation $\frac{\partial L}{\partial \lambda} = H^2$. If we again restrict to the case of a single real scalar considered above, and use the result (\ref{QQbar_eom_replacement}) which is equivalent to $\mathcal{Q} \calQbar$ on-shell, this leads to a differential equation
\begin{align}
    f'(y) = \left( f(y) + 2 y f'(y) \right)^2 \, .
\end{align}
for the function $f(y)$ appearing in the ansatz (\ref{susy_qm_ansatz}). This is a quadratic equation that can be solved for $f'(y)$ as
\begin{align}
    f'(y) = \frac{1 - 4 y f(y) - \sqrt{1 - 8 y f(y)}}{8 y^2} \, .
\end{align}
However, we note that this deformation, and flows driven by other functions of $\mathcal{Q}$ and $\calQbar$, will not be related to the usual two-dimensional $\TT$ deformation by dimensional reduction. Only the operator appearing in (\ref{susy_qm_Qsquare_flow}) has this property, and even in that case, the relationship only holds for deformations of conformal $2d$ seed theories since the derivation relies on the trace flow equation.

\section{Theories with $\mathcal{N} = 1$ Supersymmetry}\label{sec:n_equals_one}

Thus far we have focused on theories with two real supercharges, such as $2d$ field theories with $\mathcal{N} = (1, 1)$ supersymmetry or quantum mechanical theories with $\mathcal{N} = 2$ supersymmetry. However, one could carry out a totally analogous study of theories with only a single real supercharge. This would be relevant for theories with either $\mathcal{N} = (0, 1)$ or $\mathcal{N} = (1, 0)$ theories in two dimensions, which then reduce to theories with $\mathcal{N} = 1$ SUSY in $(0+1)$-dimensions.

We will not carry out an extensive analysis of the three different methods for constructing a supersymmetric $\TT$ deformation in the $\mathcal{N} = 1$ case, as we did in Sections \ref{method1} - \ref{method3} for $\mathcal{N} = 2$. However, in this section we will briefly outline some of the ingredients that would go into such an analysis, and argue that similar results hold.

\subsection{Noether Currents in $\mathcal{N} = 1$ Theories}

Consider a theory of a collection of $\mathcal{N} = 1$ superfields $X^i$ in $(0+1)$-dimensions. The $\mathcal{N} = 1$ superspace has a single anticommuting coordinate $\theta$, so the superfields $X^i$ can be expanded in components as
\begin{align}
    X^i = x^i + i \theta \psi^i \, .
\end{align}
The supercovariant derivative associated with $\theta$ is
\begin{align}
    D = \frac{\partial}{\partial \theta} - i \theta \frac{\partial}{\partial t} \, ,
\end{align}
which satisfies the algebra
\begin{align}
    \{ D, D \} = - 2 i \partial_t \, .
\end{align}
As a simple example, the free superspace Lagrangian for such a collection of $\mathcal{N} = 1$ superfields is written
\begin{align}\label{example_n_equals_one_free}
    \mathcal{A} = \frac{i}{2} \dot{X}^i D X^i \, .
\end{align}
We now carry out a version of the Noether procedure which was used to obtain expressions for the supercurrents $\mathcal{Q} , \calQbar$ in the $\mathcal{N} = 2$ case. Consider a superspace Lagrangian that depends on the $X^i$, their superspace derivatives $D X^i$, and their time derivatives $\dot{X}^i$:
\begin{align}\label{general_nequals1_action}
    S = \int \, dt \, d \theta \, \mathcal{A} ( X^i, D X^i, \dot{X}^i ) \, .
\end{align}
We note that, unlike in the $\mathcal{N} = 2$ case, the superspace Lagrangian $\mathcal{A}$ in (\ref{general_nequals1_action}) must be fermionic so that the action $S$ itself is bosonic. The variation of the superspace Lagrangian is given by
\begin{align}\label{n_equals_one_first_variation_step}
    \delta \mathcal{A} = \delta X^i \frac{\delta \mathcal{A}}{\delta X^i} + \delta ( D X^i ) \frac{\delta \mathcal{A}}{\delta ( D X^i )} + \delta \dot{X}^i \frac{\delta \mathcal{A}}{\delta \dot{X}^i} \, , 
\end{align}
The only difference in our Noether procedure is that, rather than specializing to the case of time translations $\delta X^i = ( \delta t ) \dot{X}^i$ as we did for $\mathcal{N} = 2$, we will now consider both translations along the Grassmann coordinates $\theta$ and along time. The reason for this is that we would like to construct current bilinears, which require the presence of two current-like objects such as $\mathcal{Q}$ and $\calQbar$. However, for the $\mathcal{N} = 1$ case, there is only a single current associated with time translations, and (as we will see shortly) its square does not have $\TT$ as its top component. Similarly, there is a single current associated with superspace translations, but because this current is fermionic we cannot square it to construct bilinears since the result would vanish by nilpotency.\footnote{Another way to see that we need two separate currents is that the superspace Lagrangian for $\mathcal{N} = 1$ is itself fermionic. Thus we could not have constructed a fermionic current bilinear out of a single conserved current, since the square of such a current is necessarily bosonic.}

With this motivation, we will again re-express (\ref{n_equals_one_first_variation_step}) using the product rule as before. Now we must be careful because $\delta$ does not commute with $D$ since we are allowing translations along the $\theta$ direction as well. One finds
\begin{align}\label{n_equals_one_general_variation}
    \delta \mathcal{A} &= \delta X^i \frac{\delta \mathcal{A}}{\delta X^i} + D \left( \delta X^i \right) \frac{\delta \mathcal{A}}{\delta ( D X^i )} + \left( \big[ \delta, D \big] X^i \right) \left( \frac{\delta \mathcal{A}}{\delta ( D X^i ) } \right) + \delta \dot{X}^i \frac{\delta \mathcal{A}}{\delta \dot{X}^i} \nonumber \\
    &= D \left( \delta X^i \frac{\delta \mathcal{A}}{\delta ( DX^i )} \right) + \partial_t \left( \delta X^i \frac{\delta \mathcal{A}}{\delta \dot{X}^i} \right) - \delta X^i \left( - \frac{\delta \mathcal{A}}{\delta X^i} + D \left( \frac{\delta \mathcal{A}}{\delta ( D X^i ) } \right) + \partial_t \left( \frac{\delta \mathcal{A}}{\delta \dot{X}^i} \right) \right) \nonumber \\
    &\qquad +  \left( \big[ \delta, D \big] X^i \right) \left( \frac{\delta \mathcal{A}}{\delta ( D X^i ) } \right) \, .
\end{align}
Exactly as before, in equation (\ref{susy_qm_eom}), we can read off the superspace equation of motion:
\begin{align}
    \frac{\delta \mathcal{A}}{\delta X^i} = D \left( \frac{\delta \mathcal{A}}{\delta ( D X^i ) } \right) + \partial_t \left( \frac{\delta \mathcal{A}}{\delta \dot{X}^i} \right) \, .
\end{align}
Now consider a combined superspace translation of the form $t \to t + \delta t$, $\theta \to \theta + \delta \theta$ for a commuting constant $\delta t$ and a Grassmann constant $\delta \theta$. The resulting change in the fields is
\begin{align}\label{X_variation_nequalsone}
    \delta X^i = (\delta \theta) \frac{\partial}{\partial \theta} X^i + ( \delta t ) \dot{X}^i \, .
\end{align}
Using the definition $D X^i = \frac{\partial}{\partial \theta} X^i - i \theta \dot{X}^i$, we can rewrite
\begin{align}
    \frac{\partial}{\partial \theta} X^i = D X^i + i \theta \dot{X}^i \, , 
\end{align}
and therefore repackage the variation (\ref{X_variation_nequalsone}) as
\begin{align}
    \delta X^i &= (\delta \theta) D X^i + ( \delta t + i (\delta \theta) \theta ) \dot{X}^i \nonumber \\
    &= (\delta \theta) D X^i + ( \delta \tilde{t} ) \dot{X}^i \, ,
\end{align}
where in the last step we have defined $\delta \tilde{t} \equiv \delta t + i (\delta \theta) \theta$. Likewise the variation $\delta \mathcal{A}$ of the superspace Lagrangian can be written in the same way:
\begin{align}
    \delta \mathcal{A} = ( \delta \theta ) D \mathcal{A} + ( \delta \tilde{t} ) \partial_t \mathcal{A} \, .
\end{align}
We can also compute the commutator
\begin{align}
    \big[ \delta, D \big] X^i &= \delta ( DX^i ) - D ( \delta X^i ) \nonumber \\
    &= \left( (\delta \theta) D D X^i + ( \delta \tilde{t} ) D \dot{X}^i \right) - D \left( (\delta \theta) D X^i + ( \delta \tilde{t} ) \dot{X}^i \right) \nonumber \\
    &= - i ( \delta \theta ) \dot{X}^i \, .
\end{align}

Substituting these variations into (\ref{n_equals_one_general_variation}) and going on-shell so that we can discard the equation of motion term gives
\begin{align}
    (\delta \theta) D \mathcal{A} + i ( \delta \tilde{t} ) D D \mathcal{A} &= D \left( (\delta \theta) ( D X^i ) \frac{\delta \mathcal{A}}{\delta ( D X^i ) } \right) + D D \left( (\delta \theta) ( D X^i )  \left( \frac{\delta \mathcal{A}}{\delta D D X^i} \right) \right) \nonumber \\
    &\quad + D \left( i ( \delta \tilde{t} ) ( D D X^i ) \frac{\delta \mathcal{A}}{\delta ( D X^i ) } \right) + D D \left( i ( \delta \tilde{t} ) (D D X^i ) \left( \frac{\delta \mathcal{A}}{\delta D D X^i} \right) \right) \nonumber \\
    &\quad - i ( \delta \theta ) \dot{X}^i \left( \frac{\delta \mathcal{A}}{\delta ( D X^i ) } \right) \, .
\end{align}
where we have rewritten time derivatives in terms of $D$ using $\partial_t = i D^2$. Collecting terms then gives
\begin{align}\label{nequals_one_noether_intermediate}
    0 &= - (\delta \theta) \Bigg[ D \left( ( D X^i ) \frac{\delta \mathcal{A}}{\delta ( D X^i )} - D \left( ( D X^i ) \frac{\delta \mathcal{A}}{\delta ( D D X^i )} \right) + \mathcal{A} \right) + i  \dot{X}^i \left( \frac{\delta \mathcal{A}}{\delta ( D X^i ) } \right) \Bigg] \, \nonumber \\
    &\quad + i D \left[  ( \delta \tilde{t} ) ( D D X^i ) \frac{\delta \mathcal{A}}{\delta ( D X^i ) } + D \left( ( \delta \tilde{t} ) ( D D X^i ) \frac{\delta \mathcal{A}}{\delta ( D D X^i ) } \right) \right] - i ( \delta \tilde{t} ) D D \mathcal{A}\, .
\end{align}
It is now tempting to commute the $\delta \tilde{t}$ past various instances of $D$ in the second line of (\ref{nequals_one_noether_intermediate}) and define two charge-like objects corresponding to the quantities in brackets, namely
\begin{align}\label{n_equals_one_supercurrents_defn}
    \mathcal{Q}_\theta &= ( D X^i ) \frac{\delta \mathcal{A}}{\delta ( D X^i )} - i D \left( ( D X^i ) \frac{\delta \mathcal{A}}{\delta \dot{X}^i} \right) + \mathcal{A} \, , \nonumber \\
    \mathcal{Q}_t &= - i \dot{X}^i \frac{\delta \mathcal{A}}{\delta ( D X^i ) } + D \left( \dot{X}^i \frac{\delta \mathcal{A}}{\delta \dot{X}^i } \right) - D \mathcal{A} \, .
\end{align}
One might then conclude that $D \mathcal{Q}_t$ must vanish and $D \mathcal{Q}_\theta$ must be related to the remaining term $i  \dot{X}^i \left( \frac{\delta \mathcal{A}}{\delta ( D X^i ) } \right)$, giving us one conserved charge and one object which is not conserved but which has a known source. However this manipulation is not valid because $\delta \tilde{t}$ itself depends on $\theta$ and therefore does not commute with $D$. If we first set $\delta \theta = 0$ and consider only a finite $\delta t$, then $\delta \tilde{t} = \delta t$ does commute with the $D$ operator so we can write
\begin{align}
    0 &= ( \delta t ) D \left[\left( ( D D X^i ) \frac{\delta \mathcal{A}}{\delta ( D X^i ) } + D \left( ( D D X^i ) \frac{\delta \mathcal{A}}{\delta ( D D X^i ) } \right) - D \mathcal{A} \right) \right] \, .
\end{align}
which is interpreted as a conservation equation of the form
\begin{align}
    D \mathcal{Q}_t = 0 \, ,
\end{align}
with $\mathcal{Q}_t$ defined in (\ref{n_equals_one_supercurrents_defn}). Next let us set $\delta t = 0$ and look at a fermionic translation $\delta \theta$. First we must account for the additional terms introduced when commuting $\delta \tilde{t} = i ( \delta \theta ) \theta$ past the $D$ operators. Note that
\begin{align}
    - i ( \delta \tilde{t} ) D D \mathcal{A} &= ( \delta \theta ) \theta D D \mathcal{A} \nonumber \\
    &= ( \delta \theta ) \left( - D \left( \theta D \mathcal{A} \right) + D \mathcal{A} \right) \nonumber \\
    &= ( \delta \theta ) D \left( \mathcal{A} - \theta D \mathcal{A} \right) \, .
\end{align}
Then one finds
\begin{align}
    (\delta \theta) D \mathcal{Q}_\theta &= - D \left[  ( \delta \theta ) \theta ( D D X^i ) \frac{\delta \mathcal{A}}{\delta ( D X^i ) } + D \left( ( \delta \theta ) \theta \dot{X}^i \frac{\delta \mathcal{A}}{\delta \dot{X}^i } \right) \right] + ( \delta \theta ) D \left( \mathcal{A} - \theta D \mathcal{A} \right) \, \nonumber \\
    &\qquad - i (\delta \theta) \dot{X}^i \left( \frac{\delta \mathcal{A}}{\delta ( D X^i ) } \right) \nonumber \\
    &= ( \delta \theta ) D \left[ \theta ( D D X^i ) \frac{\delta \mathcal{A}}{\delta ( D X^i ) } - D \left( \theta \dot{X}^i \frac{\delta \mathcal{A}}{\delta \dot{X}^i } \right) + \mathcal{A} - \theta D \mathcal{A} \right] - i (\delta \theta) \dot{X}^i \left( \frac{\delta \mathcal{A}}{\delta ( D X^i ) } \right) \nonumber \\
    &= ( \delta \theta ) D \left[  \theta \mathcal{Q}_t - \dot{X}^i \frac{\delta \mathcal{A}}{\delta \dot{X}^i} + \mathcal{A} \right] - i (\delta \theta) \dot{X}^i \left( \frac{\delta \mathcal{A}}{\delta ( D X^i ) } \right) \nonumber \\
    &= ( \delta \theta ) \left( \mathcal{Q}_t - \theta D \mathcal{Q}_t - D \left( \dot{X}^i \frac{\delta \mathcal{A}}{\delta \dot{X}^i} \right) + D \mathcal{A} - i \dot{X}^i \left( \frac{\delta \mathcal{A}}{\delta ( D X^i ) } \right) \right) \, .
\end{align}
Using the conservation equation $D \mathcal{Q}_t = 0$, we then have
\begin{align}\label{qtheta_non_conservation_intermediate}
    D \mathcal{Q}_\theta = \mathcal{Q}_t - D \left( \dot{X}^i \frac{\delta \mathcal{A}}{\delta \dot{X}^i} \right) + D \mathcal{A} - i \dot{X}^i \left( \frac{\delta \mathcal{A}}{\delta ( D X^i ) } \right) \, .
\end{align}
Thus we see that the ``charge'' $\mathcal{Q}_\theta$ is not an independent quantity but is in fact related to the time translation charge $\mathcal{Q}_t$, as one might expect from the intuition that the supersymmetry algebra relates successive superspace translations to time translations through $D^2 = - i \partial_t$. In particular, $\mathcal{Q}_\theta$ itself is not conserved in general. We can quantify this non-conservation by acting again on (\ref{qtheta_non_conservation_intermediate}) with $D$ and using $D \mathcal{Q}_t = 0$ to write
\begin{align}
    \partial_t \mathcal{Q}_\theta = \partial_t \left( \mathcal{A} - \dot{X}^i \frac{\delta \mathcal{A}}{\delta \dot{X}^i}  \right) - i D \left( \dot{X}^i \left( \frac{\delta \mathcal{A}}{\delta ( D X^i ) } \right) \right) \, .
\end{align}
Therefore one could define a modified charge $\widetilde{Q}_\theta$ and a correction term $\mathcal{Q}_c$ by
\begin{align}
    \widetilde{\mathcal{Q}}_\theta = \mathcal{Q}_\theta + \dot{X}^i \frac{\delta \mathcal{A}}{\delta \dot{X}^i} - \mathcal{A} \, , \qquad \mathcal{Q}_c = i  \dot{X}^i \left( \frac{\delta \mathcal{A}}{\delta ( D X^i ) } \right) \, ,
\end{align}
with the property that
\begin{align}
    \partial_t \widetilde{\mathcal{Q}}_\theta + D \mathcal{Q}_c = 0 \, .
\end{align}

\subsection{Definition of $\mathcal{Q}_\theta \mathcal{Q}_t$ Deformation and Solution for One Scalar}

To get some intuition for these objects constructed in the preceding subsection, we compute them for the free theory (\ref{example_n_equals_one_free}):
\begin{align}
    \mathcal{Q}_\theta &= i D X^i \dot{X}^i \, , \nonumber \\
    \widetilde{\mathcal{Q}}_\theta &= i D X^i \dot{X}^i \, , \nonumber \\
    \mathcal{Q}_c &= - \frac{1}{2} \dot{X}^i \dot{X}^i \, , \nonumber \\
    \mathcal{Q}_t &= \frac{1}{2} \dot{X}^i \dot{X}^i  \, .
\end{align}
Note that $\mathcal{Q}_\theta$, $\widetilde{\mathcal{Q}}_\theta$ are fermionic and $\mathcal{Q}_t$ is bosonic, as expected for Noether currents associated with Grassmann translations and time translations respectively. Therefore the product $\mathcal{Q}_\theta \mathcal{Q}_t $ is a fermion and thus an appropriate quantity to add to the Lagrangian as a deformation. In particular, for the free theory we note that the top component of $\mathcal{Q}_\theta \mathcal{Q}_t $ is proportional to $(\dot{x}^i \dot{x}^i)^2$, which is the square of the Hamiltonian. 
% Likewise, the bottom component of $\mathcal{Q}_t$ is simply $\dot{x}^i \dot{x}^i$ which is the Hamiltonian itself.
Using this intuition, we propose an $\mathcal{N} = 1$ version of the SUSY-QM deformation as
\begin{align}\label{n_equals_one_susy_flow}
    \frac{\partial \mathcal{A}}{\partial \lambda} = \frac{1}{2} \mathcal{Q}_\theta \mathcal{Q}_t \, .
\end{align}
In this proposal we do \emph{not} divide by the combination $\frac{1}{2} - 2 \lambda \mathcal{Q}_t$, as one might expect from the analogous $f(\mathcal{Q}, \calQbar)$ expression in the $\mathcal{N} = 2$ case. This may seem strange because the form of this deformation is very different than in the preceding cases that we have considered. However, we will later see that there is an equivalent rewriting of this flow equation as
\begin{align}\label{alternate_tilde_nequalsone_flow}
    \frac{\partial \mathcal{A}}{\partial \lambda} = \frac{\tilde{\mathcal{Q}}_\theta \mathcal{Q}_t}{1 + 2 \lambda \mathcal{Q}_t} \, .
\end{align}
For the class of Lagrangians that we focus on in this work, the solution to the flow equation (\ref{alternate_tilde_nequalsone_flow}) is identical to the solution of (\ref{n_equals_one_susy_flow}).  We will explore the reason for this in Section \ref{subsec:HT_form}, where we see that there is a simpler way to understand this equivalence by studying an analogous pair of deformations in the non-supersymmetric setting. For the moment, however, we will work with the first deformation (\ref{n_equals_one_susy_flow}).

First, we argue that this is on-shell equivalent to the dimensional reduction of the supercurrent-squared flow for theories with $\mathcal{N} = ( 0, 1)$ supersymmetry. In particular, we will solve the flow equation (\ref{n_equals_one_susy_flow}) for the seed theory of a single free boson and verify that it matches the dimensional reduction of the corresponding $2d$ flow. We make an ansatz for the finite-$\lambda$ deformed superspace Lagrangian of the form
\begin{align}
    \mathcal{A}^{(\lambda)} = \frac{i}{2} f ( \lambda \dot{X}^2 ) \, \dot{X} D X \, .
\end{align}
Next we compute the supercurrents. To ease notation, we define the dimensionless combination $\xi = \lambda \dot{X}^2$. Then using (\ref{n_equals_one_supercurrents_defn}) one finds
\begin{align}
    \mathcal{Q}_\theta &= i  f ( \xi ) \dot{X} D X \, , \nonumber \\
    \mathcal{Q}_t &= \frac{1}{2} \dot{X}^2 f ( \xi ) + \frac{i}{2} D \Big( \dot{X} \left( f ( \xi ) + 2 \xi f'(\xi) \right) D X \Big) - \frac{i}{2} D \left( f ( \xi ) \dot{X} D X \right) \, .
\end{align}
Here we have used that $DX$ is fermionic so $(DX)^2 = 0$. Furthermore, since $\mathcal{Q}_\theta$ is proportional to $DX$, when we construct the combination $\mathcal{Q}_\theta \mathcal{Q}_t$, any terms proportional to $DX$ in $\mathcal{Q}_t$ will drop out by nilpotency. Therefore we can write
\begin{align}\label{calQ_t}
    \mathcal{Q}_t \sim \frac{1}{2} \left( f ( \xi ) +  2 \xi f'(\xi) \right) \dot{X}^2 \, ,
\end{align}
where ``$\sim$'' means equality up to terms which will not contribute in the product $\mathcal{Q}_\theta \mathcal{Q}_t$. The flow equation (\ref{n_equals_one_susy_flow}) therefore becomes
\begin{align}
    \frac{i}{2} f'(\xi) \dot{X}^3 D X = \frac{i}{2} \cdot \left( f(\xi) \left( f ( \xi ) + 2 \xi f'(\xi) \right) \right) \, \dot{X}^3 \, D X \, ,
\end{align}
whose solution is
\begin{align}\label{calQ_theta_calQ_t_solution}
    f ( \xi ) = \frac{1}{2 \xi} \left( 1 - \sqrt{1 - 4 \xi} \right) \, .
\end{align}
Thus the full solution for the deformed superspace Lagrangian is
\begin{align}\label{nequalsone_susyqm_full_solution}
    \mathcal{A}^{(\lambda)} = \frac{i}{4 \lambda \dot{X}^2} \left( 1 - \sqrt{1 - 4 \lambda \dot{X}^2 } \right) \, \dot{X} D X \, .
\end{align}
As we mentioned around equation (\ref{alternate_tilde_nequalsone_flow}), the same flow can be acquired by deforming with another operator
\begin{equation}
\label{eq:newN=1Operator}
    \frac{\widetilde{\mathcal{Q}}_\theta \mathcal{Q}_t}{1 + 2\lambda \mathcal{Q}_t},
\end{equation}
similar to the irrelevant operators used in the previous sections. To see that (\ref{eq:newN=1Operator}) yields the same flow, we first compute
\begin{equation}
\begin{aligned}
    \widetilde{\mathcal{Q}}_\theta &= \mathcal{Q}_\theta + \dot{X} \frac{\delta \mathcal{A}}{\delta \dot{X}} - \mathcal{A} \\ 
    &= if(\xi)\dot{X} DX + \frac{i}{2}\left(f(\xi) + 2f'(\xi)\xi \right)\dot{X} DX - \frac{i}{2}f(\xi)\dot{X} DX \\
    &= i(f(\xi) + \xi f'(\xi)) \dot{X} DX.
\end{aligned}
\end{equation}
Then using the expression for $\mathcal{Q}_t$ from (\ref{calQ_t}), we have
\begin{equation}
    \frac{\widetilde{\mathcal{Q}}_\theta \mathcal{Q}_t}{1 + 2 \lambda \mathcal{Q}_t}  = \frac{\left(f(\xi) + \xi f'(\xi)\right)\left(f(\xi) + 2\xi f'(\xi)\right)}{1 + \xi \left(f(\xi) + 2\xi f'(\xi)\right)} \frac{i\dot{X}^3 DX}{2}.
\end{equation}
The flow equation $\frac{\partial \mathcal{A}}{\partial \lambda } = \frac{\widetilde{\mathcal{Q}}_\theta \mathcal{Q}_t}{1 + 2 \lambda \mathcal{Q}_t}$ leads to the following differential equation:
\begin{equation}
    f'(\xi) = \frac{(f(\xi) + 2\xi f'(\xi))(f(\xi) + \xi f'(\xi))}{1 + \xi (f(\xi) + 2\xi f'(\xi))},
\end{equation}
which has the same solution as in (\ref{calQ_theta_calQ_t_solution}) from the flow triggered by the operator $\mathcal{Q}_\theta \mathcal{Q}_t$,
\begin{equation}
    f(\xi) = \frac{1 - \sqrt{1 - 4\xi}}{2\xi}.
\end{equation}
Therefore, the operator $\frac{\tilde{\mathcal{Q}}_\theta \mathcal{Q}_t}{1 + 2\lambda \mathcal{Q}_t}$ also triggers the same $T\overline{T}$-like flow. Notice that from $\mathcal{Q}_t = D\widetilde{ \mathcal{Q}}_\theta + \mathcal{Q}_c$, we can express the operator $\frac{\widetilde{\mathcal{Q}}_\theta \mathcal{Q}_t}{1 + 2\lambda \mathcal{Q}_t}$ in terms of the conserved currents $\mathcal{Q}_c$ and $\widetilde{\mathcal{Q}}_\theta$, satisfying the conservation equation
\begin{equation}
    \partial_t \widetilde{\mathcal{Q}}_\theta + D \mathcal{Q}_c = 0.
\end{equation}
We now argue that this result is on-shell equivalent to the dimensional reduction of the solution to the supercurrent-squared flow for the corresponding $\mathcal{N} = (0, 1)$ theory in $2d$. The dimensional lift of this theory can be written as
\begin{align}\label{2d_nequalszeroone_seed}
    S = \int \, d^2 x \, d \theta \, D_+ \Phi \partial_{++} \Phi \, .
\end{align}
A superspace Noether procedure totally analogous to the one that we have used in the $\mathcal{N} = (1, 1)$ analysis of Section \ref{method1} can also be applied here. The input of this process is a superspace Lagrangian $\mathcal{A} ( D_+ \Phi, \partial_{\pm \pm} \Phi )$. The output is a conservation equation
\begin{align}\label{nequalszerooneconservation}
\begin{split}
	& \partial_\mm \mathcal{S}_\ppp + D_+ \mathcal{T}_\ppmm = 0, \\
	& \partial_\mm \mathcal{S}_\mmp + D_+ \mathcal{T}_\mmmm = 0,
\end{split}
\end{align}
where $\mathcal{S}_{\pm\pm+}$ and $\mathcal{T}_{\pm \pm - -}$ are superfields given by:
\begin{align}
\begin{split}
    & \mathcal{S}_\ppp = \frac{\delta\mathcal{A}}{\delta (\partial_\mm \Phi)}\partial_\pp \Phi, \\
    & \mathcal{S}_\mmp = \frac{\delta\mathcal{A}}{\delta (\partial_\mm \Phi)}\partial_\mm \Phi - \mathcal{A}, \\
    & \mathcal{T}_\ppmm = \frac{\delta\mathcal{A}}{\delta (D_+ \Phi)} \partial_\pp \Phi + D_+ \left( \frac{\delta\mathcal{A}}{\delta (\partial_\pp \Phi)} \partial_\pp \Phi \right) -D_+ \mathcal{A}, \\
    & \mathcal{T}_\mmmm = \frac{\delta\mathcal{A}}{\delta (D_+ \Phi)} \partial_\mm \Phi + D_+ \left( \frac{\delta\mathcal{A}}{\delta (\partial_\pp \Phi)} \partial_\mm  \Phi \right).
\end{split}
\end{align}
The $\mathcal{N} = (0, 1)$ supercurrent-squared flow is defined by
\begin{align}\label{nequalszeroonesupercurrentsquareddefn}
    \frac{\partial}{\partial \lambda} \mathcal{A}^{(\lambda)} = \mathcal{S}_\ppp \mathcal{T}_\mmmm - \mathcal{S}_\mmp \mathcal{T}_\ppmm .
\end{align}
Beginning from the seed superspace Lagrangian (\ref{2d_nequalszeroone_seed}), we make an ansatz for the finite-$\lambda$ solution:
\begin{align}
    \mathcal{A}^{(\lambda)} = f(\lambda \partial_\pp\Phi\partial_\mm\Phi)D_+\Phi\partial_\mm\Phi,
\end{align}
After evaluating the supercurrents, computing the combination of bilinears (\ref{nequalszeroonesupercurrentsquareddefn}), and simplifying the differential equation, one finds that
\begin{align}
    x \frac{\partial f}{\partial x} = - x f^2 - 2 x^2 f \frac{\partial f}{\partial x} \, ,
\end{align}
where $x = \lambda \partial_\pp\Phi\partial_\mm\Phi$. The solution is
\begin{align}
    f(x) = \frac{\sqrt{1+4x}-1}{2x}.
\end{align}
Thus the full deformed superspace Lagrangian is
\begin{align}\label{full_2d_nequalsone_soln}
    \mathcal{A}^{(\lambda)} = \frac{1}{2 \lambda \partial_\pp\Phi\partial_\mm\Phi} \left( \sqrt{ 1 + 4 \lambda \partial_\pp\Phi\partial_\mm\Phi } - 1 \right) D_+ \Phi \partial_{++} \Phi \, .
\end{align}
Upon dimensional reduction, we identify the superfield $\Phi$ with $X$ and all partial derivatives $\partial_{\pm \pm} \Phi$ are proportional to $\dot{X}$. Doing this, we see that -- up to various constant factors that can be absorbed
%into $\lambda$ and the definition of the fields 
into rescalings -- the solution (\ref{full_2d_nequalsone_soln}) exactly matches (\ref{nequalsone_susyqm_full_solution}). 

We will not perform the analogue of the analysis in Section \ref{method2}, where we dimensionally reduced the supercurrent-squared operator itself using the trace flow equation, in this $\mathcal{N} = 1$ case. However such a procedure should certainly be possible. One would identify a superfield analogue of the trace flow equation which relates $\mathcal{S}_{\pm \pm +}$ and $\mathcal{T}_{\pm \pm - -}$, and then use this to eliminate the appropriate linear combinations of these superfields that correspond to the $x$ directions. One might even expect the process to be simpler in this case, since the dimensionally reduced deformation should simply be a bilinear of the form $\mathcal{Q}_\theta \mathcal{Q}_t$ rather than a rational function of supercurrents. In particular, it appears $\mathcal{T}_{++--}$ is structurally similar to $\mathcal{Q}_t$, so one might believe that the correct dimensionally reduced deformation would be some product of $\mathcal{T}_{++--}$ with another superfield that plays the role of $\mathcal{Q}_\theta$.

% after making some modifications to these superfields $\mathcal{S}_{--+} \longrightarrow \widetilde{\mathcal{S}}_{--+}$ and  $\mathcal{T}_{++--} \longrightarrow \widetilde{\mathcal{T}}_{++--}$, the reduced deformation might be of the form $\widetilde{\mathcal{S}}_{--+} \widetilde{\mathcal{T}}_{++--}$. The reason we say that the superfields likely need to be changed to some (unspecified) decorated versions is since the conservation equations (\ref{nequalszerooneconservation}) are not of the same form as the constraints satisfied by $\mathcal{Q}_\theta$ and $\mathcal{Q}_t$, so the superfields might need to be modified in some way after reducing.

We conclude this subsection with a few comments about the relationship between $\mathcal{N} = 1$ and $\mathcal{N} = 2$ theories.

\begin{enumerate}
    \item Every SUSY-QM theory with $\mathcal{N} = 2$ SUSY can be viewed as a special case of a theory with $\mathcal{N} = 1$ supersymmetry. Therefore, one can always write the $f(\mathcal{Q}, \calQbar)$ deformation for a theory with $\mathcal{N} = 2$ supersymmetry and integrate out one of the fermionic directions to obtain a deformation in $\mathcal{N} = 1$ superspace. The resulting $\mathcal{N} = 1$ deformation should be on-shell equivalent to the combination $\mathcal{Q}_\theta \mathcal{Q}_t$ which we described in this section, since this generates the appropriate supercurrent-squared flows for $\mathcal{N} = 1$ theories. Evidence for the on-shell equivalence of these two flows in the case of $2d$ field theory was given in \cite{Chang:2018dge}; the SUSY-QM case should be similar.
    
    \item As pointed out in \cite{Combescure:2004ey}, quantum mechanical theories with $\mathcal{N} = 1$ supersymmetry are often equivalent to $\mathcal{N} = 2$ theories because they have a hidden second supersymmetry. In particular, this will be true for any $\mathcal{N} = 1$ theory with a fermion number symmetry. A second hidden supersymmetry of this form was not present in the case of a single $\mathcal{N} = 1$ superfield which we considered in this section, but it would be present in other cases (such as those with an even number of $\mathcal{N} = 1$ superfields). For those theories, one should be able to present the supercurrent-squared deformation of the theory in either $\mathcal{N} = 1$ or $\mathcal{N} = 2$ language, and we expect the results to be equivalent on-shell.
\end{enumerate}

\subsection{The $HT$ Form of the Deformation}\label{subsec:HT_form}

We now turn to the question of why our deforming operator in the case of $\mathcal{N} = 1$ supersymmetry could be written either as a bilinear $\mathcal{Q}_\theta \mathcal{Q}_t$ or a rational function of the form
\begin{align}
    \frac{\widetilde{\mathcal{Q}}_\theta \mathcal{Q}_t}{1 + 2 \lambda \mathcal{Q}_t} \, ,
\end{align}
although these two expressions appear quite different. This equivalence is related to an exact correspondence between two expressions involving the Hamiltonian which holds for $\TT$-like deformations of any kinetic seed theory. It will be simplest to discuss this correspondence in the purely bosonic context first, without any supersymmetry.

To begin, we first point out that there are two natural notions of energy in a theory of quantum mechanics. The first is the Hamiltonian $H$ of the system. Since we work in Euclidean signature and interpret the Euclidean Lagrangian as the Hamiltonian, $H$ is the object which sits under the integral sign in the action:
\begin{align}
    S_E = \int \, dt \, H \, .
\end{align}
The second notion of energy is the (Euclidean) Hilbert stress tensor $\Th$. In $(0+1)$ dimensions, there is only a single component of the stress tensor. It is defined by coupling the theory to a worldline metric $g_{tt}$, or equivalently an einbein $e_t$, and computing
\begin{align}\label{one_dim_hilbert_stress}
    \Th = - \frac{2}{\sqrt{g^{tt}}} \frac{\delta S_E}{\delta g^{tt}} = H - 2 \frac{\partial H ( g^{tt} )}{\partial g^{tt}} \Big\vert_{g^{tt} = 1} \, .
\end{align}
Here by $H ( g^{tt} )$ we mean the expression obtained by minimally coupling $H$ to a worldline metric. Since generically $\frac{\partial H ( g^{tt} )}{\partial g^{tt}} \neq 0$, the two notions of energy differ. Thus far we have been somewhat sloppy and used the symbols $H$ and $T$ interchangeably, for instance in the deformation (\ref{gross_flow_eqn}). Although this deformation is written in terms of $T$, it is more properly a flow equation for the object $H$ appearing under the integral in the Euclidean action. Therefore we will be more careful and write
\begin{align}\label{H_T_flow_one}
    \frac{\partial H}{\partial \lambda} = \frac{H^2}{\frac{1}{2} - 2 \lambda H} \, ,
\end{align}
whose solution is
\begin{align}\label{ham_sqrt_soln}
    H ( \lambda ) = \frac{1}{4 \lambda} \left( 1 - \sqrt{ 1 - 8 \lambda H_0 } \right) \, .
\end{align}
We recall that (\ref{H_T_flow_one}) was derived using the trace flow equation, which means that it is valid only for theories that descend from CFTs. In particular, it does not hold for theories with a potential. We now restrict to a particular class of theories for which (\ref{H_T_flow_one}) is valid, which we will refer to as ``kinetic seed theories.'' Explicitly, we assume that the undeformed Hamiltonian does not depend on any dimensionful scale but depends linearly on the inverse metric $g^{tt}$ when coupled to worldline gravity. For instance, the free scalar Hamiltonian
%s
\begin{align}
    H_0 ( g^{tt} ) = \dot{x}^2 = g^{tt} \partial_t x \partial_t x
\end{align}
belongs to this class of theories. Since $H_0 ( g^{tt} )$ depends linearly on the metric which is a scalar, the Hilbert stress tensor (\ref{one_dim_hilbert_stress}) associated with $H(\lambda)$ is simply
\begin{align}\label{explicit_onedim_hilbert_stress}
    \Th &= H ( \lambda ) - 2 \frac{\partial H}{\partial H_0} \frac{\partial H_0 ( g^{tt} ) }{\partial g^{tt} } \Big\vert_{g^{tt}=1} \nonumber \\
    &= H ( \lambda ) - 2 H_0 \frac{\partial H}{\partial H_0} \, .
\end{align}
We now ask whether we can express the right side of the flow equation (\ref{H_T_flow_one}) more simply in terms of $H$ and $\Th$, rather than simply $H$. One can verify by explicit calculation that, for a Hamiltonian of the form (\ref{ham_sqrt_soln}), the operator appearing in the flow is
\begin{align}
    \frac{H^2}{\frac{1}{2} - 2 \lambda H} = \frac{\left( \sqrt{ 1 - 8 \lambda H_0} - 1 \right)^2}{8 \lambda^2 \sqrt{1 - 8 \lambda H_0}} \, .
\end{align}
On the other hand, using the expression (\ref{explicit_onedim_hilbert_stress}) for the Hilbert stress tensor, one can also compute the combination
\begin{align}
    H \Th &= H \left( H - 2 H_0 \frac{\partial H}{\partial H_0} \right) \nonumber \\
    &= - \frac{ \left( \sqrt{1 - 8 \lambda H_0} - 1 \right)^2}{16 \lambda^2 \sqrt{1 - 8 \lambda H_0} } \, .
\end{align}
Therefore, for this class of theories, we conclude that
\begin{align}\label{HT_relation}
    H \Th = - \frac{1}{2} \left( \frac{H^2}{\frac{1}{2} - 2 \lambda H} \right) \, .
\end{align}
The upshot of this discussion is that, up to an overall constant which can be absorbed into the scaling of $\lambda$, we are free to deform \emph{either} by the combination $\frac{H^2}{\frac{1}{2} - 2 \lambda H}$ or by the combination $H\Th$. We refer to this latter expression as the $HT$ form of the flow equation (dropping the superscript (Hilb) for simplicity).

This $HT$ deformation has a simple interpretation.\footnote{In the holographic context, we interpret the addition of the double-trace $HT$ operator as a change in boundary conditions for the dual BF gauge theory fields. This is discussed in \cite{us:gravity}.} From the form (\ref{one_dim_hilbert_stress}) of the Hilbert stress tensor, one has
\begin{align}\label{one_dim_burgers}
    \frac{\partial H}{\partial \lambda} = H^2 - 2 H \frac{\partial H}{\partial g^{tt}} \, .
\end{align}
This is reminiscent of the form of the inviscid Burgers' equation (\ref{burgers}) for the cylinder energy levels of a $\TT$-deformed CFT in two dimensions, which we repeat:
\begin{align}
    \frac{\partial E_n}{\partial \lambda} = E_n \frac{\partial E_n}{\partial R} + \frac{1}{R} P_n^2 \, .
\end{align}
In the zero-momentum sector, the Burgers' equation (\ref{burgers}) admits an implicit solution
\begin{align}\label{2d_tt_burgers_implicit}
    E_n ( R, \lambda ) = E_n ( R + \lambda E_n ( R, \lambda ) , 0 ) \, .
\end{align}
This has the interpretation that the theory has effectively been put on a cylinder with an ``energy-dependent radius.'' That is, energy eigenstates with different energy eigenvalues see different effective geometries.

There is no straightforward analogue of the limit $P_n = 0$ in the quantum mechanical case, but if we restrict to the case of small energies so that $H^2$ is negligible compared to $H$, the equation (\ref{one_dim_burgers}) becomes
\begin{align}\label{one_dim_burgers_QM}
    \frac{\partial H}{\partial \lambda} \approx - 2 H \frac{\partial H}{\partial g^{tt}} \, ,
\end{align}
which likewise has the implicit solution
\begin{align}\label{qm_implicit_burgers_soln}
    H ( g^{tt} , \lambda ) = H ( g^{tt} - 2 \lambda H ( g^{tt} , \lambda ) , 0 ) \, .
\end{align}
This has the similar interpretation that different states in the deformed quantum mechanics theory see different effective energy-dependent metrics. Note that the relative factor of $-2$ between the rescalings in (\ref{2d_tt_burgers_implicit}) and (\ref{qm_implicit_burgers_soln}) is because the relation (\ref{HT_relation}) between $HT$ and the dimensionally reduced $\TT$ operator required us to re-scale $\lambda$ by a factor of $-\frac{1}{2}$.

We now see why there were also two equivalent ways of writing the deformation in the $\mathcal{N} = 1$ case. On-shell, the quantity $\mathcal{Q}_\theta$ is always proportional to the superspace Lagrangian $\mathcal{A}$ for deformations of a free seed theory, whereas the time translation current $\mathcal{Q}_t$ contains the Hilbert stress tensor. Therefore, the top component of their product is proportional to
\begin{align}
    \mathcal{Q}_\theta \mathcal{Q}_t \Big\vert_{\theta} \sim H \Th \, .
\end{align}
That is, the bilinear $\mathcal{Q}_\theta \mathcal{Q}_t$ is the superspace analogue of the $HT$ deformation. On the other hand, the second form of the deformation
\begin{align}
    \frac{\widetilde{Q}_\theta \mathcal{Q}_t}{1 + 2 \lambda \mathcal{Q}_t} \, ,
\end{align}
is the $\mathcal{N} = 1$ superspace analogue of the $\frac{H^2}{\frac{1}{2} - 2 \lambda H}$ form of the deformation. The fact that we obtained square root solutions to the two flows driven by $\mathcal{Q}_\theta \mathcal{Q}_t$ and $\frac{\widetilde{Q}_\theta \mathcal{Q}_t}{1 + 2 \lambda \mathcal{Q}_t}$ is therefore expected, since this is related to the statement that we likewise obtain square roots solutions in the bosonic sector using either $H T$ or $\frac{H^2}{\frac{1}{2} - 2 \lambda H}$.

\section{Discussion}
\label{sec: Discussion}

In this work, we have proposed a manifestly supersymmetric deformation of the superspace Lagrangian for a theory of $\mathcal{N} = 2$ quantum mechanics, namely
\begin{align}
    \frac{\partial \mathcal{A}}{\partial \lambda} = f(\mathcal{Q}, \calQbar) \equiv \frac{\mathcal{Q} \calQbar}{\frac{1}{2} - 2 \lambda \Dbar \mathcal{Q}} \, .
\end{align}
The conserved superfields $\mathcal{Q}, \calQbar$ are computed using a Noether prescription, for which we have given explicit formulas that apply to a class of theories involving scalar superfields $X^i$. We have also performed several non-trivial checks that this superspace deformation is on-shell equivalent to the dimensional reduction of the $\mathcal{N} = (1, 1)$ supercurrent-squared deformation of two-dimensional field theories, at least for conformally invariant seed theories. For such conformal seeds, this deformation is therefore a natural candidate for the appropriate supersymmetric version of $\TT$ for $(0+1)$-dimensional theories.

Additionally, we proposed two manifestly supersymmetric deformations for an $\mathcal{N} =1$ quantum mechanics theory
\begin{align}
    \frac{\partial \mathcal{A}}{\partial \lambda} = \frac{1}{2} \mathcal{Q}_\theta \mathcal{Q}_t \quad  \text{ and } \quad \frac{\partial \mathcal{A}}{\partial \lambda} = \frac{\widetilde{\mathcal{Q}_\theta} \mathcal{Q}_t}{1 + 2 \lambda \mathcal{Q}_t} \, .
\end{align}
Although the form of these deformation appear different, we showed that they produce the same flow equation when applied to the seed theory of a single free scalar. This flow equation matches the dimensional reduction of the $\mathcal{N} = (0 ,1)$ supercurrent-squared operator. We also interpreted the equivalence of these deformations by pointing out an analogous rewriting which holds for deformations of the bosonic sector of kinetic seed theories, namely $-\frac{1}{2} H T$ and $\frac{H^2}{\frac{1}{2} - 2\lambda H}$.

There remain several directions for future research. We will outline a few of these in the subsections that follow and make some speculative remarks about what one might expect.

\subsubsection*{\ul{\it More Supersymmetry}}

Perhaps the most obvious follow-up to this work is to exhibit a version of our superspace deformation with differing amounts of supersymmetry. For instance, it should be possible to define deformations of SUSY-QM theories which are related by dimensional reduction to the supercurrent-squared deformations of theories with $(0,2)$ or $(2,2)$ supersymmetry \cite{Chang:2019kiu,Jiang:2019hux}. The case of an $\mathcal{N}= 4$ SUSY-QM theory which descends from a $\mathcal{N} = (2, 2)$ field theory is perhaps more interesting, since such field theories are especially well-studied.

It may be that such an analysis is more amenable to a different technique for obtaining the supercurrents than the one we have used here. In the $2d$ case, such supercurrents for theories with $\mathcal{N} = (1, 1)$ supersymmetry were straightforward to compute using either a Noether procedure \cite{Chang:2018dge} or via coupling to supergravity \cite{Baggio:2018rpv}. However, in the case of $2d \, , \, \mathcal{N} = (2, 2)$ theories, it was more convenient to couple to the appropriate supergravity rather than employing a Noether approach \cite{Chang:2019kiu}. From this intuition, one might expect that the computation of supercharges for deformations of $\mathcal{N} = 4$ SUSY-QM theories might likewise be easier to perform by coupling to worldline supergravity.

It would also be interesting to understand $\TT$-type deformations in theories with even more supersymmetry, like $\mathcal{N} = 8$ or maximal SUSY.\footnote{Other deformations of QM theories with more supersymmetry, albeit not related to $\TT$, have been considered in \cite{Ivanov:2018czk,Ivanov:2019gxo,Sidorov:2014lvo,Fedoruk:2017efi}. See also \cite{Galajinsky:2020hsy,Kozyrev:2021agn,Kozyrev:2021icm} for discussions of the super-Schwarzian with more SUSY.} Such an endeavor is complicated by the absence of a conventional superspace which makes all of the supersymmetries manifest. One could of course work with a reduced superspace like $\mathcal{N} = 2$ or $\mathcal{N} = 4$ which geometrizes a subset of the supersymmetry transformations, but the action of the non-manifest SUSY generators will then be corrected order-by-order in $\lambda$ after turning on a $\TT$-like deformation.

\subsubsection*{\ul{\it Connections to Supersymmetric BF Gauge Theory}}

Another direction concerns the holographic interpretation of these results. We have emphasized that part of the motivation for considering deformations of $(0+1)$-dimensional theories whose Lagrangians take the purely-kinetic form
\begin{align}\label{purely_kinetic}
    S = \int \, dt \, g_{ij} ( X ) \dot{X}^i \dot{X}^j \, ,
\end{align}
where the $X^i$ are coordinates on a Lie group $G$, is that such theories are dual to BF gauge theories with gauge group $G$. This relationship holds with or without supersymmetry; in the SUSY case, the dual is a SUSY-BF theory and the quantum mechanics theory admits an interpretation as a particle moving on a supergroup. In the special case that the gauge group is an extension of $\mathrm{SL}(2, \mathbb{R})$, the dual BF theory is also related to JT gravity \cite{Iliesiu:2019xuh} and to other interesting theories such as SYK.

There have been some interpretations offered for the holographic interpretations of the $\TT$-like deformation of quantum mechanics in these various dual theories, at least in the non-manifestly-supersymmetric context. For instance, connections to cutoff JT gravity and to the Schwarzian have been discussed in \cite{Gross:2019ach,Gross:2019uxi,Iliesiu:2020zld,Stanford:2020qhm,Griguolo:2021wgy}, related analyses of the dual matrix models have been carried out in \cite{Rosso:2020wir,Ebert:2022gyn}, and a connection to modified boundary conditions in BF gauge theory is discussed in \cite{us:gravity}. 

It would be very interesting to extend these holographic interpretations to the case with manifest supersymmetry. In the undeformed case, the correspondence between the quantum mechanical theory of a particle moving on an $\mathrm{SL}(2, \mathbb{R})$ group manifold and the BF gauge theory with gauge group $\mathrm{SL}(2, \mathbb{R})$ is lifted to the supersymmetric setting by promoting the gauge group to either $\mathrm{OSp}( 1 \, \vert \, 2)$ for $\mathcal{N} = 1$ SUSY or $\mathrm{OSp} ( 2 \, \vert \, 2)$ for $\mathcal{N} = 2$ SUSY, as is nicely reviewed in Section 4.2 of \cite{Mertens:2018fds}. Here $\mathrm{OSp} ( N \, \vert \, 2p )$ is the orthosymplectic supergroup, a particular sub-supergroup of $\mathrm{GL}(N \, \vert \, 2p)$, which is the supergroup version of the general linear group $\mathrm{GL}(N)$.\footnote{In particular, $\mathrm{OSp} ( N \, \vert \, 2p )$ is the sub-supergroup of $\mathrm{GL}(N \, \vert \, 2p)$ which preserves a symmetric bilinear form on the bosonic elements (analogous to the orthogonal group) and preserves a symplectic form on the fermionic elements (analogous to the symplectic group); hence the name ``orthosymplectic.''} We focus on the $\mathcal{N} = 2$ case which was the main focus of this paper.\footnote{ Various aspects of the $\mathcal{N} = 1$ version of this theory, including its relationship to the super-Schwarzian and the properties of boundary-anchored Wilson lines, have been studied in \cite{Fan:2021wsb}.} This $\mathcal{N} = 2$ supersymmetric BF theory was analyzed in \cite{Astorino:2002bj,Livine:2007dx}, and its action can be written as
\begin{align}
    S_{BF}^{\mathcal{N} =2} = \int_{M} \mathrm{STr} \left( \Phi F \right) - \frac{1}{2} \oint_{\partial M} \mathrm{STr} ( \Phi \mathscr{A}_t )\, ,
\end{align}
where $\mathrm{STr}$ is the supertrace, $\Phi$ is the supersymmetric analogue of the scalar $\phi$ appearing in the usual BF Lagrangian $\mathcal{L}_{BF} = \mathrm{tr} ( \phi F )$, and $F = d \mathscr{A} + \mathscr{A} \wedge \mathscr{A}$ is the field strength of a supersymmetric gauge connection $\mathscr{A}$. In this $\mathcal{N} = 2$ case, each of $\mathscr{A}$ and $\Phi$ admit an expansion in terms of the $8$ generators of the $\mathfrak{osp} ( 2 \, \vert \, 2)$ Lie superalgebra; these consist of the usual $3$ generators of $\mathfrak{sl} (2, \mathbb{R})$, along with four fermionic generators, and one additional bosonic $\mathfrak{u} ( 1 )$ generator required by supersymmetry.

One would like to understand what modification of the bulk super-BF theory corresponds to turning on the $f(\mathcal{Q}, \calQbar)$ operator in the boundary SUSY-QM theory. We can view this question as the dimensional reduction of a related query: what has happened to a bulk $\mathrm{AdS}_3$ supergravity theory, written in Chern-Simons variables, when the dual supersymmetric field theory is deformed by supercurrent-squared? The standard intuition from $\mathrm{AdS}/\mathrm{CFT}$ is that the addition of a double-trace operator in the field theory corresponds to a modification of the boundary conditions for the bulk fields \cite{Klebanov:1999tb,Witten:2001ua}, although it is not clear that this intuition should generically apply for irrelevant double-trace deformations as opposed to relevant (or marginal) operators. In the non-supersymmetric context, it has been argued that this expectation is indeed correct, and that activating $\TT$ in a $2d$ CFT corresponds to a rotation of the sources and expectation values in the dual $\mathrm{SL} ( 2 , \mathbb{R} ) \times \mathrm{SL} ( 2, \mathbb{R} )$ Chern-Simons theory \cite{Llabres:2019jtx}. A similar rotation of boundary conditions appears in the non-supersymmetric setting of a $2d$ BF gauge theory which is dual to a boundary $(0+1)$-dimensional theory \cite{us:gravity}. It would be interesting to see whether the deformed super-BF theory, dual to a quantum mechanics theory deformed by $f(\mathcal{Q}, \calQbar)$ likewise admits such an interpretation, perhaps involving a linear mixing of the coefficient functions multiplying the $8$ generators of $\mathfrak{osp} ( 2 \, \vert \, 2 )$ in the expansions of $\Phi$ and $\mathcal{A}$.

\subsubsection*{\ul{\it Deformations of Multiple Scalars; Target Space Geometry}}

Another avenue for investigation is seeking solutions to the $f(\mathcal{Q},\calQbar)$ flow equations for theories with multiple scalars. In this work, we have only managed to find a closed-form result (\ref{susy_qm_1_scalar_soln}) for the deformed theory in the case of a single scalar, and even then, we have only found an expression which is on-shell equivalent to the full solution since we have imposed one implication of the superspace equations of motion. But of course the most interesting examples are the deformed theories of a particle moving on a higher-dimensional manifold, such as the $3$-dimensional $\mathrm{SL}(2, \mathbb{R})$ group manifold relevant for the Schwarzian theory. We have already mentioned in the analysis of the corresponding question for $2d$ field theories, around equation (\ref{finite_lambda_2d_schematic}), that solving the flow in this context is much more difficult because one expects a system of coupled PDEs for the functions multiplying the various two-fermion, four-fermion, etc. terms in the superspace Lagrangian. However, if one could find a partial or approximate solution with multiple scalars -- perhaps after going partly on-shell, as we have done here -- the result could be quite interesting.

For example, given such a solution, we could ask whether the resulting deformed theory still admits an interpretation as a point particle moving on some deformed target-space geometry. One might think not, since our intuition is that the ordinary $\TT$ flow in $2d$ generates theories which are no longer local QFTs. Analogously, one might expect that $f(\mathcal{Q}, \calQbar)$ deformed SUSY-QM theories exhibit some signature of non-locality. For instance, the particle whose position is described by the $X^i$ in the undeformed theory could become delocalized into a ``fuzzy particle'' over a length scale controlled by $\lambda$. It would be interesting to ask whether other properties of the target manifold can be probed in this case, or if the target manifold itself is changed.

On the other hand, in the undeformed theory, the Witten index of the theory is controlled by the Euler characteristic of the target space. Since our $f(\mathcal{Q}, \calQbar)$ flow is the supersymmetric extension of an $f(H)$ deformation -- which does not affect the energy eigenstates but merely modifies their energy eigenvalues -- it seems that this index remains unchanged under our deformation, which suggests that the target space topology is also unmodified.

There is some evidence that other indices cannot flow under $\TT$-like deformations. For instance, related indices like the elliptic genus have been shown not to flow under the usual $\TT$ in two dimensions if the seed theory is conformal \cite{Datta:2018thy}, and the same conclusion seems likely to hold if the undeformed theory is integrable but not conformal \cite{Ebert:2020tuy}. Nonetheless, it would be worthwhile to make this intuition precise in the SUSY-QM case, and perhaps look for other index-like quantities that do flow under $f(\mathcal{Q}, \calQbar)$ and which may admit an interpretation via target space geometry.

\subsubsection*{\ul{\it Including a Potential; Vacuum Structure}}

Another puzzle is how one should analyze $\TT$-like deformations of more general quantum mechanical theories than (\ref{purely_kinetic}), which is just a collection of scalars with a metric. The most obvious extension is to include a superpotential, writing
\begin{align}
    L = \int \, d^2 \theta \, \left( g_{ij} ( X ) \, \dot{X}^i \dot{X}^j + W ( X ) \right) \, . 
\end{align}
However, the presence of a superpotential means that this theory does not descend via dimensional reduction from a CFT in two dimensions; a generic superpotential, such as a mass term, introduces a scale which breaks conformal invariance. We therefore cannot rely on the trace flow equation to dimensionally reduce the $2d$ supercurrent-squared deformation and argue that the result is on-shell equivalent to the $f(\mathcal{Q} , \calQbar)$ deformation of Section \ref{method3}.

Without the trace flow equation, our only available technique for studying this case is to solve the corresponding flow equation in two dimensions and dimensionally reduce. As a toy example, we can perform this exercise for the non-supersymmetric theory of a single boson $\phi$ subject to a generic potential $V(\phi)$, which is presented in Appendix \ref{app:no_trace_flow}. One interesting feature of this procedure is that, at least for small momenta, the effective potential seen by a particle is schematically modified as
\begin{align}
    V ( \phi ) \longrightarrow \frac{V(\phi)}{1 - \lambda V(\phi)} \, .
\end{align}
Therefore, the potential na\"ively appears to diverge when $V(\phi)$ is of order $\frac{1}{\lambda}$. This was also discussed in the context of $\mathcal{N} = (2,2)$ theories in $2d$ in \cite{Chang:2019kiu}. We emphasize that this is a purely classical result concerning the flow equation for the Lagrangian, which does not necessarily imply anything about the Hilbert space of the deformed theory. A fully quantum analysis is needed to understand the fate of these poles. However, the possible presence of poles is quite interesting and hints at a modification of the vacuum structure of the theory. For the moment, we will allow ourselves to speculate about the physical implications of the existence of such poles if indeed they persist at the quantum level.

We mention a few explicit potentials by way of examples. For instance, suppose we begin with the harmonic oscillator potential $V_0 ( \phi ) = m^2 \phi^2$, where we take $m=1$ for simplicity. The potential deforms as follows:
\begin{align}
    \raisebox{-0.5\height}{\includegraphics[width=0.4\linewidth]{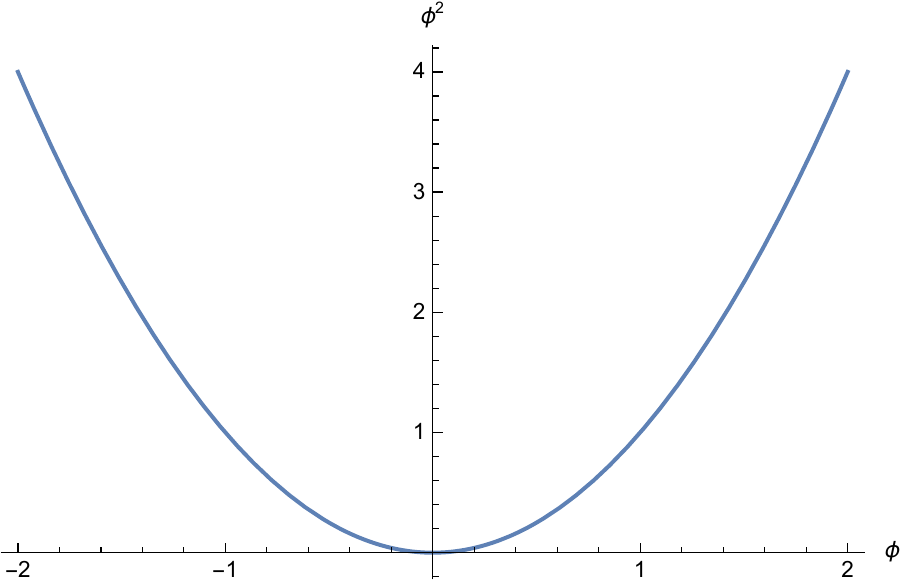}} \longrightarrow \; \raisebox{-0.5\height}{\includegraphics[width=0.4\linewidth]{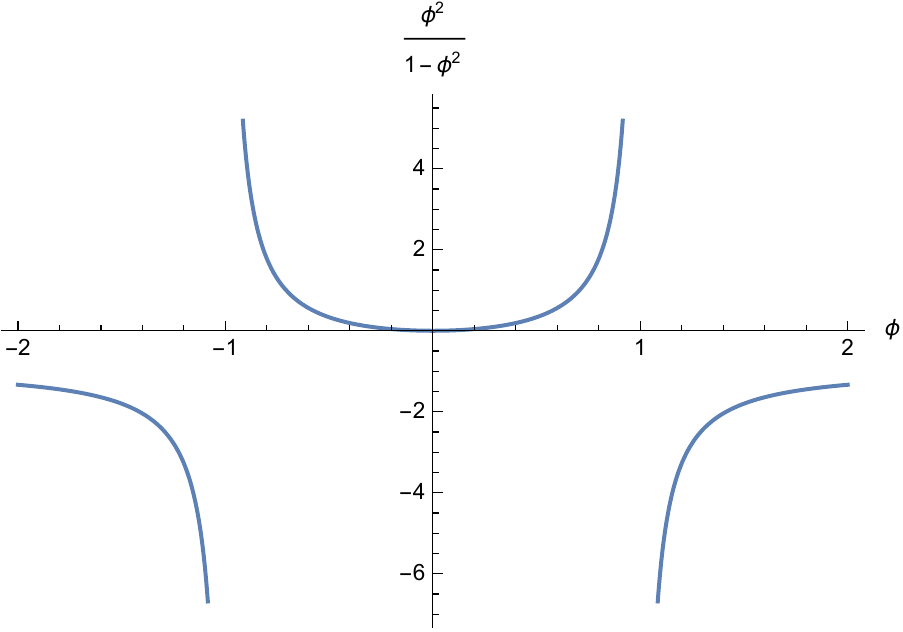}}
\end{align}
It is very natural to ask what has happened to the basis of eigenfunctions after applying this deformation. The undeformed potential is the usual harmonic oscillator, whose eigenstates are Hermite polynomials. However, the deformed potential has infinite barries at $\phi = \pm 1$. One might believe that there is a complete set of eigenfunctions for the deformed potential which are forced to vanish at $\phi = \pm 1$. The regions $| \phi | > 1$ seem to have been ``cut off'' from the theory by applying this deformation.

Another interesting case to consider is a linear potential $V ( \phi ) = \phi$.
\begin{align}
    \raisebox{-0.5\height}{\includegraphics[width=0.4\linewidth]{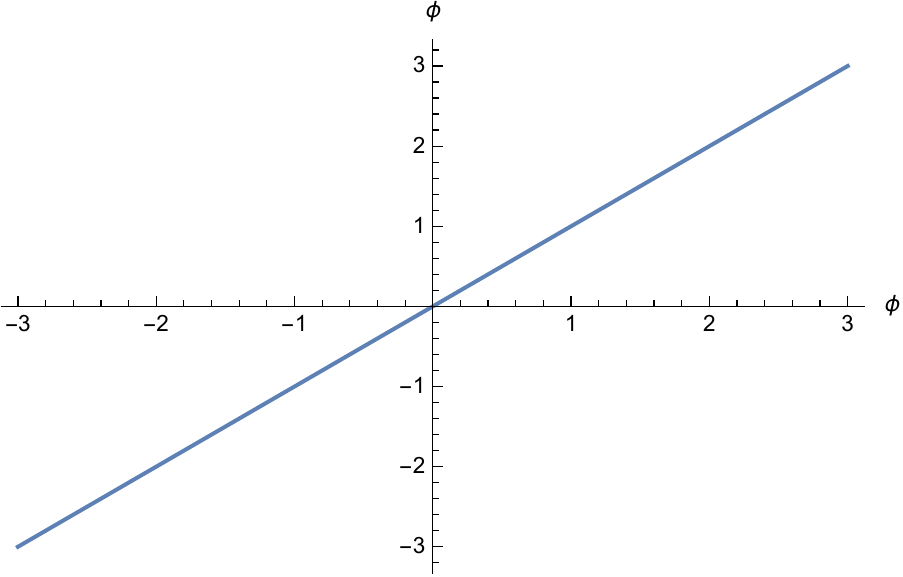}} \longrightarrow \;  \raisebox{-0.5\height}{\includegraphics[width=0.4\linewidth]{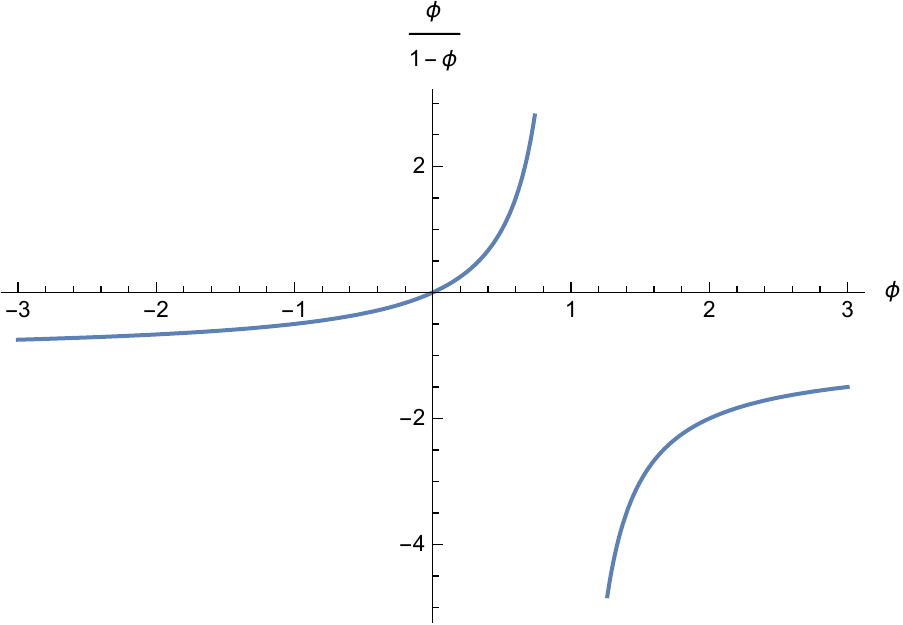}}
\end{align}
Now the change is even more drastic: the undeformed linear potential had eigenstates which were Airy functions, but they were non-normalizable because the potential was unbounded below. The deformation has now inserted a hard cutoff at $\phi = 1$. To the left of this cutoff, the potential is bounded below as $V(\phi) > -1$. Has the $\TT$ deformation ``cured'' the non-normalizability of the linear potential? If so, is there a relationship between the undeformed eigenstates (Airy functions) and the eigenstates of the deformed potential?

For a third example, the double-well potential $V(\phi) = \left( 1 - \phi^2 \right)^2$ deforms as
\begin{align}\label{double_well}
    \raisebox{-0.5\height}{\includegraphics[width=0.4\linewidth]{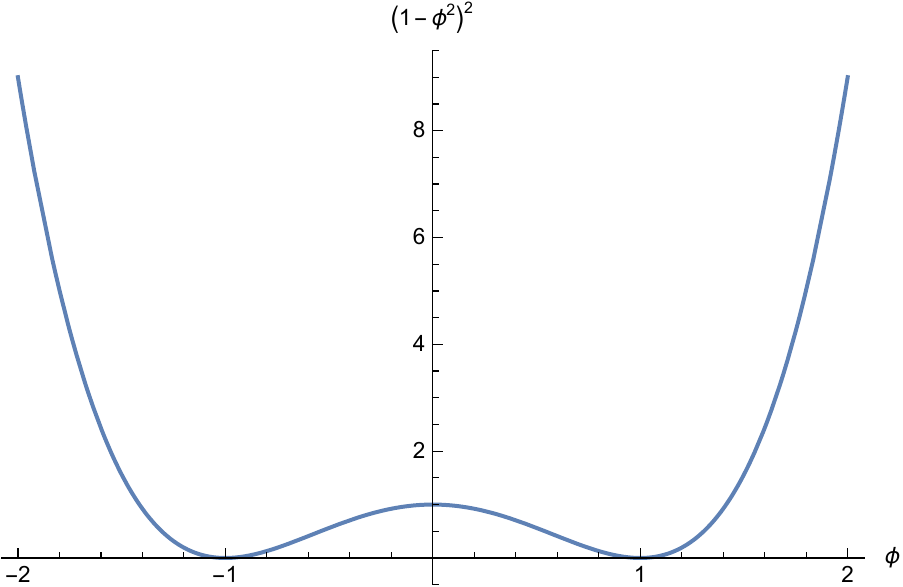}} \longrightarrow \; \raisebox{-0.5\height}{\includegraphics[width=0.4\linewidth]{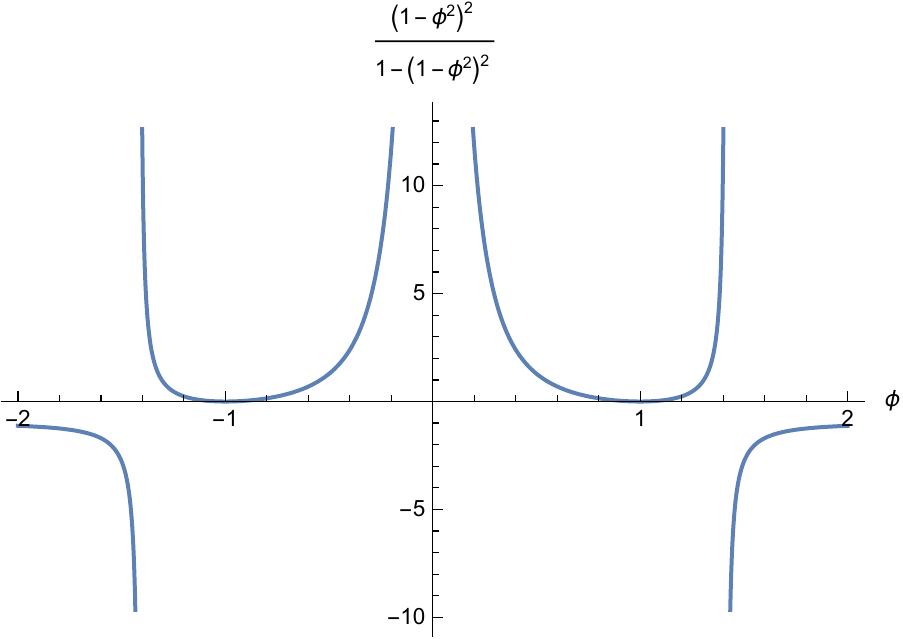}}
\end{align}
Now there is a pole at $\phi = 0$, so the two wells have become separated by an infinite potential barrier. Again, one might wonder what has happened to the Hilbert space. Is there still a complete basis of eigenfunctions, but now localized to each of the disconnected wells?

The above examples are presented only in the context of ordinary quantum mechanics without any supersymmetry. However, one might hope that the presence of some SUSY might be useful in learning about the fate of the Hilbert space after such a deformation. For instance, the spectrum of ground states in a supersymmetric theory exhibits a great deal of structure and one can extract data about it using index-like quantities. Is there some calculation in SUSY-QM theory which is sensitive to the fact that the two ground states in the double well (\ref{double_well}) may have been ``cut off'' from one another in the deformed theory? It would be exciting to gain a better understanding of the Hilbert space of these deformed quantum mechanics theories and to understand whether $\TT$ or $f(\mathcal{Q}, \calQbar)$ indeed has effects on the infrared structure of the kind described here.

\subsubsection*{\ul{\it Relation to Supersymmetric SYK}}

We mention one final future direction, along the lines of the previously mentioned question about the relationship of this deformation with super-BF theory, but which is also related to the issue of defining our $f ( \mathcal{Q}, \calQbar )$ deformation in quantum mechanics with a potential.

It is well-known that the Schwarzian or particle-on-a-group theory is also related to the SYK model of Majorana fermions with random all-to-all interactions \cite{kitaev_talk,PhysRevLett.70.3339}. The SYK model has a supersymmetric extension \cite{Fu:2016vas,Murugan:2017eto,Peng:2017spg}; for an (incomplete) collection of related works on the SYK model and supersymmetry, see \cite{Kanazawa:2017dpd,Yoon:2017gut,Hunter-Jones:2017raw,Hunter-Jones:2017crg,Narayan:2017hvh,Forste:2017apw,Garcia-Garcia:2018ruf,Kato:2018kop,Sun:2019yqp,Behrends:2019sbd,Berkooz:2020xne,Peng:2020euz,Gates:2021jdm,Peng:2016mxj,Li:2017hdt} and references therein.

The application of $\TT$-like deformations in quantum mechanics to the non-supersymmetric SYK model was carried out in \cite{Gross:2019uxi} (see also \cite{He:2021dhr}). In that case, after shifting the ground state energy of the model by a constant $E_0$, it was pointed out that there are two choices for how to perform the deformation:
\begin{enumerate}
    \item First perform the average over disorder in the undeformed model, and then deform the Hamiltonian by the desired $\TT$ or $f(H)$ operator.
    
    \item Begin by deforming the Hamiltonian by some $f(H)$ operator and then perform the disorder average in the deformed theory.
\end{enumerate}
The authors of \cite{Gross:2019uxi} point out that it is easier to do the former, since if one first deforms the Hamiltonian then this procedure will introduce higher powers of the disorder which makes the resulting disorder average difficult. Although the latter provides  a microscopic picture of physics.

It would be interesting to carry out a version of this analysis in the supersymmetric setting using the techniques developed in the present work. To do this, one should use a presentation of the supersymmetric SYK action which is written directly in $\mathcal{N} = 1$ or $\mathcal{N} = 2$ superspace, such as those developed in \cite{Fu:2016vas,Bulycheva:2018qcp}. For concreteness, let us focus on the $\mathcal{N} = 2$ case. The degrees of freedom for the $\mathcal{N} = 2$ super-SYK model are packaged into chiral superfields $\Psi, \Psib$ which obey the constraints
\begin{align}
    \Dbar \Psi = D \Psib = 0 \, .
\end{align}
The Lagrangian is a sum of a kinetic $F$-term plus a holomorphic superpotential:
\begin{gather}
    L = \int \, d \thetab \, \mathcal{A}_{\text{kin}} + \left( \int d \theta \, \mathcal{A}_{\text{potential}} + \text{c.c} \right) \, , \nonumber \\
    \mathcal{A}_{\text{kin}} = \Psib D_i \Psi \, \qquad \mathcal{A}_{\text{potential}} = C_{i_1 \cdots i_k} \Psi_{i_1} \cdots \Psi_{i_k} \, .
\end{gather}
To study the appropriate $\TT$-type deformation of such a superspace Lagrangian, one would therefore need to generalize the analysis presented in this work to allow for fermionic superfields and potentials. One could also attempt to understand deformations of SYK via the dimensional reductions of appropriate two-dimensional field theories \cite{Turiaci:2017zwd,Murugan:2017eto}, or to investigate $\TT$-like deformations in related disordered supersymmetric models \cite{Chang:2018sve,Chang:2021fmd,Chang:2021wbx}.

\vspace{10pt}

\noindent We hope to return to some of these interesting questions in future works.

\section*{Acknowledgements}

We would like to thank Per Kraus, Ruben Monten, Mukund Rangamani, Savdeep Sethi, and Gabriele Tartaglino-Mazzucchelli for helpful discussions. S.E. is supported from the Bhaumik Institute. C. F. is supported by U.S. Department of Energy grant DE-SC0009999 and by funds from the University of California. H.-Y.S. is supported from the Simons Collaborations on Ultra-Quantum Matter, which is a grant from the Simons Foundation (651440, AK). Z.S. is supported from the US Department of Energy (DOE) under cooperative research agreement DE-SC0009919 and Simons Foundation award No. 568420.

%\newpage
\appendix

\section{Change of Coordinates to Complex Supercharges}\label{app:change_to_complex}

In this Appendix, we carry out the change of variables to express our SUSY-QM deformation $f(\mathcal{Q}_+, \mathcal{Q}_-)$ of (\ref{scsquare_reduction_final}) in complex coordinates, ultimately arriving at the expression (\ref{scsquare_reduction_final_complex}) for $f(\mathcal{Q}, \calQbar)$. This is a straightforward application of the change of variables described in equations (\ref{theta_complex_change_of_variables}) - (\ref{ddbar_algebra}) of Section \ref{sec:conventions}, but because it involves some on-shell manipulations we have moved the calculation to this Appendix to avoid cluttering the main body.

We shift to complex supercovariant derivatives via
\begin{align}
    D = \frac{1}{\sqrt{2}} \left( D_+ + i D_- \right)  \, , \quad \Dbar = \frac{1}{\sqrt{2}} \left( D_+ - i D_- \right) \, , 
\end{align}
and similarly rotate the supercurrents via
\begin{align}
    \mathcal{Q} = \frac{1}{\sqrt{2}} \left( \mathcal{Q}_- + i \mathcal{Q}_+ \right)   \, , \quad \calQbar = \frac{1}{\sqrt{2}} \left( \mathcal{Q}_- - i \mathcal{Q}_+ \right) \, .
\end{align}
Note that since $\mathcal{Q}_{\pm}$ are fermionic, one has
\begin{align}\label{QQbar_conversion}
    \mathcal{Q} \calQbar &= \frac{1}{2} \left( \mathcal{Q}_-^2 - i \mathcal{Q}_- \mathcal{Q}_+ + i \mathcal{Q}_+ \mathcal{Q}_- + \mathcal{Q}_+^2 \right) \nonumber \\
    &= i \mathcal{Q}_+ \mathcal{Q}_- \, .
\end{align}
Next we compute the supercovariant derivatives. The combination $\Dbar \mathcal{Q}$ is
\begin{align}
    \Dbar \mathcal{Q} &= \frac{1}{2} \left( D_+ - i D_- \right) \left( \mathcal{Q}_- + i \mathcal{Q}_+ \right) \nonumber \\
    &= \frac{1}{2} \Big[ D_+ \mathcal{Q}_- + i D_+ \mathcal{Q}_+ - i D_- \mathcal{Q}_- + D_- \mathcal{Q}_+  \Big]  \, , 
\end{align}
or after using the conservation equation $D_+ \mathcal{Q}_- + D_- \mathcal{Q}_+ = 0$ and the on-shell condition that $D_+ \mathcal{Q}_+ = - D_- \mathcal{Q}_-$,
\begin{align}
    \Dbar \mathcal{Q} = i D_+ \mathcal{Q}_+ \, .
\end{align}
Likewise,
\begin{align}
    D \calQbar &= \frac{1}{2} \left( D_+ + i D_- \right) \left( \mathcal{Q}_- - i \mathcal{Q}_+ \right) \nonumber \\
    &= \frac{1}{2} \left( D_+ \mathcal{Q}_- - i D_+ \mathcal{Q}_+ + i D_- \mathcal{Q}_- + D_- \mathcal{Q}_+ \right) \, ,
\end{align}
and again this can be written on-shell as
\begin{align}\label{DpQp_to_Dbar_Q}
    D \calQbar = - i D_+ \mathcal{Q}_+ .
\end{align}
Thus we see that the new complex supercurrents satisfy the conservation equation $\Dbar \mathcal{Q} + D \calQbar = 0$, since
\begin{align}\label{new_complex_conservation}
    \Dbar \mathcal{Q} + D \calQbar = i D_+ \mathcal{Q}_+ - i D_+ \mathcal{Q}_+ = 0
\end{align}
when the equations of motion are satisfied.

We now return to the expression $f(\mathcal{Q}_+, \mathcal{Q}_-)$ defining our deformation, which can now be written in terms of complex coordinates as
\begin{align}\label{first_step_to_QQbar}
    \int \, dt \, d \theta^+ \, d \theta^- \, \frac{\mathcal{Q}_+ \mathcal{Q}_-}{4 \lambda D_+ \mathcal{Q}_+ + 1} = \int \, dt \, d \theta^+ \, d \theta^- \, \frac{-i \mathcal{Q} \calQbar}{- 4 i \lambda \Dbar \mathcal{Q} + 1 } \, .
\end{align}
We would now like to eliminate the factors of $i$ that have appeared in (\ref{first_step_to_QQbar}). One factor arises from the change of measure via $d \theta \, d \thetab = i \, d \theta^+ \, d \theta^-$. A second factor arises because, as pointed out in the discussion below equation (\ref{susy_Q_def}), there is a relative factor of $i$ arising between the natural expressions appearing in the Noether procedures which define $\mathcal{Q}, \calQbar$ as opposed to $\mathcal{Q}_+, \mathcal{Q}_-$. Therefore, to obtain an appropriate matching, we will re-scale
\begin{align}
    \mathcal{Q} \longrightarrow - i \mathcal{Q} \, , \qquad \calQbar \longrightarrow - i \calQbar \, .
\end{align}
After incorporating these two factors, we find
\begin{align}
    \int \, dt \, d \theta^+ \, d \theta^- f ( \mathcal{Q}_+ , \mathcal{Q}_- ) = \int \, dt \, d \theta \, d \thetab \frac{\mathcal{Q} \calQbar}{- 4 \lambda \Dbar \mathcal{Q} + 1} \, .
\end{align}
Finally, we scale out an overall factor of $\frac{1}{2}$ to write
\begin{align}
    f ( \mathcal{Q}_+ , \mathcal{Q}_- ) &\sim \frac{\mathcal{Q} \calQbar}{\frac{1}{2} - 2 \lambda \Dbar \mathcal{Q}} \, \nonumber \\
    &\equiv f ( \mathcal{Q} , \calQbar ) \, ,
\end{align}
where $\sim$ indicates proportionality on-shell (as we have used conservation equations to relate $D_+ \mathcal{Q}_+$ to $\Dbar \mathcal{Q}$). We chose to rescale by this prefactor in order to make the right side more closely match equation (\ref{gross_flow_eqn}). This is the form quoted in equation (\ref{scsquare_reduction_final_complex}).

\section{Dimensional Reduction without Trace Flow Equation}\label{app:no_trace_flow}

As we have pointed out in the main body of this paper, the deformation (\ref{gross_flow_eqn}) is very convenient for deforming quantum mechanical theories that descend from $2d$ CFTs via dimensional reduction. However, for theories with a potential, the trace flow equation \eqref{trace_flow} fails and we cannot use this expression for the reduced $\TT$ deformation. In this case, our only recourse is to directly study the $\TT$-deformed field theory in two dimensions, then compactify one spatial direction on a circle and truncate to the lowest Fourier mode.

In this Appendix, we will obtain the Hamiltonian for such a theory by first solving the $2d$ flow equation and then performing the circle compactification only at the final step. Suppose we begin with an undeformed Lagrangian
\begin{align}
    \mathcal{L}_E ( \lambda = 0 , \phi ) = \frac{1}{2} \partial^\mu \phi \partial_\mu \phi + V (\phi ), 
\end{align}
with a positive sign on the potential because we work in Euclidean signature for now. The deformed Lagrangian at finite $\lambda$ appears in equation (2.8) of \cite{Bonelli:2018kik} (see also \cite{Cavaglia:2016oda}) as
\begin{align}
    \mathcal{L}_E ( \lambda, \phi ) = - \frac{1}{2 \lambda} \left(\frac{1 - 2 \lambda V}{1 - \lambda V}\right) + \frac{1}{2 \lambda} \sqrt{ \left(\frac{1 - 2 \lambda V}{1 - \lambda V}\right)^2 + 2 \lambda \left(\frac{ \partial^\mu \phi \partial_\mu \phi + 2 V }{1 - \lambda V} \right) } .
\end{align}
Again, here the metric appearing in the $\partial^\mu \phi \partial_\mu \phi$ contraction is $\tensor{\delta}{^\mu_\nu}$ because we are in Euclidean signature. The prescription for rotating back to Minkowski signature is to multiply the Lagrangian by an overall minus sign, then to invert the sign on the time derivative of $\phi$, giving
\begin{align}
    \mathcal{L}_M ( \lambda, \phi ) = \frac{1}{2 \lambda} \left( \frac{1 - 2 \lambda V}{1 - \lambda V} \right) - \frac{1}{2 \lambda} \sqrt{ \left(\frac{1 - 2 \lambda V}{1 - \lambda V}\right)^2 + 2 \lambda \bigg(\frac{ \phi^{\prime 2}  - \dot{\phi}^2 + 2 V }{1 - \lambda V}\bigg) } .
    \label{deformed_minkowski}
\end{align}
Here we used $\dot{\phi} = \frac{\partial \phi}{\partial t}$ and $\phi' = \frac{\partial \phi}{\partial x}$. We can study the behavior of $\mathcal{L}_M$ in a few limits:
\begin{align}
    \mathcal{L}_M ( \lambda \to 0 , \phi ) &= \frac{1}{2} \dot{\phi}^2 - \frac{1}{2} \phi^{\prime 2} - V ( \phi ) , \nonumber \\
    \mathcal{L}_M ( \lambda, \phi ) \Big\vert_{\dot{\phi} = \phi' = 0} &=- \frac{V(\phi)}{1 - \lambda V(\phi)} , \nonumber \\
    \mathcal{L}_M ( \lambda , \phi ) \Big\vert_{V=0} &= \frac{1}{2 \lambda} \bigg( 1 - \sqrt{1 + 2 \lambda \big( \phi^{\prime 2} - \dot{\phi}^2 \big) } \bigg) \sim \mathcal{L}_{\text{Nambu-Goto}} .
\end{align}
To write this as a Hamiltonian, we will resort to Legendre transform. The conjugate momentum to $\phi$ is
\begin{align}
    \Pi = \frac{\partial \mathcal{L}}{\partial \dot{\phi}} = \frac{\dot{\phi}}{\sqrt{1 - 2 \lambda \left( 1 - \lambda V \right) \big( \dot{\phi}^2 - \phi^{\prime 2} \big) } } .
    \label{conjugate_momentum}
\end{align}
The relation (\ref{conjugate_momentum}) can be inverted to find
\begin{align}
    \dot{\phi} = \Pi \cdot \sqrt{ \frac{ 1 + 2 \lambda ( 1 - \lambda V ) \phi^{\prime 2}}{1 + 2 \lambda ( 1 -  \lambda V ) \Pi^2 } } .
    \label{phidot_to_pi}
\end{align}
The Hamiltonian is then defined by
\begin{align}
    \mathcal{H} = \Pi \dot{\phi} - \mathcal{L} , 
\end{align}
after replacing all instances of $\dot{\phi}$ with $\Pi$ using (\ref{phidot_to_pi}). This gives
\begin{align}
    \hspace{-10pt}\mathcal{H} = \phi^{\prime 2} \cdot &\sqrt{ \frac{1 + 2 \lambda ( 1 -  \lambda V ) \Pi^2 }{1 + 2 \lambda ( 1 - \lambda V ) \phi^{\prime 2} }} \\
    &+ \frac{1}{2 \lambda ( 1 - \lambda V ) } \cdot \sqrt{ \frac{1 + 2 \lambda ( 1 -  \lambda V ) \Pi^2 }{1 + 2 \lambda ( 1 - \lambda V ) \phi^{\prime 2} }} + \frac{V}{1 - \lambda V} - \frac{1}{2 \lambda ( 1 - \lambda V ) } \, .
    \label{2d_ham}
\end{align}
The dependence on $\Pi^2$ is masked by the terms involving square roots. Near $\lambda = 0$, (\ref{2d_ham}) is
\begin{align}
    \mathcal{H} = \frac{1}{2} \Pi^2 + \frac{1}{2} \phi^{\prime 2} + V ( \phi ) + \mathcal{O} ( \lambda ) , 
\end{align}
which is the expected Hamiltonian for a scalar field with a potential.

Next we would like to put the coordinate $x$ on a circle of radius $R$, Fourier-expand the $x$-dependence of $\phi(x,t)$, and integrate the Hamiltonian $H$ over the circle to obtain a quantum-mechanical Hamiltonian associated with the modes $\phi^{(n)} ( t )$. We expand $ \phi ( x, t ) $ in modes as
\begin{align}
    \phi ( x, t ) = \sum_{n=0}^{\infty} \left( \phi^{(n)}_c ( t ) \cos \left( \frac{2  \pi n}{R} \, x \right) + \phi^{(n)}_s ( t ) \sin \left( \frac{2  \pi  n}{R} \, x \right) \right) . 
    \label{phi_fourier}
\end{align}
Inserting (\ref{phi_fourier}) into (\ref{2d_ham}) and integrating over the circle would, in principle, leave us with a Hamiltonian for infinitely many interacting particles $\phi^{(n)}_{c} ( t )$ and $\phi^{(n)}_{s}$ in quantum mechanics. Such an analysis seems intractable in general, so for simplicity, let us restrict to the zero-momentum sector\footnote{Restricting to the zero-momentum sector also allowed us to use the implicit solution (\ref{2d_tt_burgers_implicit}) to the inviscid Burgers' equation, which was pointed out in \cite{Cavaglia:2016oda}.}
\begin{align}
    \phi ( x , t ) \equiv \phi ( t ) .
\end{align}
This gives us a Hamiltonian
\begin{align}
    H = \frac{ \sqrt{ 1 +2 \lambda ( 1 -  \lambda V ) \Pi^2 }}{2 \lambda \left( 1 - \lambda V \right)} + \frac{V}{1 - \lambda V} - \frac{1}{2 \lambda ( 1 - \lambda V ) } .
    \label{reduced_hamiltonian}
\end{align}
For small $\lambda$, \eqref{reduced_hamiltonian} looks like
\begin{align}
    H = \frac{1}{2} \Pi^2 + V ( \phi ) + \lambda \left( V ( \phi )^2 - \frac{1}{4} \Pi^4 \right) + \frac{1}{4} \lambda^2 \left( 4 V ( \phi )^3 + \Pi^4 V(\phi) + \Pi^6 \right) + \mathcal{O} ( \lambda^3 ) , 
\end{align}
The leading term is the usual Hamiltonian $H = \frac{p^2}{2m} + V$ if we identify $p = \Pi$, $m = 1$. But this usual Hamiltonian receives an infinite series of corrections, which affect both the kinetic and potential terms (and mix them). The purely kinetic part of \eqref{reduced_hamiltonian} reduces to
\begin{align}
    H \Big\vert_{V = 0} = \frac{-1 + \sqrt{1 + 2 \lambda \Pi^2} }{2 \lambda} , 
\end{align}
which is a $(0+1)$-dimensional analogue of the Nambu-Goto action. If we alternatively set $\Pi=0$ and consider the pure potential piece, from \eqref{reduced_hamiltonian} we find
\begin{align}\label{V_over_one_minus_lambda_V}
    H \Big\vert_{\Pi = 0} = \frac{V(\phi)}{1 - \lambda V(\phi)} \, .
\end{align}
This looks identical to the result of deforming a pure-potential Hamiltonian by the function $f(H) = H^2$, rather than the more complicated operator (\ref{gross_flow_eqn}) which is equivalent to $\TT$ for theories that descend from deformations of $2d$ CFTs. To be explicit, if we consider the flow equation
\begin{align}
    \frac{\partial H}{\partial \lambda} = H^2 \, ,
\end{align}
with initial condition $H(0) = H_0$, then the solution is trivially
\begin{align}
    H ( \lambda ) = \frac{H_0}{1 - \lambda H_0} \, .
\end{align}
At low momentum where the kinetic term can be neglected and the undeformed Hamiltonian is approximately the pure potential $H = V(\phi)$, this is exactly (\ref{V_over_one_minus_lambda_V}). In particular, the Hamiltonian diverges when $V(\phi) = \frac{1}{\lambda}$. This is purely a classical statement about the solution to an $f(H)$-type flow equation, which is not necessarily indicative of the structure of the quantum theory, but we will nonetheless make some speculative remarks about this pole in Section \ref{sec: Discussion}.

%\newpage
%\bibliographystyle{amsunsrt-ensp}
\bibliographystyle{utphys}
\bibliography{ref}

\end{document}